\documentclass[reprint,twocolumn,superscriptaddress,showpacs,showkeys,amsmath,amssymb,floatfix,aps,prb]{revtex4-1}

\usepackage{graphicx}
\usepackage{dcolumn}
\usepackage{bm}
\usepackage{amsmath}
\usepackage{placeins}
\usepackage{color}
\usepackage{hyperref}
\usepackage{multirow}

\begin{document}

\title{\texorpdfstring{Single and bilayer graphene on the topological insulator Bi$_2$Se$_3$: Electronic and spin-orbit properties from first-principles}{}}

\author{Klaus Zollner}
	\email{klaus.zollner@physik.uni-regensburg.de}
	\affiliation{Institute for Theoretical Physics, University of Regensburg, 93040 Regensburg, Germany}	
\author{Jaroslav Fabian}
	\affiliation{Institute for Theoretical Physics, University of Regensburg, 93040 Regensburg, Germany}
\date{\today}

\begin{abstract}
We present a detailed study of the electronic 
and spin-orbit properties of single and bilayer graphene
in proximity to the topological insulator Bi$_2$Se$_3$. 
Our approach is based on first-principles calculations, combined with symmetry
derived model Hamiltonians that capture the low-energy band properties.
We consider single and bilayer graphene on 1--3 quintuple layers of Bi$_2$Se$_3$ and extract
orbital and proximity induced spin-orbit coupling (SOC) parameters. 
We find that graphene gets significantly hole doped (350 meV), but the linear dispersion is preserved.
The proximity induced SOC parameters are about 1 meV in magnitude, and are of valley-Zeeman type.
The induced SOC depends weakly on the number of quintuple layers of Bi$_2$Se$_3$.
We also study the effect of a transverse electric field, that is applied across heterostructures
of single and bilayer graphene above 1 quintuple layer of Bi$_2$Se$_3$.
Our results show that band offsets, as well as proximity 
induced SOC parameters can be tuned by the field. 
Most interesting is the case of bilayer graphene, in which  the band gap, originating from the intrinsic dipole of the heterostructure, can be closed and 
reopened again, with inverted band character. The switching of the strong proximity SOC from 
the conduction to the valence band realizes a spin-orbit valve. 
Additionally, we find a giant increase of the proximity induced SOC of about 200\%, when we decrease the interlayer distance between graphene and Bi$_2$Se$_3$ by only 10\%. 
Finally, for a different substrate material Bi$_2$Te$_2$Se, band offsets are significantly different, with the graphene Dirac point located at the Fermi level, while the induced SOC strength stays similar in magnitude compared to the Bi$_2$Se$_3$ substrate.
\end{abstract}

\pacs{}
\keywords{spintronics, graphene, heterostructures, proximity spin-orbit coupling, topological insulator}
\maketitle

\section{Introduction}
Van der Waals (vdW) heterostructures \cite{Geim2013:Nat,Novoselov2016:Sci,Duong2017:ACS} 
and emerging proximity effects \cite{Zutic2019:MT} 
are an ideal platform to 
induce tailored properties in two-dimensional (2D) materials. 
Prominent 2D material examples are semimetallic single layer graphene \cite{Neto2009:RMP} (SLG), 
semiconducting transition-metal dichalcogenides \cite{Wang2012:NN} (TMDCs), 
and insulating hexagonal boron-nitride \cite{Catellani1987:PRB} (hBN). 
Recently, also superconductors \cite{Frindt1972:PRL} (NbSe$_2$), 
and ferro- and antiferromagnets \cite{Dillon1965:JAP, Huang2017:Nat, McGuire2015:CM, Wiedenmann1981:SSC,Carteaux1995:JPCM, Gong2017:Nat}
(CrI$_3$, Cr$_2$Ge$_2$Te$_6$, MnPSe$_3$)
have been added to the list of 2D materials. 
Within this ever-expanding field of vdW structures, there already are subfields, 
such as valleytronics \cite{Langer2018:Nat,Zhong2017:SA,Chellappan2018:NF,Schaibley2016:NRM}, 
straintronics \cite{Lin2018:APL, Roldan2015:JPCM,Fang2018:PRB}, twistronics \cite{David2019:PRB,Li2019:PRB},
and spintronics \cite{Zutic2004:RMP,Han2014:NN,Fabian2007:APS}, wherein
several major achievements have been made, for example
optical spin injection \citep{Avsar2017:ACS} in SLG or 
tunable valley polarization in a TMDC \cite{Seyler2018:NL}, which are only possible
due to vdW heterostructures and proximity effects. 

Another important large class of materials are the three-dimensional (3D) topological 
insulators such as \cite{Zhang2009:NP} Bi$_2$Se$_3$, Bi$_2$Te$_3$, and Sb$_2$Te$_3$, 
which are also layered crystals, consisting of quintuple layers (QLs) of 
alternating chalcogen (Se, Te) and pnictogen (Bi, Sb) atoms, which are held together by vdW forces. 
However, the characteristic Dirac states with spin-momentum locking \cite{Hsieh2009:Nat} emerge only when the top and bottom surfaces of the topological insulator decouple, occurring at already 5--6 QLs, as demonstrated
by angle resolved photoemission spectroscopy \cite{Zhang2010:NP} and 
first-principles calculations \cite{Liu2010:PRB,Yazyev2010:PRL,Park2010:PRL}.
Since each QL is about 1~nm in thickness, these materials are in between the 2D and 3D regime, 
depending on how many QLs one investigates. 
Nevertheless, they are important for practical applications \cite{Tian2017:Mat}, due to their 
topologically protected \cite{Hasan2010:RMP, Zhang2009:PRL} and well conducting surface states \cite{Koirala2015:NL}, and for proximity induced phenomena \cite{Song2018:NL, Khokhriakov2018:SA,Jafarpisheh2018:PRB, Zhang2014:PRL}, 
since strong SOC is present.
When the topological insulators act as a substrate, the 2D regime (1--2 QLs) is sufficient, as proximity effects are of short range nature.

Recently, the interface engineering of 2D materials has become an important topic \cite{Hesjedal2016:NM,Zhao2017:AFM}. 
Experimentalists and theorists are searching for material combinations with novel properties. 
Graphene, due to its extremely high electron 
mobility \cite{Banszerus2015:SA} and intrinsically small SOC \cite{Gmitra2009:PRB}, is perfectly suited for spintronics. 
In addition, this monolayer carbon sheet can be efficiently manipulated by short range proximity effects.

One can induce strong SOC, as well as magnetism in SLG \cite{Gmitra2015:PRB, Gmitra2016:PRB,Song2018:NL, Zollner2016:PRB, Jafarpisheh2018:PRB,Khokhriakov2018:SA}.
Similar to a TMDC \cite{Gmitra2015:PRB, Gmitra2016:PRB}, a topological insulator strongly enhances the rather weak 
intrinsic SOC of SLG from 10~$\mu$eV \cite{Gmitra2009:PRB}, to about 1--2~meV 
\cite{ Song2018:NL, Jafarpisheh2018:PRB}. 
Phase coherent transport measurements of SLG on Bi$_{1.5}$Sb$_{0.5}$Te$_{1.7}$Se$_{1.3}$ have shown
Dyakonov-Perel type spin relaxation with proximity induced SOC of at least 2.5~meV \cite{Jafarpisheh2018:PRB}. 
First-principles calculations of SLG on Bi$_2$Se$_3$ have found either pure intrinsic or valley-Zeeman type SOC in the meV range, depending on the twist angle \cite{Song2018:NL}. As a consequence of the large induced SOC in SLG, the spin lifetimes of 
electrons significantly decrease \cite{Cummings2017:PRL, Song2018:NL}, from nanoseconds
down to the picosecond range. Giant spin lifetime 
anisotropies, the ratio of out-of-plane to in-plane spin lifetimes, 
can be achieved \cite{Cummings2017:PRL, Song2018:NL}.

Bilayer graphene (BLG) is even more interesting, since only the
layer closest to the proximitizing material gets modified, allowing for highly 
efficient tuning of the proximity properties by gating and doping. 
Recent studies have shown short range proximity induced exchange or SOC in BLG on Cr$_2$Ge$_2$Te$_6$
or WSe$_2$ \cite{Zollner2017:NJP,Gmitra2017:PRL, Khoo2017:NL}. 
Due to the unique and tunable low energy band structure of BLG, all-electrical control of 
spin relaxation and polarization can be achieved in such heterostructures. 
The proposed spin-orbit and exchange valve effects \cite{Zollner2017:NJP,Gmitra2017:PRL} in proximitized BLG can lead to new opportunities for spintronics devices. 
So far, there are only few experimental studies \cite{Zalic2017:PRB, Song2010:APL, Dang2010:NL, Steinberg2015:PRB} 
of BLG on Bi$_2$Se$_3$. 
However, there is no detailed experimental and theoretical study of the electronic and spin-orbit properties of BLG interfaced with topological insulators. 

The open questions we would like to address are as follows. 
In the case of SLG on Bi$_2$Se$_3$, when valley-Zeeman SOC is predicted to be present \cite{Song2018:NL}
for 1 QL, how does the presence of more QLs influence this interesting result? (The result is interesting, since in the topological substrate the spin-orbit fields are in-plane, while the induced valley-Zeeman fields in graphene are out-of-plane). Another important question 
is, what is the influence of the topological insulator on BLG? What are the band offsets, 
doping levels, and orbital and spin-orbit proximity effects? How does the interlayer distance between SLG and Bi$_2$Se$_3$ affect the magnitude of proximity SOC? Also, 
can an electric field tune SOC in SLG and BLG
in proximity to the topological insulator? Strong hole doping of the SLG on Bi$_2$Se$_3$ is predicted \cite{Song2018:NL}.
Can one find a different topological insulator with a better band alignment, such that the graphene Dirac point is near E$_F$?

In this article we investigate these questions using first-principles calculations of SLG and BLG on the 
topological insulator Bi$_2$Se$_3$. 
We study the proximity induced SOC in SLG and BLG, originating from the topological insulator, by varying the number of QLs of Bi$_2$Se$_3$ from 1--3. 
Symmetry-derived low energy tight-binding model Hamiltonians for SLG and BLG are fitted to the first-principles band structures, to extract orbital and 
spin-orbit parameters of the proximitized materials. 
Our results show, that the dispersion of SLG (BLG) is preserved, but strong hole doping appears, as the Dirac point is about 350~meV (200~meV) above the Fermi level. 
The proximity induced SOC is about 1~meV in magnitude, but with opposite sign for A and B sublattice, the so called valley-Zeeman type.
We find that the intrinsic SOC parameters increase by about 10\%, for every QL of Bi$_2$Se$_3$ that we add, up to 3QLs. As proximity effects are short ranged, we expect this increase to saturate.

Furthermore, we study the effect of a transverse electric field on the low energy band parameters, for
1QL of Bi$_2$Se$_3$ proximitizing SLG and BLG. 
The electric field can tune the SOC parameters, which can have significant impact for tuning spin lifetimes in SLG and BLG.
In addition, the surface states of the topological insulator, as well as the Dirac points of SLG and BLG, can be tuned with respect to the Fermi level, by the field.
Most interesting is the BLG case, in which only the low energy conduction band is strongly spin-orbit split for zero field, as a consequence of short range proximity effects, 
and atom and layer localized energy states.
The tuning of the orbital gap of BLG, by gradually increasing the field, 
leads to a gap closing and subsequent reopening, now with a strongly spin-orbit split valence band.  
Consequently, a spin-orbit valve effect can be realized, 
similar to BLG on a TMDC \cite{Gmitra2017:PRL}.
An interlayer distance study, between SLG and the Bi$_2$Se$_3$ substrate, shows 
a giant increase of the proximity induced SOC of about 200\%, when we decrease the interlayer distance by only 10\%.
Furthermore, an atomically modified substrate Bi$_{2}$Te$_{2}$Se, leads to a
significantly different band alignment and enhanced proximity SOC in SLG. 
The Dirac point of SLG, as well as the surface states of the topological insulator, are now located at the Fermi level.
Especially this system holds promise, for the simultaneous study of two very different spin-orbit fields: in-plane spin-momentum locking from the topological insulator and out-of-plane proximity induced spin-orbit field from SLG.

\section{Geometry \& Computational Details}

\begin{figure}[!htb]
 \includegraphics[width=.99\columnwidth]{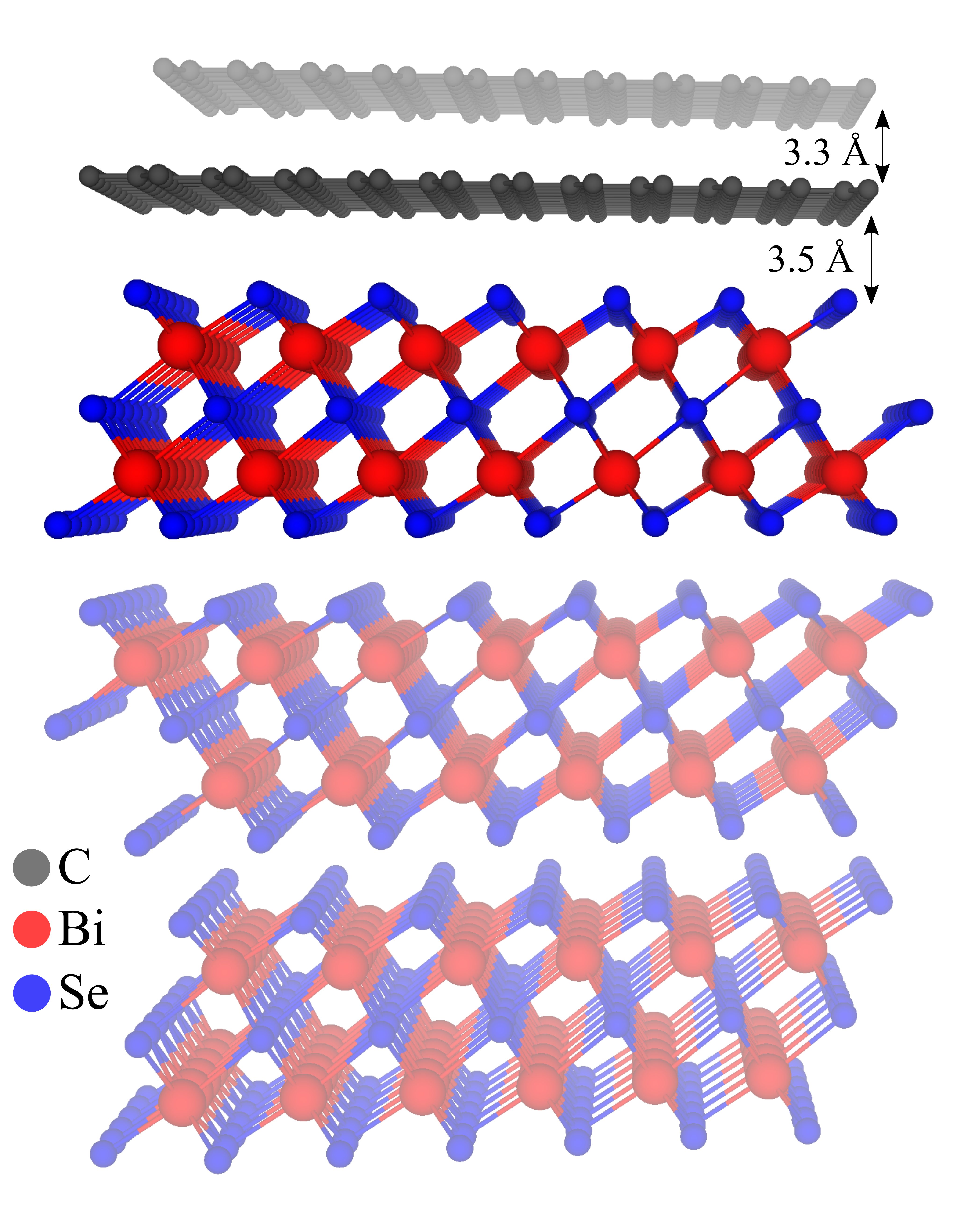}
 \caption{(Color online) Geometry of SLG or BLG above 1--3 QLs of Bi$_2$Se$_3$.
 We highlighted the SLG/1QL structure, which is the minimal geometry we consider.
 Different colors correspond to different atomic species. 
 }\label{Fig:struct}
\end{figure}

For the calculation of SLG and BLG on the topological insulator Bi$_2$Se$_3$, 
we consider $5\times 5$ supercells of SLG and BLG (in Bernal stacking) 
on top of $3\times 3$ supercells of Bi$_2$Se$_3$.
Initial atomic structures are set up with the Atomic Simulation Environment (ASE) \cite{ASE}.
We marginally stretch the lattice constant of graphene \cite{Neto2009:RMP} to $a = 2.486$~\AA~and 
leave the Bi$_2$Se$_3$ lattice constants \cite{Nakajima1963:JPCS} unchanged with $a = 4.143$~\AA~and $c = 28.636$~\AA, using
the atomic parameters $(u,v) = (0.4008, 0.2117)$.
We consider only geometries without relaxation, using interlayer distances of $3.5$~\AA~between the lowest graphene layer and 
the topmost QL of Bi$_2$Se$_3$, in agreement with 
recent studies \cite{Song2018:NL}, and an interlayer distance of $3.3$~\AA~for 
the BLG, in agreement with experiment \cite{Baskin1955:PR}.
In Fig. \ref{Fig:struct} we show the geometry of SLG or BLG on top of 1--3 QLs of Bi$_2$Se$_3$, visualized with VESTA \cite{VESTA}.
Compared to Ref. \cite{Song2018:NL}, we only study what they call the 'large unit cell', where no twist angle between the materials is present.

The electronic structure calculations are performed by 
density functional theory (DFT)~\cite{Hohenberg1964:PRB} with {Quantum ESPRESSO}~\cite{Giannozzi2009:JPCM}.
Self-consistent calculations are performed with the $k$-point sampling of 
$9\times 9\times 1$.
Only for the largest heterostructures, when 3QLs of Bi$_2$Se$_3$ are considered, 
a smaller $k$-point sampling of $6\times 6\times 1$ is used, due to computational limitations.
We use an energy cutoff for charge density of $500$~Ry, and
the kinetic energy cutoff for wavefunctions is $60$~Ry for the relativistic pseudopotentials 
with the projector augmented wave method \cite{Kresse1999:PRB} with the 
Perdew-Burke-Ernzerhof exchange correlation functional \cite{Perdew1996:PRL}.
We also add vdW corrections \cite{Grimme2006:JCC,Barone2009:JCC}, and
Dipole corrections \cite{Bengtsson1999:PRB} are included to 
get correct band offsets and internal electric fields.
In order to simulate quasi-2D systems, we add a vacuum of at least $24$~\AA, 
to avoid interactions between periodic images in our slab geometry.

In contrast to the GW method, the generalized-gradient-approximiation (GGA) used here is certainly not the most accurate available choice to describe 
3D topological insulators \cite{Aguilera2013:PRB, Nechaev2013:PRB, Yazyev2012:PRB}. 
Consequently, the charge transfer between SLG (BLG) and the topological insulator, as well as doping and proximity SOC can be different in more sophisticated calculations.
Still, the GGA does show the important band structure features 
and it allows to make predictions for proximity effects in SLG/Bi$_2$Se$_3$ heterostructures.
Furthermore, the systems we consider are too large, making GW calculations computationally inaccessible. 
In our GGA-based study, we believe that proximity effects can be captured quite nicely on a qualitative and semi-quantitative level. 
For example, calculations \cite{Song2018:NL}, also using GGA have been used to analyze and interpret recent experiments of graphene/topological insulator structures \cite{Jafarpisheh2018:PRB}.

\section{Model Hamiltonian}
It has been shown that a topological insulator induces strong proximity SOC in SLG \cite{Song2018:NL,Jafarpisheh2018:PRB}, on the order of 1~meV.
Depending on the exact geometry (twist angle), either pure intrinsic or valley-Zeeman type SOC can be realized \cite{Song2018:NL}.
The valley-Zeeman SOC has also been observed in SLG/TMDC heterostructures \cite{Gmitra2015:PRB, Gmitra2016:PRB}.
For BLG on a TMDC, even a spin-valve effect is proposed to be present \cite{Gmitra2017:PRL}.
We want to analyze in detail the influence of the topological insulator Bi$_2$Se$_3$ 
on the low energy bands of SLG and BLG, in the previously mentioned (non-twisted) supercell configuration, where valley-Zeeman SOC has been found.

\subsection{Graphene}

The band structure of proximitized SLG can be modeled by 
symmetry-derived Hamiltonians \cite{Kochan2017:PRB, Zollner2016:PRB, Zollner2019:PRB, Gmitra2015:PRB, Gmitra2016:PRB, Sante2019:PRB,Kane2005:PRL}. 
For our heterostructures, the effective low energy Hamiltonian is
\begin{flalign}
\label{Eq:Hamiltonian}
&\mathcal{H}_\textrm{SLG} = \mathcal{H}_{0}+\mathcal{H}_{\Delta}+\mathcal{H}_{\textrm{I}}+\mathcal{H}_{\textrm{R}}+\mathcal{H}_{\textrm{PIA}}+E_D,\\
&\mathcal{H}_{0} = \hbar v_{\textrm{F}}(\tau k_x \sigma_x - k_y \sigma_y)\otimes s_0, \\
&\mathcal{H}_{\Delta} =\Delta \sigma_z \otimes s_0,\\
&\mathcal{H}_{\textrm{I}} = \tau (\lambda_{\textrm{I}}^\textrm{A} \sigma_{+}+\lambda_{\textrm{I}}^\textrm{B} \sigma_{-})\otimes s_z,\\
&\mathcal{H}_{\textrm{R}} = -\lambda_{\textrm{R}}(\tau \sigma_x \otimes s_y + \sigma_y \otimes s_x),\\
&\mathcal{H}_{\textrm{PIA}} = a(\lambda_{\textrm{PIA}}^\textrm{A} \sigma_{+}-\lambda_{\textrm{PIA}}^\textrm{B} 
\sigma_{-})\otimes (k_x s_y - k_y s_x). 
\end{flalign}
Here $v_{\textrm{F}}$ is the Fermi velocity and the in-plane wave vector 
components $k_x$ and $k_y$ are measured from $\pm$K, 
corresponding to the valley index $\tau = \pm 1$.
The Pauli spin matrices are $s_i$, 
acting on spin space ($\uparrow, \downarrow$), and $\sigma_i$ are pseudospin 
matrices, acting on sublattice space (C$_\textrm{A}$, C$_\textrm{B}$), 
with $i = \{ 0,x,y,z \}$. For shorter notation, we introduce $\sigma_{\pm} = \frac{1}{2}(\sigma_z \pm \sigma_0)$. 
The lattice constant of pristine graphene is $a$ and 
the staggered potential gap is $\Delta$.
The parameters $\lambda_{\textrm{I}}^\textrm{A}$ and 
$\lambda_{\textrm{I}}^\textrm{B}$ 
describe the sublattice resolved intrinsic SOC, 
$\lambda_{\textrm{R}}$ stands for the Rashba SOC, 
and $\lambda_{\textrm{PIA}}^\textrm{A}$ and 
$\lambda_{\textrm{PIA}}^\textrm{B}$ are the sublattice 
resolved pseudospin-inversion asymmetry (PIA) SOC parameters. 
The basis states are $|\Psi_{\textrm{A}}, \uparrow\rangle$, 
$|\Psi_{\textrm{A}}, \downarrow\rangle$, $|\Psi_{\textrm{B}}, \uparrow\rangle$, 
and $|\Psi_{\textrm{B}}, \downarrow\rangle$, resulting in 
four eigenvalues $\varepsilon_{1/2}^{\textrm{CB/VB}}$.
Note that the model Hamiltonian describes the dispersion relative to the graphene Dirac point.
For the first-principles results doping can occur, shifting the Fermi level off the Dirac point. 
Therefore we introduce another parameter $E_{\textrm{D}}$, which 
generates a shift of the global model band structure. We call it the Dirac point energy.

\subsection{Bilayer graphene}

We wish to describe the low energy band structure of the proximitized BLG in the vicinity of the K and K' valleys.
Therefore, we introduce the following Hamiltonian derived from symmetry \cite{Konschuh2012:PRB}, where we keep 
only the most relevant terms 
\begin{widetext}
\begin{flalign}
& \mathcal{H}_\textrm{BLG} = \mathcal{H}_{\textrm{orb}} + \mathcal{H}_{\textrm{soc}}+E_D,\\
& \mathcal{H}_{\textrm{orb}} = \begin{pmatrix}
\Delta+V & \gamma_0 f(\bm{k}) & \gamma_4 f^{*}(\bm{k}) & \gamma_1 \\
\gamma_0 f^{*}(\bm{k}) & V & \gamma_3 f(\bm{k}) & \gamma_4 f^{*}(\bm{k}) \\
 \gamma_4 f(\bm{k}) & \gamma_3 f^{*}(\bm{k}) & -V & \gamma_0 f(\bm{k}) \\
 \gamma_1 & \gamma_4 f(\bm{k}) & \gamma_0 f^{*}(\bm{k}) & \Delta-V
\end{pmatrix} \otimes s_0,\\
& \mathcal{H}_{\textrm{soc}} = \begin{pmatrix}
\tau \lambda_{\textrm{I}}^\textrm{A1}s_z & \textrm{i}(\lambda_0+2\lambda_{\textrm{R}})s_{-}^{\tau} & 0 & 0\\
-\textrm{i}(\lambda_0+2\lambda_{\textrm{R}})s_{+}^{\tau} & -\tau \lambda_{\textrm{I}}^\textrm{B1}s_z & 0 & 0\\
0 & 0 & \tau \lambda_{\textrm{I}}^\textrm{A2}s_z & -\textrm{i}(\lambda_0-2\lambda_{\textrm{R}})s_{-}^{\tau} \\
0 & 0 & \textrm{i}(\lambda_0-2\lambda_{\textrm{R}})s_{+}^{\tau} & -\tau \lambda_{\textrm{I}}^\textrm{B2}s_z
\end{pmatrix}.
\end{flalign}
\end{widetext}
We use the linearized version for the nearest-neighbor structural function
$f(\bm{k}) = -\frac{\sqrt{3}a}{2}(k_x-\textrm{i}k_y)$, with the graphene lattice constant
$a$ and the Cartesian components of the wave vector $k_x$ and $k_y$ measured 
from $\pm$K for the valley index $\tau = \pm 1$. 
Parameters $\gamma$ describe intra- and interlayer hoppings of the BLG, 
when the lower (upper) graphene layer is placed in potential $V$ ($-V$). The parameter
$\Delta$ describes the asymmetry in the energy shift of the bonding and antibonding states.
The parameters $\lambda_{\textrm{I}}$ 
describe the intrinsic SOC of the corresponding layer and sublattice.
The combination of parameters $\lambda_0$ and $\lambda_{\textrm{R}}$ describe 
the global and local breaking of space inversion symmetry. 
For a more detailed description of the parameters, 
we refer the reader to Ref. \cite{Konschuh2012:PRB}.
The basis is $|\Psi_{\textrm{A1}}, \uparrow\rangle$, $|\Psi_{\textrm{A1}}, \downarrow\rangle$, 
$|\Psi_{\textrm{B1}}, \uparrow\rangle$, $|\Psi_{\textrm{B1}}, \downarrow\rangle$, 
$|\Psi_{\textrm{A2}}, \uparrow\rangle$, $|\Psi_{\textrm{A2}}, \downarrow\rangle$, 
$|\Psi_{\textrm{B2}}, \uparrow\rangle$, and $|\Psi_{\textrm{B2}}, \downarrow\rangle$.
Similar to the SLG case, doping can occur, and again we denote the energy shift by the Dirac point energy $E_{\textrm{D}}$.

\section{Band structure, fit results, and spin-orbit fields}

\begin{figure*}[!htb]
 \includegraphics[width=.99\textwidth]{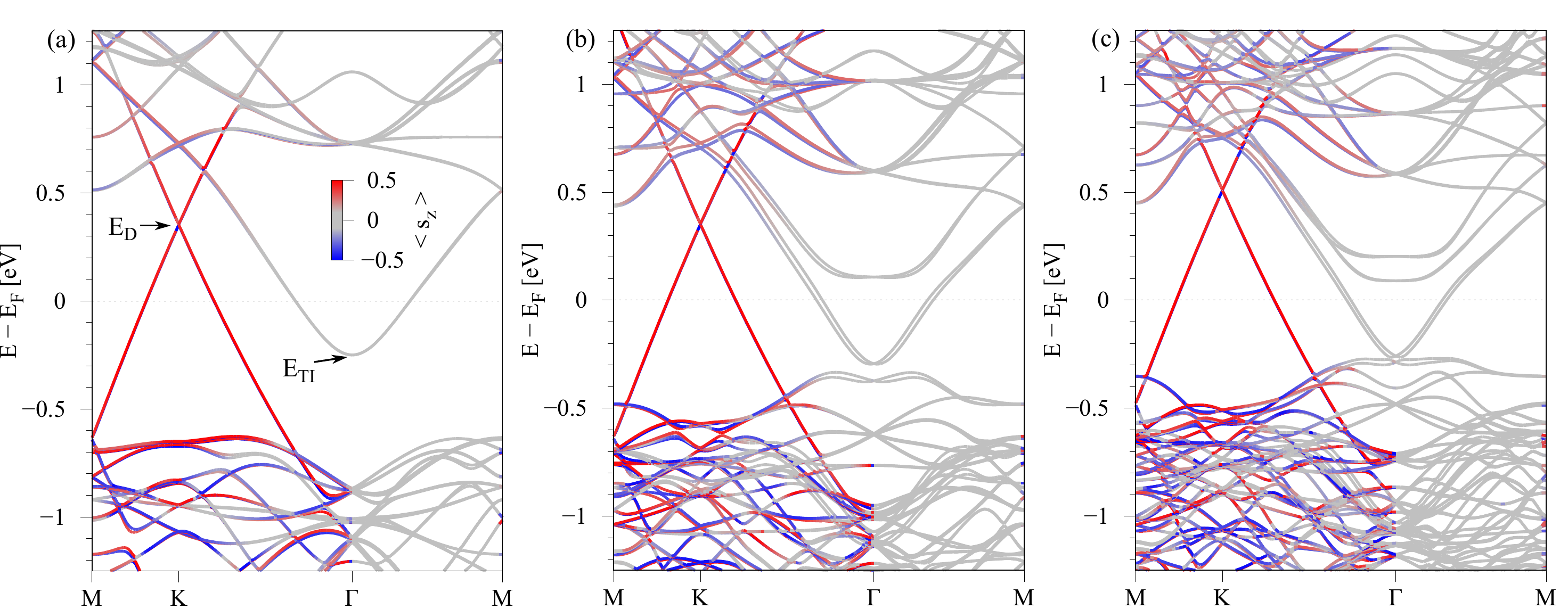}
 \caption{(Color online) Calculated band structures of SLG on (a) one (b) two and
 (c) three QLs of Bi$_2$Se$_3$. The color is the $s_z$ expectation value.
 In (a), we define the Dirac point energy $E_{\textrm{D}}$ and the \textit{doping energy} of the topological insulator 
 $E_{\textrm{TI}}$.
 }\label{Fig:bandstructures}
\end{figure*}

Here, we analyze the dependence of the proximity SOC in SLG and BLG on the number of QLs. 
We show the full calculated band structures, as well as a 
zoom to the low energy bands originating from SLG or BLG, being
proximitized by the topological insulator, and fit the individual model Hamiltonians. 
The orbital and spin-orbit fit parameters are summarized in a tabular form.

\subsection{Graphene}

\begin{figure}[!htb]
 \includegraphics[width=.99\columnwidth]{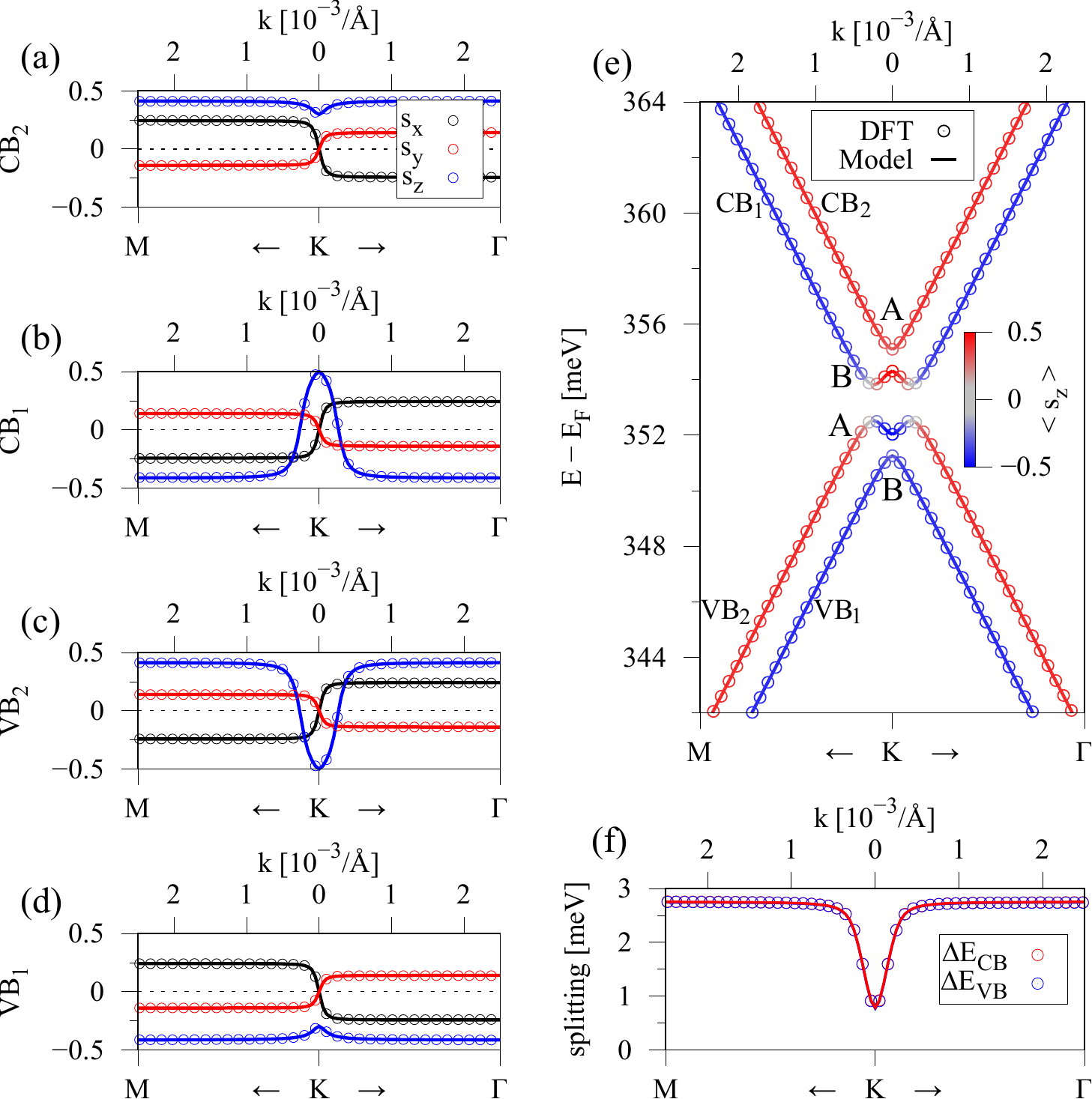}
 \caption{(Color online) Calculated low energy band properties (symbols) for SLG/Bi$_2$Se$_3$ for 1QL, with 
 a fit to the model Hamiltonian $\mathcal{H}_\textrm{SLG}$ (solid lines).
 (a)-(d) The spin expectation values of the four low energy bands. 
 (e) The low energy band structure of proximitzed SLG. The color is the $s_z$ expectation value. 
 Letters A and B indicate the pseudospin of the bands at the K point. 
 (f) The splitting of the valence (conduction) band in blue (red).
 }\label{Fig:fit_1QL}
\end{figure}

\begin{figure}[!htb]
\centering
 \includegraphics[width=.99\columnwidth]{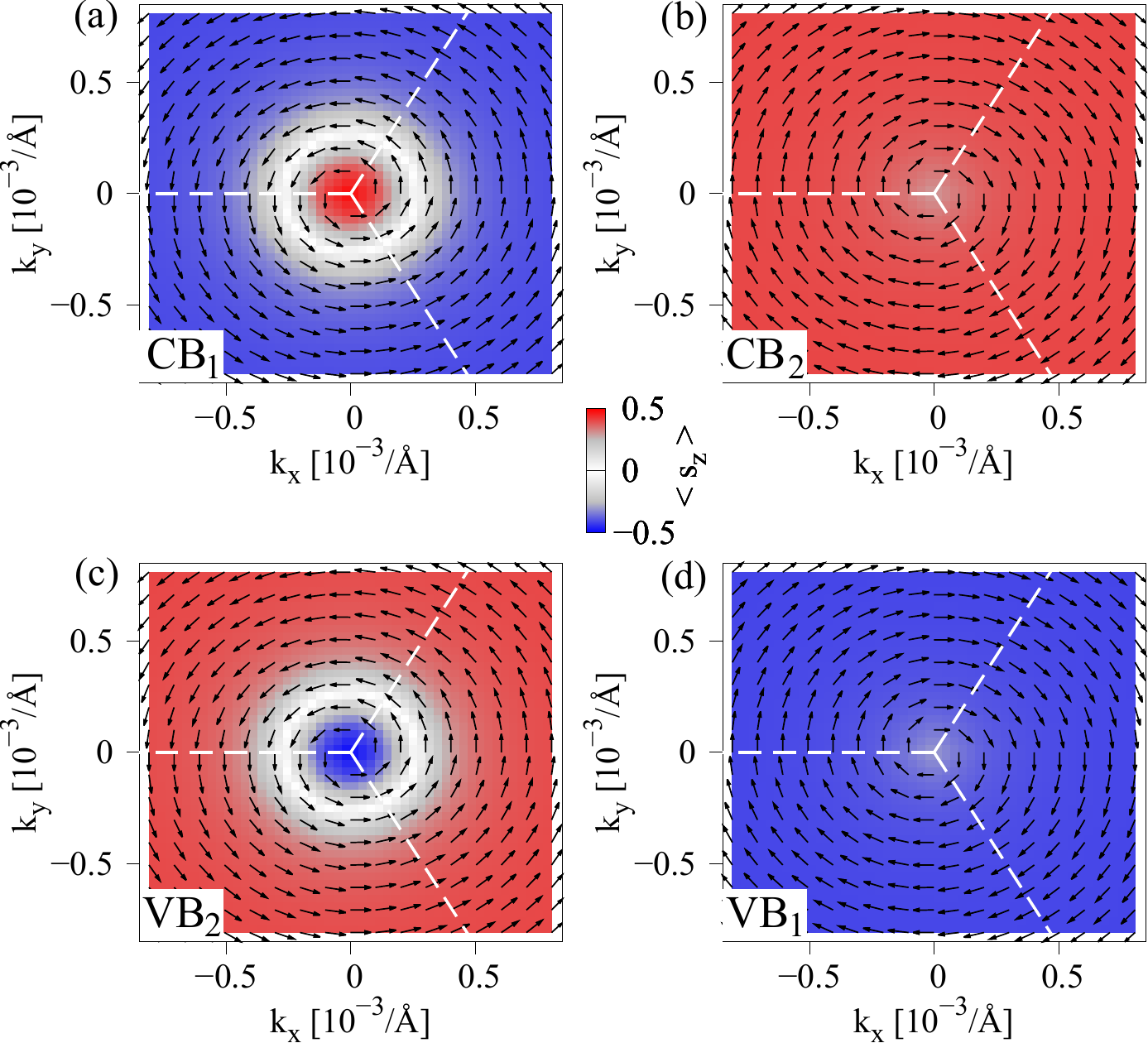}
 \caption{(Color online) First-principles calculated spin-orbit fields around the K point of the bands
 (a) $\varepsilon_{1}^{\textrm{CB}}$, (b) $\varepsilon_{2}^{\textrm{CB}}$, 
 (c) $\varepsilon_{2}^{\textrm{VB}}$, and (d) $\varepsilon_{1}^{\textrm{VB}}$, 
 for the SLG/Bi$_2$Se$_3$ stack with 1QL, 
 corresponding to the four low energy bands in Fig. \ref{Fig:fit_1QL}(e).
 The dashed white lines represent the edge of the Brillouin zone. }
 \label{Fig:grp_SOF}
\end{figure}

\begin{figure}[!htb]
\centering
 \includegraphics[width=.99\columnwidth]{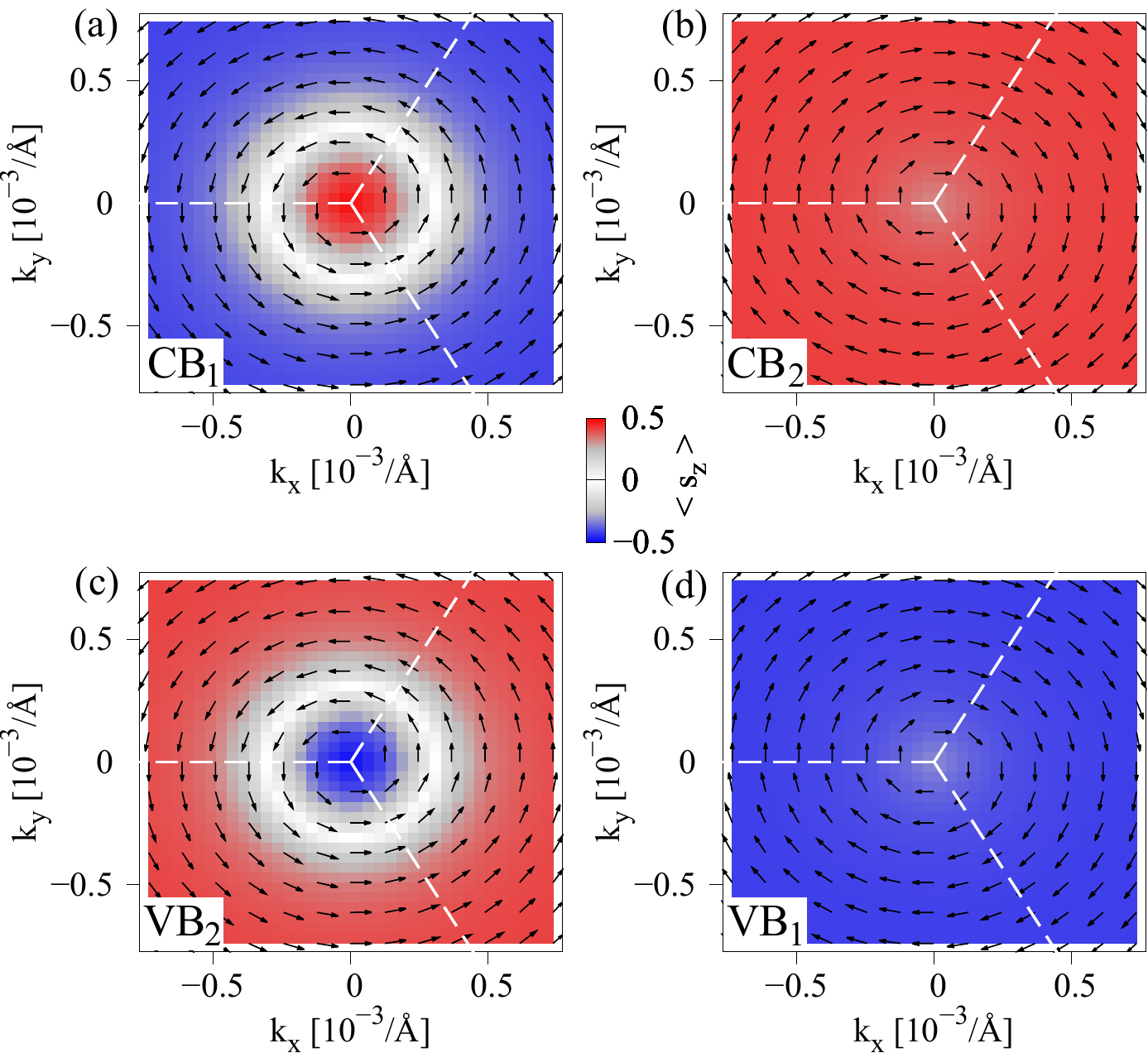}
 \caption{(Color online) First-principles calculated spin-orbit fields around the K point of the bands
 (a) $\varepsilon_{1}^{\textrm{CB}}$, (b) $\varepsilon_{2}^{\textrm{CB}}$, 
 (c) $\varepsilon_{2}^{\textrm{VB}}$, and (d) $\varepsilon_{1}^{\textrm{VB}}$, 
 for the SLG/Bi$_2$Se$_3$ stack with 3QLs.
 The dashed white lines represent the edge of the Brillouin zone.}
 \label{Fig:grp_SOF_3QL}
\end{figure}

In Fig. \ref{Fig:bandstructures} we show the full band structures of SLG above one, 
two, and three QLs of Bi$_2$Se$_3$. 
At the K point, one can recognize the Dirac bands originating from bare SLG \cite{Gmitra2009:PRB}, 
while at the $\Gamma$ point, the surface states of the topological insulator form. 
By gradually increasing the number of QLs, the Dirac bands of Bi$_2$Se$_3$ start to form. 
We do not show a zoom to the surface bands of Bi$_2$Se$_3$, as we are mainly interested in proximity induced SOC in SLG and BLG. 
However, one would see a pair of linear bands at the $\Gamma$ point, 
each originating from one surface of the topological insulator, which still hybridize for 3QLs only, exhibiting a gap.
At first glance the bands of SLG seem to be not affected by adding more QLs. 
We find that SLG gets hole doped, as the Dirac point is shifted roughly $350$~meV above the Fermi level. Only for 3QLs, the Dirac point is at about $500$~meV, which happens to appear due to the thicker topological insulator, consistent with recent calculations \cite{Song2018:NL}.

\begin{table*}[!htb]
\begin{ruledtabular}
\caption{\label{tab:fit_graphene} Fit parameters of Hamiltonian $\mathcal{H}_\textrm{SLG}$ 
for the SLG/Bi$_2$Se$_3$ stacks for different number of QLs. 
The Fermi velocity $v_{\textrm{F}}$, gap parameter $\Delta$, 
 Rashba SOC parameter $\lambda_{\textrm{R}}$, 
 intrinsic SOC parameters $\lambda_{\textrm{I}}^\textrm{A}$ and $\lambda_{\textrm{I}}^\textrm{B}$, 
 and PIA SOC parameters $\lambda_{\textrm{PIA}}^\textrm{A}$ and $\lambda_{\textrm{PIA}}^\textrm{B}$.
 The Dirac point energy $E_{\textrm{D}}$, as defined in Fig. \ref{Fig:bandstructures}(a).}
\begin{tabular}{l  c  c c  c  c   c  c  c}
 QLs & $v_{\textrm{F}}/10^5 [\frac{\textrm{m}}{\textrm{s}}]$ & 
$\Delta$~[$\mu$eV]& $\lambda_{\textrm{R}}$~[meV] & $\lambda_{\textrm{I}}^\textrm{A}$~[meV] &
$\lambda_{\textrm{I}}^\textrm{B}$~[meV] & $\lambda_{\textrm{PIA}}^\textrm{A}$~[meV] & 
$\lambda_{\textrm{PIA}}^\textrm{B}$~[meV] & $E_{\textrm{D}}$ [meV] \\
\hline
1 & 8.134 & 0.6 & -0.771 & 1.142 & -1.135 & 0.465 & 0.565 & 353.2\\
2 & 8.131 & 1.1 & -0.691 & 1.221 & -1.211 & 1.834 & 1.733 & 352.1\\
3 & 8.126 & 0.4 & -0.827 & 1.343 & -1.330 & 2.904 & 2.976 & 509.1\\
1\footnote{with additional hBN above SLG, parameter $\Delta$ in meV} &  8.110 & 21.32 & -0.799 & 1.638 & -1.517 & 2.708 & 2.433 & 341.2 
\end{tabular}
\end{ruledtabular}
\end{table*}

In Fig. \ref{Fig:fit_1QL} we show the low energy band properties for SLG/Bi$_2$Se$_3$ for 1QL, 
with a fit to the model Hamiltonian. 
We find that the model dispersion, energy splittings, and spin expectation values
agree very well with the first-principles data.
In Tab. \ref{tab:fit_graphene} we summarize the fit 
parameters of Hamiltonian $\mathcal{H}_\textrm{SLG}$ 
for the SLG/Bi$_2$Se$_3$ stacks for different number of QLs.
We find that the fit parameters are almost independent on the number of QLs, indicating that
only the closest QL is mainly responsible for proximity SOC. 
However, the intrinsic SOC parameters gradually increase from about $1.1$~meV, for 1QL, to $1.3$~meV, for 3QLs.
Such an increasing (and saturating) behavior of the induced SOC, with respect to the number of QLs, has already been reported for other interface configurations, yielding much larger SOC parameters \cite{Jin2013:PRB}.
Also the PIA SOC parameters increase with the number of QLs, 
while the Rashba SOC stays roughly the same.  
Similar to SLG/TMDC heterostructures \cite{Gmitra2015:PRB,Gmitra2016:PRB}, we find staggered intrinsic SOC, 
i.e., $\lambda_{\textrm{I}}^\textrm{A} \approx -\lambda_{\textrm{I}}^\textrm{B}$. 
In analogy to the SLG/WSe$_2$ heterostructure \cite{Gmitra2016:PRB,Frank2018:PRL}, we find an inverted band
structure, as the gap $\Delta$, being in the $\mu$eV range, is much smaller than the spin splittings of the bands. As a consequence, SLG proximitized by the topological insulator could host protected pseudohelical states \cite{Frank2018:PRL}.
All our results are in agreement with a recent study of SLG on 1QL of Bi$_2$Se$_3$ \cite{Song2018:NL}.
In particular, the SOC parameters are of the same magnitude (1 meV) and are also of valley-Zeeman type, for their 'large unit cell' calculation. 
In addition, we have calculated a heterostructure consisting of hBN/SLG/Bi$_2$Se$_3$ for 1QL, 
which is especially interesting for comparison to experimental studies where 
graphene is protected by the insulating hBN and proximitized by the topological insulator. 
The fit parameters are summarized in Tab. \ref{tab:fit_graphene}. 
In agreement with Ref. \cite{Zollner2019:PRB}, the orbital gap parameter $\Delta$ of the proximitized SLG is about 20~meV, 
much larger than for the case without hBN. Also the intrinsic SOC is a bit larger, 
while the rest of the fit parameters are barely different. 

To further analyze the low energy bands, we have calculated the 
spin-orbit fields, see Fig. \ref{Fig:grp_SOF}, of the four low energy bands
$\varepsilon_{1/2}^{\textrm{VB/CB}}$, corresponding to Fig. \ref{Fig:fit_1QL}(e).
The spin-orbit fields of the two outer (inner) bands rotate clockwise (counter-clockwise), 
being a clear signature of Rashba SOC. 
The spin-orbit fields are isotropic around the K point and show no signs of trigonal warping. 
Only the two inner bands change their $s_z$ expectation value, due to the inverted band structure.
The induced spin-orbit fields point out-of-plane, 
while the topological insulator surface states have in-plane spin-orbit fields. 

As a comparative case, we also show the calculated spin-orbit fields for graphene above 3QLs, 
see Fig. \ref{Fig:grp_SOF_3QL}, where the PIA SOC parameters are significantly larger, 
see Tab. \ref{tab:fit_graphene}, than in the 1QL case. 
However, comparing the 1QL and 3QL cases, we do not notice any difference 
in the spin-orbit fields, as the magnitude (meV) of PIA SOC is still comparable in the two cases, 
and the other SOC parameters are almost unchanged. 
Only a much larger PIA SOC can significantly change the low energy band properties \cite{Kochan2017:PRB}.

\subsection{Bilayer graphene}
\begin{figure*}[!htb]
 \includegraphics[width=.98\textwidth]{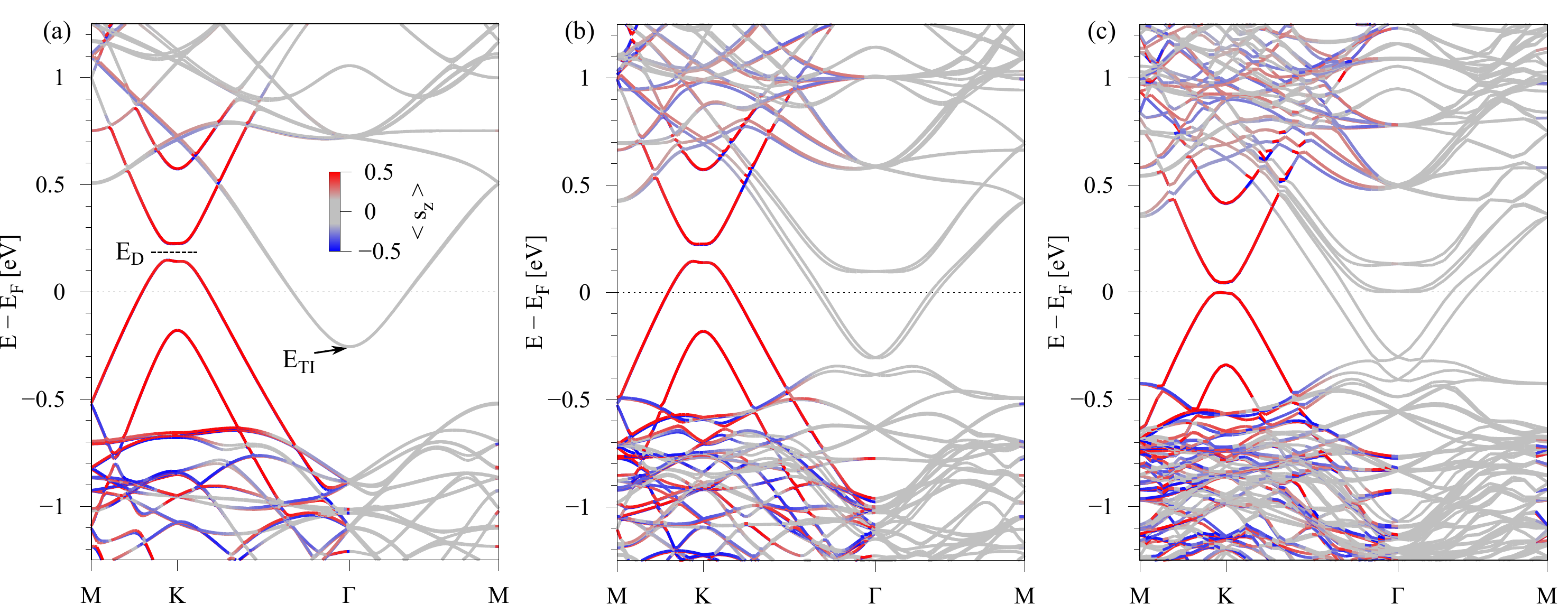}
 \caption{(Color online) Calculated band structures of BLG on (a) one (b) two and
 (c) three QLs of Bi$_2$Se$_3$. The color is the $s_z$ expectation value.
 In (a), we define the Dirac point energy $E_{\textrm{D}}$ and the \textit{doping energy} of the topological insulator 
 $E_{\textrm{TI}}$.
 }\label{Fig:bandstructures_BLG}
\end{figure*}

\begin{figure}[!htb]
 \includegraphics[width=.99\columnwidth]{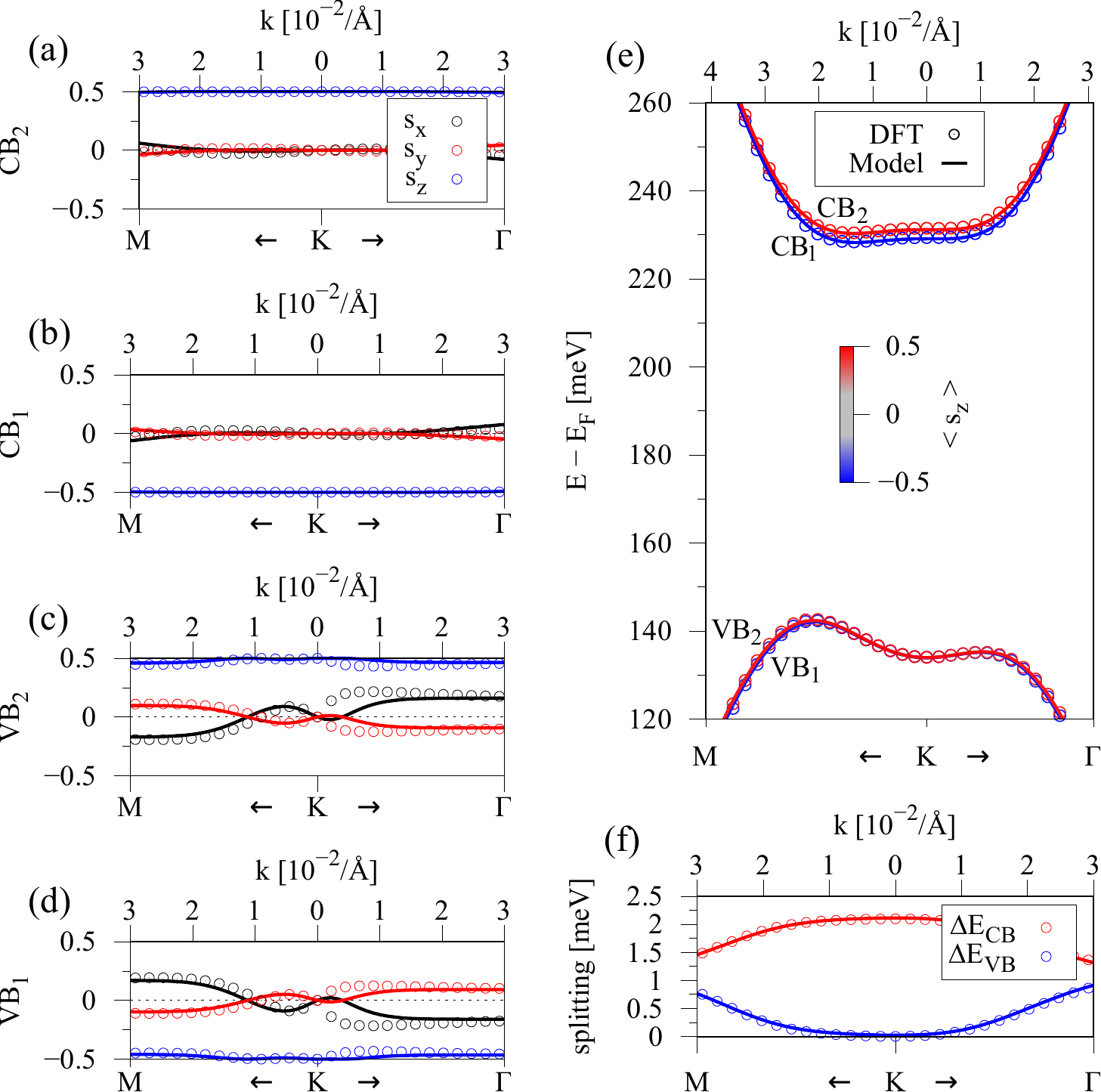}
 \caption{(Color online) Calculated low energy band properties (symbols) for BLG/Bi$_2$Se$_3$ for 1QL, with 
 a fit to the model Hamiltonian $\mathcal{H}_\textrm{BLG}$ (solid lines). 
 (a)-(d) The spin expectation values of the four low energy bands.
 (e) The low energy band structure of proximitized BLG. The color is the $s_z$ expectation value. 
 (f) The splitting of the valence (conduction) band in blue (red).
 }\label{Fig:fit_1QL_BLG}
\end{figure}

\begin{figure}[!htb]
\centering
 \includegraphics[width=.99\columnwidth]{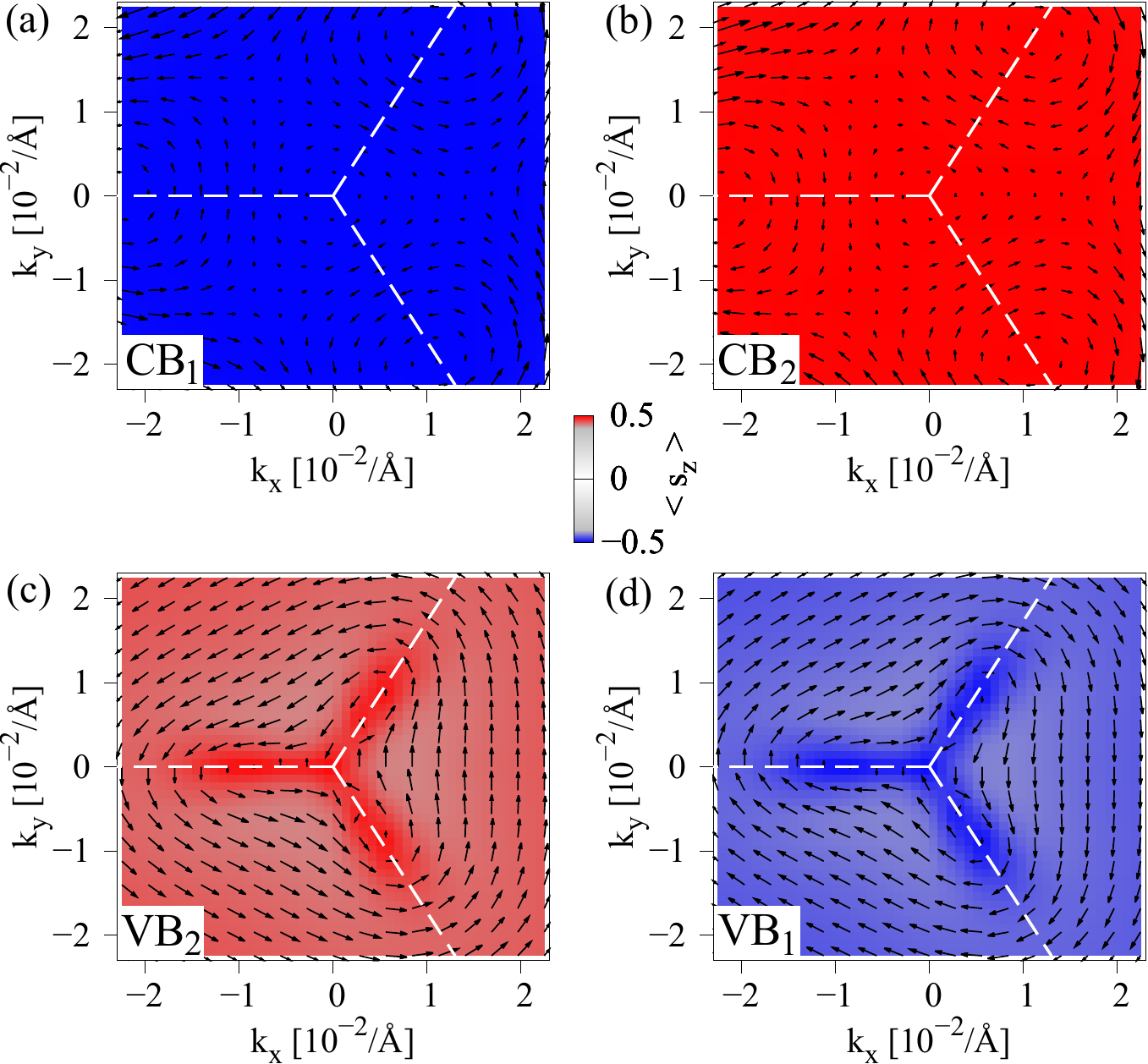}
 \caption{(Color online) First-principles calculated spin-orbit fields around the K point of the bands
 (a) $\varepsilon_{1}^{\textrm{CB}}$, (b) $\varepsilon_{2}^{\textrm{CB}}$, 
 (c) $\varepsilon_{2}^{\textrm{VB}}$, and (d) $\varepsilon_{1}^{\textrm{VB}}$, 
 for the BLG/Bi$_2$Se$_3$ stack with 1QL, 
 corresponding to the four low energy bands in Fig. \ref{Fig:fit_1QL_BLG}(e).
 The dashed white lines represent the edge of the Brillouin zone. }
 \label{Fig:blg_SOF}
\end{figure}

In Fig. \ref{Fig:bandstructures_BLG} we show the full band structures of BLG above one, 
two, and three QLs of Bi$_2$Se$_3$. 
At the K point, one can recognize the parabolic bands originating from bare BLG \cite{Konschuh2012:PRB},
while at the $\Gamma$ point, the surface states of the topological insulator form, just as for the SLG case. 
By gradually increasing the number of QLs, the Dirac bands of Bi$_2$Se$_3$ start to form. 
The BLG bands exhibit a sizable band gap.
Additionally, the BLG gets hole doped by about 200~meV, for one and two QLs, whereas the Fermi level is in the band gap of BLG for three QLs.
Experiments have reported about 380~meV of hole doping in BLG on Bi$_2$Se$_3$ \cite{Zalic2017:PRB}, 
which may be attributed to the additional SiO$_2$ substrate.

In Fig. \ref{Fig:fit_1QL_BLG} we show the low energy band properties for BLG/Bi$_2$Se$_3$ for 1QL, with a fit to the model Hamiltonian $\mathcal{H}_\textrm{BLG}$.
As for the SLG case, we find very good agreement of our model and the 
first-principles dispersion, energy splittings, and spin expectation values for 
the proximitized BLG.  
In contrast to SLG, the BLG exhibits a large band gap of about 80 meV.
As a consequence of the Bi$_2$Se$_3$ substrate, and the resulting induced dipole field, 
the two graphene layers are at a different potential $V$. 
Since the low energy bands are formed by the non-dimer carbon atoms of BLG, a band gap opens. 
In this case, the low energy conduction band is formed by the graphene layer closer to the Bi$_2$Se$_3$. 
The reason is that the conduction band is strongly spin split (2 meV) around the K point, due to proximity induced SOC. 
Since proximity induced phenomena are short range effects, only the closest graphene layer is affected. 
The band structure of BLG on Bi$_2$Se$_3$ is very similar to the one of BLG on WSe$_2$ \cite{Gmitra2017:PRL}, 
where a spin-orbit valve effect has been proposed. 

In Tab. \ref{tab:fit_BLG} we summarize the fit parameters of Hamiltonian $\mathcal{H}_\textrm{BLG}$ 
for the BLG/Bi$_2$Se$_3$ stacks for different number of QLs.
We find that the fit parameters are almost independent on the number of QLs, indicating that
only the closest QL is mainly responsible for proximity SOC, 
similar to what we have found from the SLG case. 
For the fit, we assume for simplicity the pristine graphene intrinsic SOC  
parameters \cite{Gmitra2009:PRB} for the top layer 
($\lambda_{\textrm{I}}^\textrm{A2} = \lambda_{\textrm{I}}^\textrm{B2} = 12~\mu$eV). 
In agreement to the monolayer case, we find staggered intrinsic SOC, 
i.e., $\lambda_{\textrm{I}}^\textrm{A1} \approx -\lambda_{\textrm{I}}^\textrm{B1}$.
Consequently, also proximitized BLG can exhibit topologically protected phases, 
as recently shown \cite{Alsharari2018:PRB}.
Surprisingly, the gap parameter $V$ diminishes by about 50\%, for 3QLs and the Fermi level is now
located within the band gap of the BLG. 

\begin{table*}[!htb]
\begin{ruledtabular}
\begin{tabular}{l c c c c c c c c c c c c c}
\multirow{2}{*}{QLs} & $\gamma_0$ &  $\gamma_1$ & $\gamma_3$ & $\gamma_4$ & $V$ & $\Delta$ & 
 $\lambda_{\textrm{I}}^\textrm{A1}$ & $\lambda_{\textrm{I}}^\textrm{B1}$ & $\lambda_{\textrm{I}}^\textrm{A2}$ & 
 $\lambda_{\textrm{I}}^\textrm{B2}$ &  $\lambda_{0}$ & $\lambda_{\textrm{R}}$  
 &  $E_{\textrm{D}}$\\
 & [eV] & [meV] & [meV] & [meV] & [meV] & [meV] & [meV] & [meV] 
& [meV] & [meV] & [meV] & [meV] & [meV] \\
\hline
1 & 2.513 & 373.5 & -274.9 & -165.3 & 41.60 & 13.79 & -1.056 & 1.170 & 0.012 & 0.012 & -0.433 & -0.273  & 183.3\\
2 & 2.512 & 373.5 & -275.5 & -165.6 & 43.12 & 13.79 & -1.117 & 1.301 & 0.012 & 0.012 & -0.167 & -0.199  & 181.4\\
3 & 2.513 & 373.9 & -264.9 & -163.5 & 23.63 & 13.54 & -1.030 & 1.168 & 0.012 & 0.012 & -0.190 & -0.188  & 19.95 \\
\end{tabular}
\end{ruledtabular}
\caption{\label{tab:fit_BLG} Fit parameters of Hamiltonian $\mathcal{H}_\textrm{BLG}$ 
for the BLG/Bi$_2$Se$_3$ stacks for different number of QLs.
Intra- and interlayer hoppings $\gamma$ of the BLG, 
when the lower (upper) GRP layer is placed in potential $V$ ($-V$). 
The asymmetry in the energy shift of the bonding and antibonding states $\Delta$.
Intrinsic SOC parameters of the corresponding layer and sublattice $\lambda_{\textrm{I}}$. 
Global and local space inversion symmetry breaking SOC parameters $\lambda_0$ and $\lambda_{\textrm{R}}$.}
\end{table*}

By looking at the calculated spin-orbit fields, see Fig. \ref{Fig:blg_SOF}, we find that the 
two proximity spin-orbit split conduction bands are strongly $s_z$ polarized. 
In contrast, the two valence bands exhibit a typical Rashba type spin-orbit field. 
In the case of SLG on TMDCs, the induced strong staggered SOC (opposite sign on the two sublattices) seems to be a consequence of the $s_z$ polarized spin-valley locked states from the TMDC, getting imprinted on the SLG (or BLG). 
In our investigated case, the surface states of the topological insulator have spin-momentum locking, however with in-plane ($s_x$ and $s_y$) spin components. 
Still, the induced spin-orbit fields are out-of-plane, similar to the SLG/TMDC case. 

So far, there is no complete microscopic understanding why in these two cases (SLG/Bi$_2$Se$_3$ and SLG/TMDC), the induced SOC is of valley-Zeeman type.
However, in the case of SLG/TMDC structures, there are already first approaches discussing
how the induced SOC depends on the twist angle and the position of the SLG Dirac point with respect to the TMDC band edges \cite{David2019:PRB}.

\section{Transverse electric field}
We have seen that the parameters and proximity SOC are marginally affected by adding more QLs. We now 
study the effect of a transverse electric field only for 1QL of Bi$_2$Se$_3$. 
Can we tune proximity SOC by the electric field?
Is a spin-orbit valve effect present in BLG?

\subsection{Graphene}
\begin{figure}[!htb]
 \includegraphics[width=.99\columnwidth]{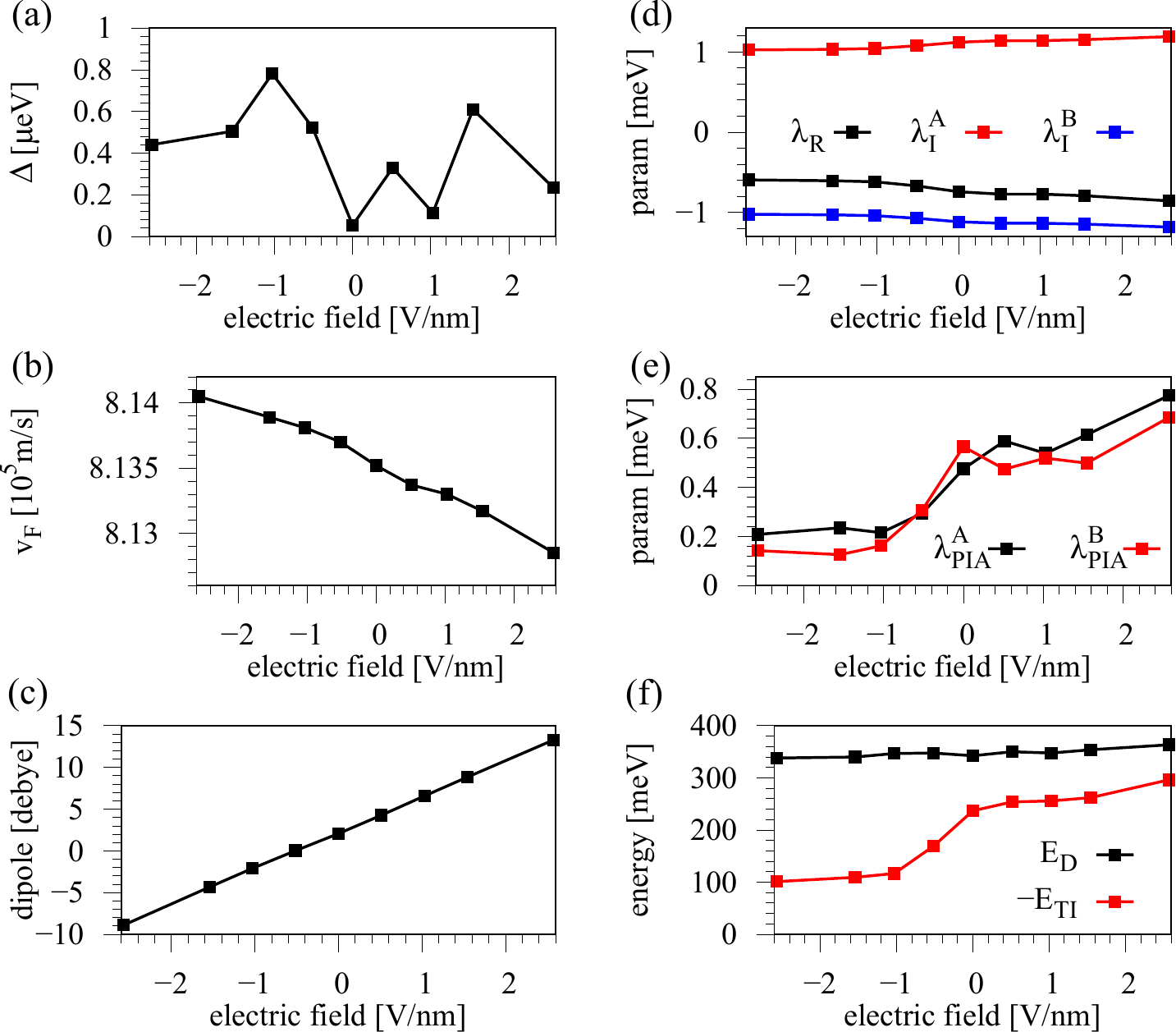}
 \caption{(Color online) Fit parameters of Hamiltonian $\mathcal{H}_\textrm{SLG}$ 
for the SLG/Bi$_2$Se$_3$ stack with 1QL as a function of a transverse electric field. 
(a) The gap parameter $\Delta$, (b) the Fermi velocity $v_{\textrm{F}}$, (c) the dipole of the structure, 
(d) Rashba SOC parameter $\lambda_{\textrm{R}}$, intrinsic SOC parameters $\lambda_{\textrm{I}}^\textrm{A}$ 
and $\lambda_{\textrm{I}}^\textrm{B}$, (e) PIA SOC parameters $\lambda_{\textrm{PIA}}^\textrm{A}$ and $\lambda_{\textrm{PIA}}^\textrm{B}$, and (f) the Dirac point energy $E_{\textrm{D}}$ and the doping energy of the topological insulator 
 $E_{\textrm{TI}}$, as defined in Fig. \ref{Fig:bandstructures}(a).
   }\label{Fig:Efield}
\end{figure}

In Fig. \ref{Fig:Efield} we show the fit parameters of Hamiltonian $\mathcal{H}_\textrm{SLG}$ 
for the SLG/Bi$_2$Se$_3$ stack as function of a transverse electric field. 
We find that the dipole of the structure grows linearly with applied electric field. 
The Fermi velocity $v_{\textrm{F}}$ is only slightly affected by the field, but shows a linear dependence.
The gap parameter $\Delta$, reflecting the sublattice symmetry breaking, stays tiny in magnitude without any noticeable trend. 
The intrinsic as well as Rashba SOC parameters grow in magnitude, when changing the field
from negative to positive amplitude. 
While the intrinsic SOC parameters, $\lambda_{\textrm{I}}^\textrm{A}$ 
and $\lambda_{\textrm{I}}^\textrm{B}$, can be changed from about 1 to 1.2~meV, the Rashba SOC
parameter $\lambda_{\textrm{R}}$ changes from about 0.6 to 0.9~meV in magnitude, when 
tuning the electric field from $-2.5$~V/nm to 2.5~V/nm.
A roughly linear trend can also be observed in the PIA SOC parameters, which can be tuned from 0 up to 0.8~meV.
The Dirac point energy $E_{\textrm{D}}$ stays at the same position, with respect to the Fermi level, as the field changes. 
The doping energy of the topological insulator $E_{\textrm{TI}}$ decreases, when a negative electric field is applied. 
Such a field tunability of the SOC parameters, especially the Rashba one, can lead to a giant control of spin-relaxation times and anisotropies in SLG \cite{Zollner2019:PRB,Song2018:NL}.

\subsection{Bilayer graphene}
\begin{figure}[!htb]
 \includegraphics[width=.99\columnwidth]{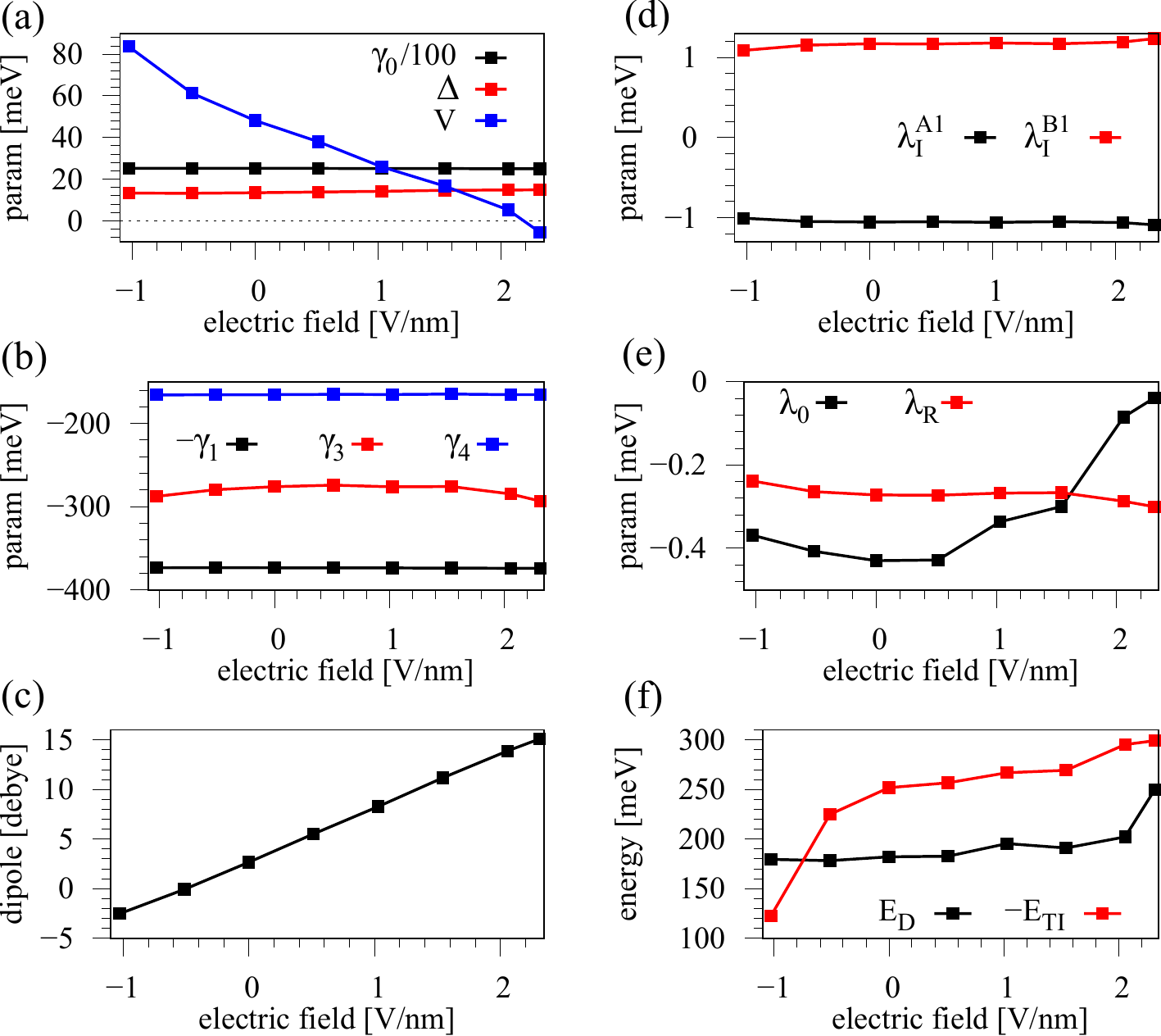}
 \caption{(Color online) Fit parameters of Hamiltonian $\mathcal{H}_\textrm{BLG}$ 
for the BLG/Bi$_2$Se$_3$ stack with 1QL as a function of a transverse electric field. 
(a) The nearest neighbor hopping parameter $\gamma_0$, the asymmetry in the energy shift of the bonding and antibonding states $\Delta$, and the SLG layer potential $V$. (b) The 
intra- and interlayer hoppings $\gamma_1$, $\gamma_3$, and $\gamma_4$. (c) The dipole of the structure. 
(d) The two proximity modified intrinsic SOC parameters $\lambda_{\textrm{I}}^\textrm{A1}$ 
and $\lambda_{\textrm{I}}^\textrm{B1}$. (e) The global and local space inversion symmetry breaking SOC parameters $\lambda_0$ and $\lambda_{\textrm{R}}$. (f) The Dirac point energy $E_{\textrm{D}}$ and the doping energy of the topological insulator 
 $E_{\textrm{TI}}$, as defined in Fig. \ref{Fig:bandstructures_BLG}(a).
   }\label{Fig:Efield_BLG}
\end{figure}

Most interesting is the case of BLG under the influence of an electric field. 
As already mentioned, the band structure is very similar to the one of BLG on WSe$_2$ \cite{Gmitra2017:PRL}, 
where a spin-orbit valve effect has been proposed. 
In Fig. \ref{Fig:Efield_BLG} we show the fit parameters of Hamiltonian $\mathcal{H}_\textrm{BLG}$ 
for the BLG/Bi$_2$Se$_3$ stack as a function of a transverse electric field. 
We find that the dipole of the structure grows linearly with applied electric field, 
similar to the SLG case.
Most of the orbital and spin-orbit parameters stay more or less constant as a transverse electric field is applied. 
However, the field can for example tune the doping energy of the topological insulator $E_{\textrm{TI}}$, and also 
slightly tune the interlayer hopping amplitude $\gamma_3$. 

\begin{figure}[!htb]
 \includegraphics[width=.99\columnwidth]{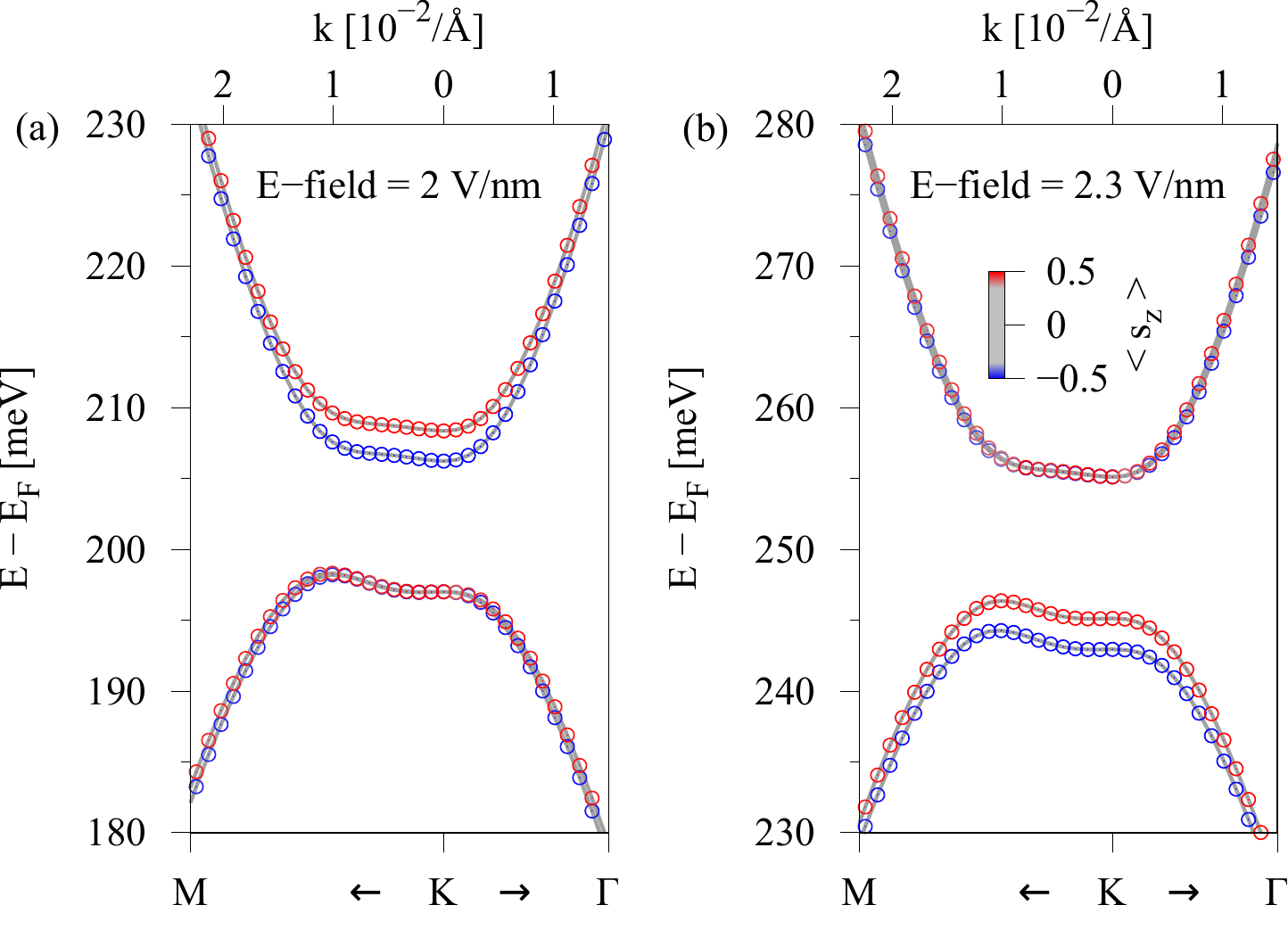}
 \caption{(Color online) Calculated low energy bands for BLG/Bi$_2$Se$_3$ for 1QL, with
applied transverse electric field of (a) 2~V/nm and (b) 2.3~V/nm.
   }\label{Fig:bands_Efields}
\end{figure}

Surprisingly, the parameter $\lambda_0$ drastically decreases in magnitude for positive fields. 
This means that the electric field, felt by one graphene layer due to the presence of the other, in this proximity
effect set-up, diminishes with applied external transverse field. 
Important for the previously mentioned spin-orbit valve effect is the closing of the orbital gap, and the subsequent reopening with \textit{inverted} band structure. 
Indeed, such a situation can be realized, for a transverse electric field around 2.1~V/nm, as the two graphene layers
are at the same potential, when the paramter $V$ goes through zero. 
For an electric field of 2~V/nm, the band structure is shown in Fig. \ref{Fig:bands_Efields}(a). 
The orbital gap of the BLG bands is about 10~meV, and the conduction band is strongly spin-orbit split. 
For a field of 2.3~V/nm, the band structure is inverted, see Fig. \ref{Fig:bands_Efields}(b), 
and the valence band is now strongly spin-orbit split. 
The spin-orbit split bands are always localized on the graphene layer closest to the substrate. 
In Fig. \ref{tab:fit_BLG}(f), we can see that the band splittings near the K point for conduction and valence band differ by orders of magnitude. 
Important for the spin-orbit valve effect is, that spin relaxation depends quadratically on the magnitude of the band splittings. 
Consequently, the spin relaxation for electrons (holes) is large (small) for an electric field of about 2~V/nm, and vice versa for a field of about 2.3~V/nm.

\section{Additional considerations}
Our DFT calculations, together with the extracted fit parameters, can give an insight 
on the magnitude of the induced SOC in SLG, and are helpful to analyze and interpret experimental data in 
SLG/topological insulator heterostructures, as recently proven \cite{Song2018:NL,Jafarpisheh2018:PRB}. 
However, there are still some open questions, that we would like to address in the following. 
Is the chosen interlayer distance of $3.5$~\AA~reasonable for the studied heterostructures? 
How does proximity SOC depend on it?
How does the proximity SOC depend on the composition of the topological insulator? 
Can we tune the Dirac states of SLG down to the Fermi level, by using a different topological insulator?

\subsection{Distance study}
\begin{figure}[!htb]
\centering
 \includegraphics[width=.99\columnwidth]{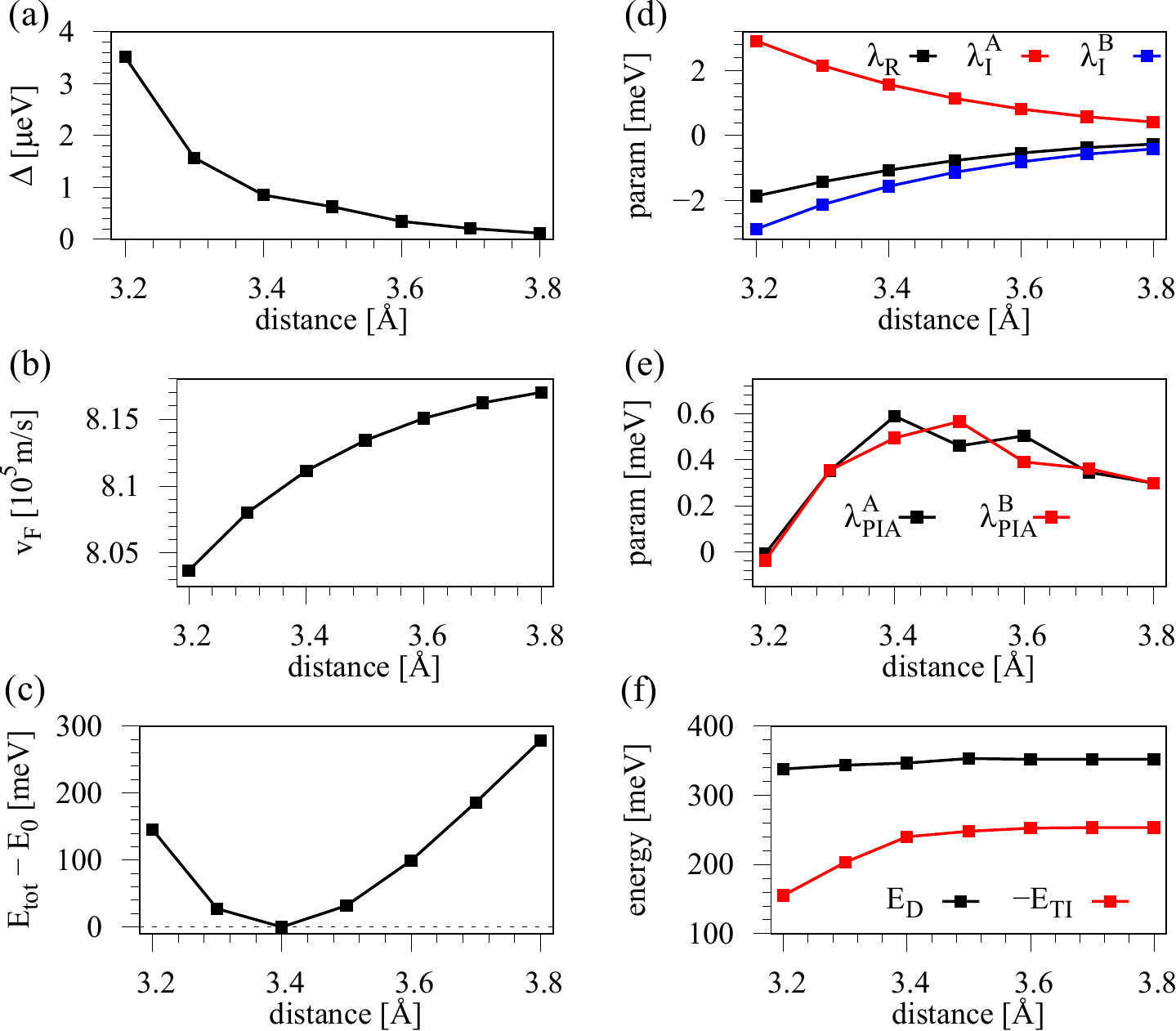}
 \caption{(Color online) Fit parameters of Hamiltonian $\mathcal{H}_\textrm{SLG}$ 
for the SLG/Bi$_2$Se$_3$ stack with 1QL as a function of the interlayer distance. 
(a) The gap parameter $\Delta$, (b) the Fermi velocity $v_{\textrm{F}}$, 
(c) the total energy with respect to the minimal energy $E_0$, 
(d) Rashba SOC parameter $\lambda_{\textrm{R}}$, intrinsic SOC parameters $\lambda_{\textrm{I}}^\textrm{A}$ 
and $\lambda_{\textrm{I}}^\textrm{B}$, (e) PIA SOC parameters $\lambda_{\textrm{PIA}}^\textrm{A}$ and $\lambda_{\textrm{PIA}}^\textrm{B}$, and (f) the Dirac point energy $E_{\textrm{D}}$ and the doping energy of the topological insulator 
 $E_{\textrm{TI}}$, as defined in Fig. \ref{Fig:bandstructures}(a).
   }\label{Fig:distance}
\end{figure}

It has been shown that proximity SOC and exchange effects can be significantly enhanced when decreasing the interlayer distance between materials \cite{Frank2016:PRB,Zollner2019:PRB,Zhang2018:PRB,Zhang2015:PRB}.
Here, we look into the proximity effects, when we modify the 
interlayer distance between SLG and Bi$_2$Se$_3$. Similar results can be expected for the BLG case.

In Fig. \ref{Fig:distance} we show the fit parameters of Hamiltonian $\mathcal{H}_\textrm{SLG}$ 
for the SLG/Bi$_2$Se$_3$ stack with 1QL as function of the interlayer 
distance between SLG and Bi$_2$Se$_3$. 
Our study shows, that the lowest energy is achieved for an interlayer distance of $3.4$~\AA, very close to our chosen 
distance of $3.5$~\AA~according to Ref. \cite{Song2018:NL}. 
Note that the energetically most favorable interlayer distance depends on the specific DFT input and the chosen vdW corrections.
The proximity-induced intrinsic and Rashba SOC parameters show the expected behavior and decrease in magnitude, 
as we increase the interlayer distance, see Fig. \ref{Fig:distance}(d). 
The increase of the parameters is giant, about 200\% when we decrease the interlayer distance by only about 10\%. 

Surprisingly, the PIA SOC parameters first increase with increasing the interlayer distance from $3.2~\textrm{\AA}$ to $3.4~\textrm{\AA}$. 
A possible explanation might be the position of the SLG Dirac point with respect to the bands of the topological insulator. 
Looking at the band structure Fig. \ref{Fig:bandstructures}(a), we see that there are bands of the topological insulator anti-crossing with the SLG Dirac bands, at around 0.6~eV. 
With increasing the distance, the Dirac point of SLG stays roughly at the same energy, while the bands of the topological insulator shift down in energy; compare the energies $E_{\textrm{D}}$ and $E_{\textrm{TI}}$ in Fig. \ref{Fig:distance}(f). 
Consequently, the anti-crossing bands move closer towards the SLG Dirac point and we need larger PIA parameter to capture the low energy bands. 
For interlayer distances, larger than $3.4~\textrm{\AA}$, the two energies $E_{\textrm{D}}$ and $E_{\textrm{TI}}$ barely change anymore and the PIA SOC parameters decrease in magnitude as expected.

\subsection{\texorpdfstring{Bi$_2$Te$_2$Se substrate}{}}
\begin{figure*}[!htb]
\centering
 \includegraphics[width=.99\textwidth]{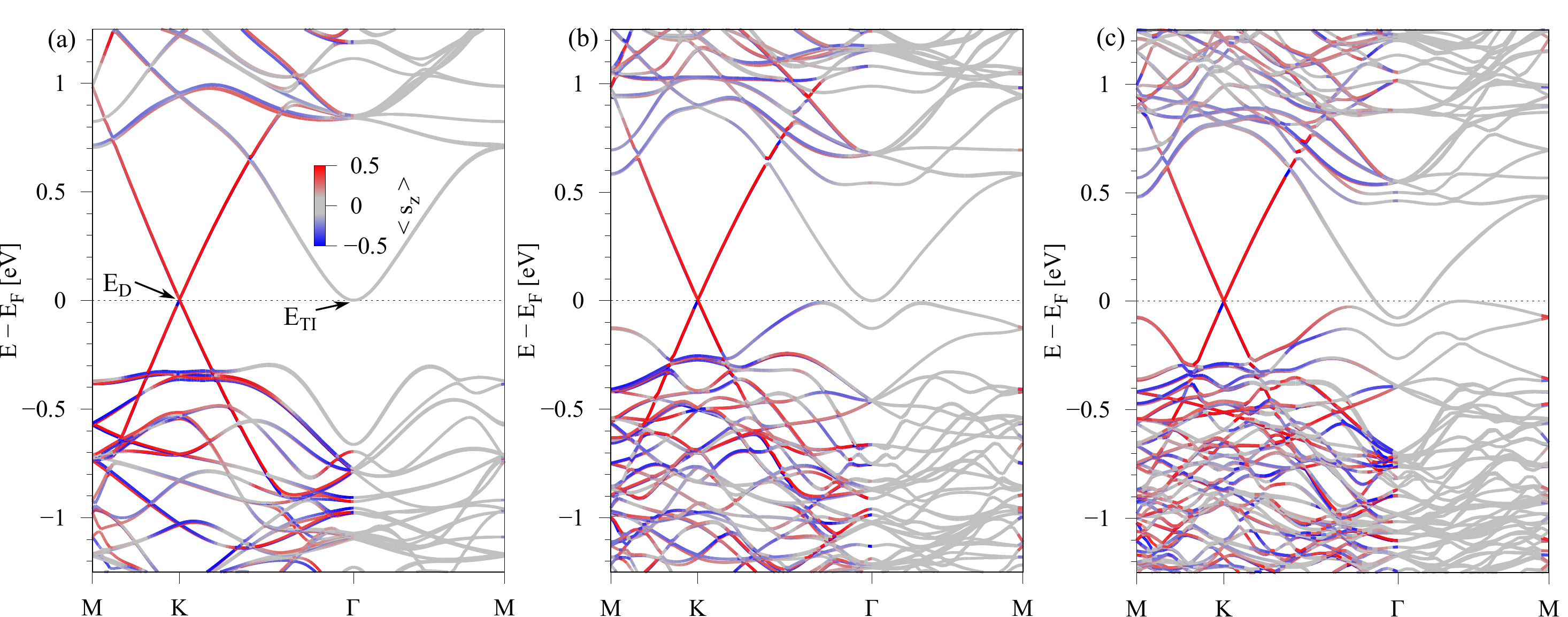}
 \caption{(Color online) Calculated band structures of SLG on (a) one (b) two and
 (c) three QLs of Bi$_2$Te$_2$Se. The color corresponds to the $s_z$ expectation value.
  In (a), we define the Dirac point energy $E_{\textrm{D}}$ and the \textit{doping energy} of the topological insulator $E_{\textrm{TI}}$.
 }\label{Fig:bandstructures_BST}
\end{figure*}

\begin{table*}[!htb]
\begin{ruledtabular}
\begin{tabular}{l  c  c c  c  c   c  c  c}
 QLs & $v_{\textrm{F}}/10^5 [\frac{\textrm{m}}{\textrm{s}}]$ & 
$\Delta$~[$\mu$eV]& $\lambda_{\textrm{R}}$~[meV] & $\lambda_{\textrm{I}}^\textrm{A}$~[meV] &
$\lambda_{\textrm{I}}^\textrm{B}$~[meV] & $\lambda_{\textrm{PIA}}^\textrm{A}$~[meV] & 
$\lambda_{\textrm{PIA}}^\textrm{B}$~[meV] & $E_{\textrm{D}}$ [meV] \\
\hline
1 & 8.123 & 0.3 & -0.669 & 1.353 & -1.351 & -1.091 & -1.209 & 4.0\\
2 & 8.105 & 0.4 & -0.487 & 1.446 & -1.441 & 1.317 & 1.350 & 2.4\\
3 & 8.109 & 1.8 & -0.521 & 1.466 & -1.460 & 1.581 & 1.399 & -0.7\\
\end{tabular}
\end{ruledtabular}
\caption{\label{tab:fit_graphene_BTS} Fit parameters of Hamiltonian $\mathcal{H}_\textrm{SLG}$ 
for the SLG/Bi$_2$Te$_2$Se stacks for different number of QLs. 
The Fermi velocity $v_{\textrm{F}}$, gap parameter $\Delta$, 
 Rashba SOC parameter $\lambda_{\textrm{R}}$, 
 intrinsic SOC parameters $\lambda_{\textrm{I}}^\textrm{A}$ and $\lambda_{\textrm{I}}^\textrm{B}$, 
 and PIA SOC parameters $\lambda_{\textrm{PIA}}^\textrm{A}$ and $\lambda_{\textrm{PIA}}^\textrm{B}$.
 The Dirac point energy $E_{\textrm{D}}$, as defined in Fig. \ref{Fig:bandstructures_BST}(a).}
\end{table*}

Experimentally, various atomic compositions are used to create the artificial topological insulator crystals Bi$_{2-x}$Sb$_{x}$Te$_{3-y}$Se$_y$, 
with some portions $x$ and $y$. The reason is that the unintentional intrinsic defect doping of the 
topological insulator can be compensated such that the Dirac surface states, with in-plane spin-momentum locking, 
are located near the Fermi level, and bulk transport can be suppressed \cite{Ren2011:PRB, Arakane2012:NC}.
Recent transport measurements have shown giant proximity SOC in SLG on 
Bi$_{1.5}$Sb$_{0.5}$Te$_{1.7}$Se$_{1.3}$, of at least 2.5~meV \cite{Jafarpisheh2018:PRB}. 
Here we show and discuss the first-principles calculated results for SLG on Bi$_2$Te$_2$Se, where we replace the outermost
Se atoms of each QL by Te atoms in Fig. \ref{Fig:struct}.

In Fig. \ref{Fig:bandstructures_BST}, we show the full band structures of SLG above one, 
two, and three QLs of Bi$_2$Te$_2$Se. We find that the
SLG Dirac point is now located near the Fermi level.  
Indeed, tuning the constituents of the topological insulator by $x$ and $y$, different band alignments can be formed. 
This case is also interesting for BLG, as the spin-orbit valve active bands could be shifted down to the Fermi level, by using a different topological insulator. 
In contrast to the case of Bi$_2$Se$_3$, the surface states of Bi$_2$Te$_2$Se are not yet gapless for 3QLs, but they are located much closer to the system Fermi level, compare Figs. \ref{Fig:bandstructures_BST} and \ref{Fig:bandstructures}.

In Tab. \ref{tab:fit_graphene_BTS} we summarize the fit parameters 
of Hamiltonian $\mathcal{H}_\textrm{SLG}$ 
for the SLG/Bi$_2$Te$_2$Se stacks for different number of QLs.
Again, we find that the fit parameters are almost independent on the number of QLs, indicating that
only the closest QL is mainly responsible for proximity SOC. 
The proximity-induced intrinsic SOC parameters are about 20\% larger, than for the Bi$_2$Se$_3$ substrate, and increase by about 10\% for each QL that we add. 
The origin of the strong induced SOC in SLG is due to the nearest Se or Te atoms of the topological insulator. 
Because Te atoms have stronger atomic SOC than Se atoms, also the proximity induced SOC is enhanced.

\section{Summary}
In summary, we have shown that SLG and BLG on the topological insulator Bi$_2$Se$_3$ experience significant hole doping (350~meV for SLG, 200~meV for BLG) and strong proximity induced SOC of about 1-2~meV, giant compared to the small intrinsic SOC of the pristine SLG and BLG. Most surprising is, that the induced SOC is also of valley-Zeeman type, similar to TMDC substrates.
The induced spin-orbit fields point mainly out-of-plane, even though the topological insulator hosts surface states with in-plane spin-momentum locking.

As we increase the number of QLs of Bi$_2$Se$_3$, below the proximitized SLG, the induced SOC
can be increased by about 10\%, each time a QL is added up to 3QLs. 
We expect the SOC to saturate as 
proximity induced phenomena are usually short range effects. 
In addition we show that an externally applied transverse electric field 
can tune band offsets and SOC parameters in SLG.
This tunability of SOC by electric fields, can have significant impact on the spin relaxation properties. 
For the BLG, a spin-orbit valve effect can be realized, similar to the recently studied case of a TMDC substrate. In particular, we find that without applied electric field, the BLG states exhibit a band gap and a strongly spin-orbit split conduction band. For a moderate and experimentally accessible field of about 2.3~V/nm, the band structure can be inverted, now with strongly spin-orbit split valence band, offering the possibility to fully electrically control the magnitude of spin relaxation of electrons and holes in BLG. 

Furthermore, in the case of SLG on 1QL of Bi$_2$Se$_3$, we have shown that a
small decrease of the interlayer distance by only 10\%, can strongly enhance proximity SOC by about 200\%, allowing to tailor the magnitude of SOC by external pressure. 
Finally, we have extracted the orbital and SOC parameters for SLG on a different topological insulator 
Bi$_2$Te$_2$Se where the SLG is essentially undoped and still experiences strong valley-Zeeman SOC, even larger than in Bi$_2$Se$_3$.
Experimentally, the tunability of the SLG doping level can be controlled by varying
the constituents, $x$ and $y$, of the topological insulator  Bi$_{2-x}$Sb$_{x}$Te$_{3-y}$Se$_y$.
Tuning both Dirac states, from the SLG and the topological insulator, to the Fermi level allows to study the interplay of two very distinct spin-orbit fields at the same time.

\acknowledgments
This work was supported by DFG SPP 1666, SFB 1277 (A09 and B07), 
the European Unions Horizon 2020 research and innovation program under Grant No. 785219.

\bibliography{paper}

\begin{thebibliography}{84}%
\makeatletter
\providecommand \@ifxundefined [1]{%
 \@ifx{#1\undefined}
}%
\providecommand \@ifnum [1]{%
 \ifnum #1\expandafter \@firstoftwo
 \else \expandafter \@secondoftwo
 \fi
}%
\providecommand \@ifx [1]{%
 \ifx #1\expandafter \@firstoftwo
 \else \expandafter \@secondoftwo
 \fi
}%
\providecommand \natexlab [1]{#1}%
\providecommand \enquote  [1]{``#1''}%
\providecommand \bibnamefont  [1]{#1}%
\providecommand \bibfnamefont [1]{#1}%
\providecommand \citenamefont [1]{#1}%
\providecommand \href@noop [0]{\@secondoftwo}%
\providecommand \href [0]{\begingroup \@sanitize@url \@href}%
\providecommand \@href[1]{\@@startlink{#1}\@@href}%
\providecommand \@@href[1]{\endgroup#1\@@endlink}%
\providecommand \@sanitize@url [0]{\catcode `\\12\catcode `\$12\catcode
  `\&12\catcode `\#12\catcode `\^12\catcode `\_12\catcode `\%12\relax}%
\providecommand \@@startlink[1]{}%
\providecommand \@@endlink[0]{}%
\providecommand \url  [0]{\begingroup\@sanitize@url \@url }%
\providecommand \@url [1]{\endgroup\@href {#1}{\urlprefix }}%
\providecommand \urlprefix  [0]{URL }%
\providecommand \Eprint [0]{\href }%
\providecommand \doibase [0]{http://dx.doi.org/}%
\providecommand \selectlanguage [0]{\@gobble}%
\providecommand \bibinfo  [0]{\@secondoftwo}%
\providecommand \bibfield  [0]{\@secondoftwo}%
\providecommand \translation [1]{[#1]}%
\providecommand \BibitemOpen [0]{}%
\providecommand \bibitemStop [0]{}%
\providecommand \bibitemNoStop [0]{.\EOS\space}%
\providecommand \EOS [0]{\spacefactor3000\relax}%
\providecommand \BibitemShut  [1]{\csname bibitem#1\endcsname}%
\let\auto@bib@innerbib\@empty
\bibitem [{\citenamefont {Geim}\ and\ \citenamefont
  {Grigorieva}(2013)}]{Geim2013:Nat}%
  \BibitemOpen
  \bibfield  {author} {\bibinfo {author} {\bibfnamefont {A.~K.}\ \bibnamefont
  {Geim}}\ and\ \bibinfo {author} {\bibfnamefont {I.~V.}\ \bibnamefont
  {Grigorieva}},\ }\href {\doibase 10.1038/nature12385} {\bibfield  {journal}
  {\bibinfo  {journal} {Nature}\ }\textbf {\bibinfo {volume} {499}},\ \bibinfo
  {pages} {419} (\bibinfo {year} {2013})},\ \Eprint
  {http://arxiv.org/abs/1307.6718} {1307.6718} \BibitemShut {NoStop}%
\bibitem [{\citenamefont {Novoselov}\ \emph {et~al.}(2016)\citenamefont
  {Novoselov}, \citenamefont {Mishchenko}, \citenamefont {Carvalho},\ and\
  \citenamefont {{Castro Neto}}}]{Novoselov2016:Sci}%
  \BibitemOpen
  \bibfield  {author} {\bibinfo {author} {\bibfnamefont {K.~S.}\ \bibnamefont
  {Novoselov}}, \bibinfo {author} {\bibfnamefont {A.}~\bibnamefont
  {Mishchenko}}, \bibinfo {author} {\bibfnamefont {A.}~\bibnamefont
  {Carvalho}}, \ and\ \bibinfo {author} {\bibfnamefont {A.~H.}\ \bibnamefont
  {{Castro Neto}}},\ }\href {\doibase 10.1126/science.aac9439} {\bibfield
  {journal} {\bibinfo  {journal} {Science}\ }\textbf {\bibinfo {volume}
  {353}},\ \bibinfo {pages} {aac9439} (\bibinfo {year} {2016})},\ \Eprint
  {http://arxiv.org/abs/arXiv:1411.1235v1} {arXiv:1411.1235v1} \BibitemShut
  {NoStop}%
\bibitem [{\citenamefont {Duong}\ \emph {et~al.}(2017)\citenamefont {Duong},
  \citenamefont {Yun},\ and\ \citenamefont {Lee}}]{Duong2017:ACS}%
  \BibitemOpen
  \bibfield  {author} {\bibinfo {author} {\bibfnamefont {D.~L.}\ \bibnamefont
  {Duong}}, \bibinfo {author} {\bibfnamefont {S.~J.}\ \bibnamefont {Yun}}, \
  and\ \bibinfo {author} {\bibfnamefont {Y.~H.}\ \bibnamefont {Lee}},\ }\href
  {\doibase 10.1021/acsnano.7b07436} {\bibfield  {journal} {\bibinfo  {journal}
  {ACS Nano}\ }\textbf {\bibinfo {volume} {11}},\ \bibinfo {pages} {11803}
  (\bibinfo {year} {2017})}\BibitemShut {NoStop}%
\bibitem [{\citenamefont {{\v{Z}}uti{\'{c}}}\ \emph {et~al.}(2019)\citenamefont
  {{\v{Z}}uti{\'{c}}}, \citenamefont {Matos-Abiague}, \citenamefont {Scharf},
  \citenamefont {Dery},\ and\ \citenamefont {Belashchenko}}]{Zutic2019:MT}%
  \BibitemOpen
  \bibfield  {author} {\bibinfo {author} {\bibfnamefont {I.}~\bibnamefont
  {{\v{Z}}uti{\'{c}}}}, \bibinfo {author} {\bibfnamefont {A.}~\bibnamefont
  {Matos-Abiague}}, \bibinfo {author} {\bibfnamefont {B.}~\bibnamefont
  {Scharf}}, \bibinfo {author} {\bibfnamefont {H.}~\bibnamefont {Dery}}, \ and\
  \bibinfo {author} {\bibfnamefont {K.}~\bibnamefont {Belashchenko}},\ }\href
  {\doibase 10.1016/j.mattod.2018.05.003} {\bibfield  {journal} {\bibinfo
  {journal} {Mater. Today}\ }\textbf {\bibinfo {volume} {22}},\ \bibinfo
  {pages} {85} (\bibinfo {year} {2019})}\BibitemShut {NoStop}%
\bibitem [{\citenamefont {{Castro Neto}}\ \emph {et~al.}(2009)\citenamefont
  {{Castro Neto}}, \citenamefont {Guinea}, \citenamefont {Peres}, \citenamefont
  {Novoselov},\ and\ \citenamefont {Geim}}]{Neto2009:RMP}%
  \BibitemOpen
  \bibfield  {author} {\bibinfo {author} {\bibfnamefont {A.~H.}\ \bibnamefont
  {{Castro Neto}}}, \bibinfo {author} {\bibfnamefont {F.}~\bibnamefont
  {Guinea}}, \bibinfo {author} {\bibfnamefont {N.~M.~R.}\ \bibnamefont
  {Peres}}, \bibinfo {author} {\bibfnamefont {K.~S.}\ \bibnamefont
  {Novoselov}}, \ and\ \bibinfo {author} {\bibfnamefont {A.~K.}\ \bibnamefont
  {Geim}},\ }\href {\doibase 10.1103/RevModPhys.81.109} {\bibfield  {journal}
  {\bibinfo  {journal} {Rev. Mod. Phys.}\ }\textbf {\bibinfo {volume} {81}},\
  \bibinfo {pages} {109} (\bibinfo {year} {2009})}\BibitemShut {NoStop}%
\bibitem [{\citenamefont {Wang}\ \emph {et~al.}(2012)\citenamefont {Wang},
  \citenamefont {Kalantar-Zadeh}, \citenamefont {Kis}, \citenamefont
  {Coleman},\ and\ \citenamefont {Strano}}]{Wang2012:NN}%
  \BibitemOpen
  \bibfield  {author} {\bibinfo {author} {\bibfnamefont {Q.~H.}\ \bibnamefont
  {Wang}}, \bibinfo {author} {\bibfnamefont {K.}~\bibnamefont
  {Kalantar-Zadeh}}, \bibinfo {author} {\bibfnamefont {A.}~\bibnamefont {Kis}},
  \bibinfo {author} {\bibfnamefont {J.~N.}\ \bibnamefont {Coleman}}, \ and\
  \bibinfo {author} {\bibfnamefont {M.~S.}\ \bibnamefont {Strano}},\ }\href
  {\doibase 10.1038/nnano.2012.193} {\bibfield  {journal} {\bibinfo  {journal}
  {Nat. Nanotechnol.}\ }\textbf {\bibinfo {volume} {7}},\ \bibinfo {pages}
  {699} (\bibinfo {year} {2012})},\ \Eprint {http://arxiv.org/abs/1205.1822}
  {1205.1822} \BibitemShut {NoStop}%
\bibitem [{\citenamefont {Catellani}\ \emph {et~al.}(1987)\citenamefont
  {Catellani}, \citenamefont {Posternak}, \citenamefont {Baldereschi},\ and\
  \citenamefont {Freeman}}]{Catellani1987:PRB}%
  \BibitemOpen
  \bibfield  {author} {\bibinfo {author} {\bibfnamefont {A.}~\bibnamefont
  {Catellani}}, \bibinfo {author} {\bibfnamefont {M.}~\bibnamefont
  {Posternak}}, \bibinfo {author} {\bibfnamefont {A.}~\bibnamefont
  {Baldereschi}}, \ and\ \bibinfo {author} {\bibfnamefont {A.~J.}\ \bibnamefont
  {Freeman}},\ }\href {\doibase 10.1103/PhysRevB.36.6105} {\bibfield  {journal}
  {\bibinfo  {journal} {Phys. Rev. B}\ }\textbf {\bibinfo {volume} {36}},\
  \bibinfo {pages} {6105} (\bibinfo {year} {1987})}\BibitemShut {NoStop}%
\bibitem [{\citenamefont {Frindt}(1972)}]{Frindt1972:PRL}%
  \BibitemOpen
  \bibfield  {author} {\bibinfo {author} {\bibfnamefont {R.~F.}\ \bibnamefont
  {Frindt}},\ }\href {\doibase 10.1103/PhysRevLett.28.299} {\bibfield
  {journal} {\bibinfo  {journal} {Phys. Rev. Lett.}\ }\textbf {\bibinfo
  {volume} {28}},\ \bibinfo {pages} {299} (\bibinfo {year} {1972})}\BibitemShut
  {NoStop}%
\bibitem [{\citenamefont {Dillon}\ and\ \citenamefont
  {Olson}(1965)}]{Dillon1965:JAP}%
  \BibitemOpen
  \bibfield  {author} {\bibinfo {author} {\bibfnamefont {J.~F.}\ \bibnamefont
  {Dillon}}\ and\ \bibinfo {author} {\bibfnamefont {C.~E.}\ \bibnamefont
  {Olson}},\ }\href {\doibase 10.1063/1.1714194} {\bibfield  {journal}
  {\bibinfo  {journal} {J. Appl. Phys.}\ }\textbf {\bibinfo {volume} {36}},\
  \bibinfo {pages} {1259} (\bibinfo {year} {1965})}\BibitemShut {NoStop}%
\bibitem [{\citenamefont {Huang}\ \emph {et~al.}(2017)\citenamefont {Huang},
  \citenamefont {Clark}, \citenamefont {Navarro-Moratalla}, \citenamefont
  {Klein}, \citenamefont {Cheng}, \citenamefont {Seyler}, \citenamefont
  {Zhong}, \citenamefont {Schmidgall}, \citenamefont {McGuire}, \citenamefont
  {Cobden}, \citenamefont {Yao}, \citenamefont {Xiao}, \citenamefont
  {Jarillo-Herrero},\ and\ \citenamefont {Xu}}]{Huang2017:Nat}%
  \BibitemOpen
  \bibfield  {author} {\bibinfo {author} {\bibfnamefont {B.}~\bibnamefont
  {Huang}}, \bibinfo {author} {\bibfnamefont {G.}~\bibnamefont {Clark}},
  \bibinfo {author} {\bibfnamefont {E.}~\bibnamefont {Navarro-Moratalla}},
  \bibinfo {author} {\bibfnamefont {D.~R.}\ \bibnamefont {Klein}}, \bibinfo
  {author} {\bibfnamefont {R.}~\bibnamefont {Cheng}}, \bibinfo {author}
  {\bibfnamefont {K.~L.}\ \bibnamefont {Seyler}}, \bibinfo {author}
  {\bibfnamefont {D.}~\bibnamefont {Zhong}}, \bibinfo {author} {\bibfnamefont
  {E.}~\bibnamefont {Schmidgall}}, \bibinfo {author} {\bibfnamefont {M.~A.}\
  \bibnamefont {McGuire}}, \bibinfo {author} {\bibfnamefont {D.~H.}\
  \bibnamefont {Cobden}}, \bibinfo {author} {\bibfnamefont {W.}~\bibnamefont
  {Yao}}, \bibinfo {author} {\bibfnamefont {D.}~\bibnamefont {Xiao}}, \bibinfo
  {author} {\bibfnamefont {P.}~\bibnamefont {Jarillo-Herrero}}, \ and\ \bibinfo
  {author} {\bibfnamefont {X.}~\bibnamefont {Xu}},\ }\href {\doibase
  10.1038/nature22391} {\bibfield  {journal} {\bibinfo  {journal} {Nature}\
  }\textbf {\bibinfo {volume} {546}},\ \bibinfo {pages} {270} (\bibinfo {year}
  {2017})}\BibitemShut {NoStop}%
\bibitem [{\citenamefont {McGuire}\ \emph {et~al.}(2015)\citenamefont
  {McGuire}, \citenamefont {Dixit}, \citenamefont {Cooper},\ and\ \citenamefont
  {Sales}}]{McGuire2015:CM}%
  \BibitemOpen
  \bibfield  {author} {\bibinfo {author} {\bibfnamefont {M.~A.}\ \bibnamefont
  {McGuire}}, \bibinfo {author} {\bibfnamefont {H.}~\bibnamefont {Dixit}},
  \bibinfo {author} {\bibfnamefont {V.~R.}\ \bibnamefont {Cooper}}, \ and\
  \bibinfo {author} {\bibfnamefont {B.~C.}\ \bibnamefont {Sales}},\ }\href
  {\doibase 10.1021/cm504242t} {\bibfield  {journal} {\bibinfo  {journal}
  {Chem. Mater.}\ }\textbf {\bibinfo {volume} {27}},\ \bibinfo {pages} {612}
  (\bibinfo {year} {2015})}\BibitemShut {NoStop}%
\bibitem [{\citenamefont {Wiedenmann}\ \emph {et~al.}(1981)\citenamefont
  {Wiedenmann}, \citenamefont {Rossat-Mignod}, \citenamefont {Louisy},
  \citenamefont {Brec},\ and\ \citenamefont {Rouxel}}]{Wiedenmann1981:SSC}%
  \BibitemOpen
  \bibfield  {author} {\bibinfo {author} {\bibfnamefont {A.}~\bibnamefont
  {Wiedenmann}}, \bibinfo {author} {\bibfnamefont {J.}~\bibnamefont
  {Rossat-Mignod}}, \bibinfo {author} {\bibfnamefont {A.}~\bibnamefont
  {Louisy}}, \bibinfo {author} {\bibfnamefont {R.}~\bibnamefont {Brec}}, \ and\
  \bibinfo {author} {\bibfnamefont {J.}~\bibnamefont {Rouxel}},\ }\href
  {\doibase 10.1016/0038-1098(81)90253-2} {\bibfield  {journal} {\bibinfo
  {journal} {Solid State Commun.}\ }\textbf {\bibinfo {volume} {40}},\ \bibinfo
  {pages} {1067} (\bibinfo {year} {1981})}\BibitemShut {NoStop}%
\bibitem [{\citenamefont {Carteaux}\ \emph {et~al.}(1995)\citenamefont
  {Carteaux}, \citenamefont {Brunet}, \citenamefont {Ouvrard},\ and\
  \citenamefont {Andre}}]{Carteaux1995:JPCM}%
  \BibitemOpen
  \bibfield  {author} {\bibinfo {author} {\bibfnamefont {V.}~\bibnamefont
  {Carteaux}}, \bibinfo {author} {\bibfnamefont {D.}~\bibnamefont {Brunet}},
  \bibinfo {author} {\bibfnamefont {G.}~\bibnamefont {Ouvrard}}, \ and\
  \bibinfo {author} {\bibfnamefont {G.}~\bibnamefont {Andre}},\ }\href
  {\doibase 10.1088/0953-8984/7/1/008} {\bibfield  {journal} {\bibinfo
  {journal} {J. Phys.: Condens. Mat.}\ }\textbf {\bibinfo {volume} {7}},\
  \bibinfo {pages} {69} (\bibinfo {year} {1995})}\BibitemShut {NoStop}%
\bibitem [{\citenamefont {Gong}\ \emph {et~al.}(2017)\citenamefont {Gong},
  \citenamefont {Li}, \citenamefont {Li}, \citenamefont {Ji}, \citenamefont
  {Stern}, \citenamefont {Xia}, \citenamefont {Cao}, \citenamefont {Bao},
  \citenamefont {Wang}, \citenamefont {Wang}, \citenamefont {Qiu},
  \citenamefont {Cava}, \citenamefont {Louie}, \citenamefont {Xia},\ and\
  \citenamefont {Zhang}}]{Gong2017:Nat}%
  \BibitemOpen
  \bibfield  {author} {\bibinfo {author} {\bibfnamefont {C.}~\bibnamefont
  {Gong}}, \bibinfo {author} {\bibfnamefont {L.}~\bibnamefont {Li}}, \bibinfo
  {author} {\bibfnamefont {Z.}~\bibnamefont {Li}}, \bibinfo {author}
  {\bibfnamefont {H.}~\bibnamefont {Ji}}, \bibinfo {author} {\bibfnamefont
  {A.}~\bibnamefont {Stern}}, \bibinfo {author} {\bibfnamefont
  {Y.}~\bibnamefont {Xia}}, \bibinfo {author} {\bibfnamefont {T.}~\bibnamefont
  {Cao}}, \bibinfo {author} {\bibfnamefont {W.}~\bibnamefont {Bao}}, \bibinfo
  {author} {\bibfnamefont {C.}~\bibnamefont {Wang}}, \bibinfo {author}
  {\bibfnamefont {Y.}~\bibnamefont {Wang}}, \bibinfo {author} {\bibfnamefont
  {Z.~Q.}\ \bibnamefont {Qiu}}, \bibinfo {author} {\bibfnamefont {R.~J.}\
  \bibnamefont {Cava}}, \bibinfo {author} {\bibfnamefont {S.~G.}\ \bibnamefont
  {Louie}}, \bibinfo {author} {\bibfnamefont {J.}~\bibnamefont {Xia}}, \ and\
  \bibinfo {author} {\bibfnamefont {X.}~\bibnamefont {Zhang}},\ }\href
  {\doibase 10.1038/nature22060} {\bibfield  {journal} {\bibinfo  {journal}
  {Nature}\ }\textbf {\bibinfo {volume} {546}},\ \bibinfo {pages} {265}
  (\bibinfo {year} {2017})}\BibitemShut {NoStop}%
\bibitem [{\citenamefont {Langer}\ \emph {et~al.}(2018)\citenamefont {Langer},
  \citenamefont {Schmid}, \citenamefont {Schlauderer}, \citenamefont {Gmitra},
  \citenamefont {Fabian}, \citenamefont {Nagler}, \citenamefont
  {Sch{\"{u}}ller}, \citenamefont {Korn}, \citenamefont {Hawkins},
  \citenamefont {Steiner}, \citenamefont {Huttner}, \citenamefont {Koch},
  \citenamefont {Kira},\ and\ \citenamefont {Huber}}]{Langer2018:Nat}%
  \BibitemOpen
  \bibfield  {author} {\bibinfo {author} {\bibfnamefont {F.}~\bibnamefont
  {Langer}}, \bibinfo {author} {\bibfnamefont {C.~P.}\ \bibnamefont {Schmid}},
  \bibinfo {author} {\bibfnamefont {S.}~\bibnamefont {Schlauderer}}, \bibinfo
  {author} {\bibfnamefont {M.}~\bibnamefont {Gmitra}}, \bibinfo {author}
  {\bibfnamefont {J.}~\bibnamefont {Fabian}}, \bibinfo {author} {\bibfnamefont
  {P.}~\bibnamefont {Nagler}}, \bibinfo {author} {\bibfnamefont
  {C.}~\bibnamefont {Sch{\"{u}}ller}}, \bibinfo {author} {\bibfnamefont
  {T.}~\bibnamefont {Korn}}, \bibinfo {author} {\bibfnamefont {P.~G.}\
  \bibnamefont {Hawkins}}, \bibinfo {author} {\bibfnamefont {J.~T.}\
  \bibnamefont {Steiner}}, \bibinfo {author} {\bibfnamefont {U.}~\bibnamefont
  {Huttner}}, \bibinfo {author} {\bibfnamefont {S.~W.}\ \bibnamefont {Koch}},
  \bibinfo {author} {\bibfnamefont {M.}~\bibnamefont {Kira}}, \ and\ \bibinfo
  {author} {\bibfnamefont {R.}~\bibnamefont {Huber}},\ }\href {\doibase
  10.1038/s41586-018-0013-6} {\bibfield  {journal} {\bibinfo  {journal}
  {Nature}\ }\textbf {\bibinfo {volume} {557}},\ \bibinfo {pages} {76}
  (\bibinfo {year} {2018})}\BibitemShut {NoStop}%
\bibitem [{\citenamefont {Zhong}\ \emph {et~al.}(2017)\citenamefont {Zhong},
  \citenamefont {Seyler}, \citenamefont {Linpeng}, \citenamefont {Cheng},
  \citenamefont {Sivadas}, \citenamefont {Huang}, \citenamefont {Schmidgall},
  \citenamefont {Taniguchi}, \citenamefont {Watanabe}, \citenamefont {McGuire},
  \citenamefont {Yao}, \citenamefont {Xiao}, \citenamefont {Fu},\ and\
  \citenamefont {Xu}}]{Zhong2017:SA}%
  \BibitemOpen
  \bibfield  {author} {\bibinfo {author} {\bibfnamefont {D.}~\bibnamefont
  {Zhong}}, \bibinfo {author} {\bibfnamefont {K.~L.}\ \bibnamefont {Seyler}},
  \bibinfo {author} {\bibfnamefont {X.}~\bibnamefont {Linpeng}}, \bibinfo
  {author} {\bibfnamefont {R.}~\bibnamefont {Cheng}}, \bibinfo {author}
  {\bibfnamefont {N.}~\bibnamefont {Sivadas}}, \bibinfo {author} {\bibfnamefont
  {B.}~\bibnamefont {Huang}}, \bibinfo {author} {\bibfnamefont
  {E.}~\bibnamefont {Schmidgall}}, \bibinfo {author} {\bibfnamefont
  {T.}~\bibnamefont {Taniguchi}}, \bibinfo {author} {\bibfnamefont
  {K.}~\bibnamefont {Watanabe}}, \bibinfo {author} {\bibfnamefont {M.~A.}\
  \bibnamefont {McGuire}}, \bibinfo {author} {\bibfnamefont {W.}~\bibnamefont
  {Yao}}, \bibinfo {author} {\bibfnamefont {D.}~\bibnamefont {Xiao}}, \bibinfo
  {author} {\bibfnamefont {K.-M.~C.}\ \bibnamefont {Fu}}, \ and\ \bibinfo
  {author} {\bibfnamefont {X.}~\bibnamefont {Xu}},\ }\href {\doibase
  10.1126/sciadv.1603113} {\bibfield  {journal} {\bibinfo  {journal} {Sci.
  Adv.}\ }\textbf {\bibinfo {volume} {3}},\ \bibinfo {pages} {e1603113}
  (\bibinfo {year} {2017})}\BibitemShut {NoStop}%
\bibitem [{\citenamefont {Bussolotti}\ \emph {et~al.}(2018)\citenamefont
  {Bussolotti}, \citenamefont {Kawai}, \citenamefont {Ooi}, \citenamefont
  {Chellappan}, \citenamefont {Thian}, \citenamefont {Pang},\ and\
  \citenamefont {Goh}}]{Chellappan2018:NF}%
  \BibitemOpen
  \bibfield  {author} {\bibinfo {author} {\bibfnamefont {F.}~\bibnamefont
  {Bussolotti}}, \bibinfo {author} {\bibfnamefont {H.}~\bibnamefont {Kawai}},
  \bibinfo {author} {\bibfnamefont {Z.~E.}\ \bibnamefont {Ooi}}, \bibinfo
  {author} {\bibfnamefont {V.}~\bibnamefont {Chellappan}}, \bibinfo {author}
  {\bibfnamefont {D.}~\bibnamefont {Thian}}, \bibinfo {author} {\bibfnamefont
  {A.~L.~C.}\ \bibnamefont {Pang}}, \ and\ \bibinfo {author} {\bibfnamefont
  {K.~E.~J.}\ \bibnamefont {Goh}},\ }\href {\doibase 10.1088/2399-1984/aac9d7}
  {\bibfield  {journal} {\bibinfo  {journal} {Nano Futur.}\ }\textbf {\bibinfo
  {volume} {2}},\ \bibinfo {pages} {032001} (\bibinfo {year}
  {2018})}\BibitemShut {NoStop}%
\bibitem [{\citenamefont {Schaibley}\ \emph {et~al.}(2016)\citenamefont
  {Schaibley}, \citenamefont {Yu}, \citenamefont {Clark}, \citenamefont
  {Rivera}, \citenamefont {Ross}, \citenamefont {Seyler}, \citenamefont {Yao},\
  and\ \citenamefont {Xu}}]{Schaibley2016:NRM}%
  \BibitemOpen
  \bibfield  {author} {\bibinfo {author} {\bibfnamefont {J.~R.}\ \bibnamefont
  {Schaibley}}, \bibinfo {author} {\bibfnamefont {H.}~\bibnamefont {Yu}},
  \bibinfo {author} {\bibfnamefont {G.}~\bibnamefont {Clark}}, \bibinfo
  {author} {\bibfnamefont {P.}~\bibnamefont {Rivera}}, \bibinfo {author}
  {\bibfnamefont {J.~S.}\ \bibnamefont {Ross}}, \bibinfo {author}
  {\bibfnamefont {K.~L.}\ \bibnamefont {Seyler}}, \bibinfo {author}
  {\bibfnamefont {W.}~\bibnamefont {Yao}}, \ and\ \bibinfo {author}
  {\bibfnamefont {X.}~\bibnamefont {Xu}},\ }\href {\doibase
  10.1038/natrevmats.2016.55} {\bibfield  {journal} {\bibinfo  {journal} {Nat.
  Rev. Mater.}\ }\textbf {\bibinfo {volume} {1}},\ \bibinfo {pages} {1}
  (\bibinfo {year} {2016})}\BibitemShut {NoStop}%
\bibitem [{\citenamefont {Lin}\ \emph {et~al.}(2018)\citenamefont {Lin},
  \citenamefont {Lei}, \citenamefont {Subramanian}, \citenamefont {Briggs},
  \citenamefont {Wang}, \citenamefont {Lo}, \citenamefont {Yalon},
  \citenamefont {Lloyd}, \citenamefont {Wu}, \citenamefont {Koski},
  \citenamefont {Clark}, \citenamefont {Das}, \citenamefont {Wallace},
  \citenamefont {Kuech}, \citenamefont {Bunch}, \citenamefont {Li},
  \citenamefont {Chen}, \citenamefont {Pop}, \citenamefont {Crespi},
  \citenamefont {Robinson},\ and\ \citenamefont {Terrones}}]{Lin2018:APL}%
  \BibitemOpen
  \bibfield  {author} {\bibinfo {author} {\bibfnamefont {Z.}~\bibnamefont
  {Lin}}, \bibinfo {author} {\bibfnamefont {Y.}~\bibnamefont {Lei}}, \bibinfo
  {author} {\bibfnamefont {S.}~\bibnamefont {Subramanian}}, \bibinfo {author}
  {\bibfnamefont {N.}~\bibnamefont {Briggs}}, \bibinfo {author} {\bibfnamefont
  {Y.}~\bibnamefont {Wang}}, \bibinfo {author} {\bibfnamefont {C.~L.}\
  \bibnamefont {Lo}}, \bibinfo {author} {\bibfnamefont {E.}~\bibnamefont
  {Yalon}}, \bibinfo {author} {\bibfnamefont {D.}~\bibnamefont {Lloyd}},
  \bibinfo {author} {\bibfnamefont {S.}~\bibnamefont {Wu}}, \bibinfo {author}
  {\bibfnamefont {K.}~\bibnamefont {Koski}}, \bibinfo {author} {\bibfnamefont
  {R.}~\bibnamefont {Clark}}, \bibinfo {author} {\bibfnamefont
  {S.}~\bibnamefont {Das}}, \bibinfo {author} {\bibfnamefont {R.~M.}\
  \bibnamefont {Wallace}}, \bibinfo {author} {\bibfnamefont {T.}~\bibnamefont
  {Kuech}}, \bibinfo {author} {\bibfnamefont {J.~S.}\ \bibnamefont {Bunch}},
  \bibinfo {author} {\bibfnamefont {X.}~\bibnamefont {Li}}, \bibinfo {author}
  {\bibfnamefont {Z.}~\bibnamefont {Chen}}, \bibinfo {author} {\bibfnamefont
  {E.}~\bibnamefont {Pop}}, \bibinfo {author} {\bibfnamefont {V.~H.}\
  \bibnamefont {Crespi}}, \bibinfo {author} {\bibfnamefont {J.~A.}\
  \bibnamefont {Robinson}}, \ and\ \bibinfo {author} {\bibfnamefont
  {M.}~\bibnamefont {Terrones}},\ }\href {\doibase 10.1063/1.5042598}
  {\bibfield  {journal} {\bibinfo  {journal} {APL Mater.}\ }\textbf {\bibinfo
  {volume} {6}} (\bibinfo {year} {2018}),\ 10.1063/1.5042598}\BibitemShut
  {NoStop}%
\bibitem [{\citenamefont {Rold{\'{a}}n}\ \emph {et~al.}(2015)\citenamefont
  {Rold{\'{a}}n}, \citenamefont {Castellanos-Gomez}, \citenamefont
  {Cappelluti},\ and\ \citenamefont {Guinea}}]{Roldan2015:JPCM}%
  \BibitemOpen
  \bibfield  {author} {\bibinfo {author} {\bibfnamefont {R.}~\bibnamefont
  {Rold{\'{a}}n}}, \bibinfo {author} {\bibfnamefont {A.}~\bibnamefont
  {Castellanos-Gomez}}, \bibinfo {author} {\bibfnamefont {E.}~\bibnamefont
  {Cappelluti}}, \ and\ \bibinfo {author} {\bibfnamefont {F.}~\bibnamefont
  {Guinea}},\ }\href {\doibase 10.1088/0953-8984/27/31/313201} {\bibfield
  {journal} {\bibinfo  {journal} {J. Phys. Condens. Matter}\ }\textbf {\bibinfo
  {volume} {27}},\ \bibinfo {pages} {313201} (\bibinfo {year} {2015})},\
  \Eprint {http://arxiv.org/abs/1504.07926} {1504.07926} \BibitemShut {NoStop}%
\bibitem [{\citenamefont {Fang}\ \emph {et~al.}(2018)\citenamefont {Fang},
  \citenamefont {Carr}, \citenamefont {Cazalilla},\ and\ \citenamefont
  {Kaxiras}}]{Fang2018:PRB}%
  \BibitemOpen
  \bibfield  {author} {\bibinfo {author} {\bibfnamefont {S.}~\bibnamefont
  {Fang}}, \bibinfo {author} {\bibfnamefont {S.}~\bibnamefont {Carr}}, \bibinfo
  {author} {\bibfnamefont {M.~A.}\ \bibnamefont {Cazalilla}}, \ and\ \bibinfo
  {author} {\bibfnamefont {E.}~\bibnamefont {Kaxiras}},\ }\href {\doibase
  10.1103/PhysRevB.98.075106} {\bibfield  {journal} {\bibinfo  {journal} {Phys.
  Rev. B}\ }\textbf {\bibinfo {volume} {98}},\ \bibinfo {pages} {075106}
  (\bibinfo {year} {2018})},\ \Eprint {http://arxiv.org/abs/1709.07510}
  {1709.07510} \BibitemShut {NoStop}%
\bibitem [{\citenamefont {David}\ \emph {et~al.}(2019)\citenamefont {David},
  \citenamefont {Rakyta}, \citenamefont {Korm\'anyos},\ and\ \citenamefont
  {Burkard}}]{David2019:PRB}%
  \BibitemOpen
  \bibfield  {author} {\bibinfo {author} {\bibfnamefont {A.}~\bibnamefont
  {David}}, \bibinfo {author} {\bibfnamefont {P.}~\bibnamefont {Rakyta}},
  \bibinfo {author} {\bibfnamefont {A.}~\bibnamefont {Korm\'anyos}}, \ and\
  \bibinfo {author} {\bibfnamefont {G.}~\bibnamefont {Burkard}},\ }\href
  {\doibase 10.1103/PhysRevB.100.085412} {\bibfield  {journal} {\bibinfo
  {journal} {Phys. Rev. B}\ }\textbf {\bibinfo {volume} {100}},\ \bibinfo
  {pages} {085412} (\bibinfo {year} {2019})}\BibitemShut {NoStop}%
\bibitem [{\citenamefont {Li}\ and\ \citenamefont
  {Koshino}(2019)}]{Li2019:PRB}%
  \BibitemOpen
  \bibfield  {author} {\bibinfo {author} {\bibfnamefont {Y.}~\bibnamefont
  {Li}}\ and\ \bibinfo {author} {\bibfnamefont {M.}~\bibnamefont {Koshino}},\
  }\href {\doibase 10.1103/PhysRevB.99.075438} {\bibfield  {journal} {\bibinfo
  {journal} {Phys. Rev. B}\ }\textbf {\bibinfo {volume} {99}},\ \bibinfo
  {pages} {075438} (\bibinfo {year} {2019})}\BibitemShut {NoStop}%
\bibitem [{\citenamefont {\ifmmode \check{Z}\else
  \v{Z}\fi{}uti\ifmmode~\acute{c}\else \'{c}\fi{}}\ \emph
  {et~al.}(2004)\citenamefont {\ifmmode \check{Z}\else
  \v{Z}\fi{}uti\ifmmode~\acute{c}\else \'{c}\fi{}}, \citenamefont {Fabian},\
  and\ \citenamefont {Das~Sarma}}]{Zutic2004:RMP}%
  \BibitemOpen
  \bibfield  {author} {\bibinfo {author} {\bibfnamefont {I.}~\bibnamefont
  {\ifmmode \check{Z}\else \v{Z}\fi{}uti\ifmmode~\acute{c}\else \'{c}\fi{}}},
  \bibinfo {author} {\bibfnamefont {J.}~\bibnamefont {Fabian}}, \ and\ \bibinfo
  {author} {\bibfnamefont {S.}~\bibnamefont {Das~Sarma}},\ }\href {\doibase
  10.1103/RevModPhys.76.323} {\bibfield  {journal} {\bibinfo  {journal} {Rev.
  Mod. Phys.}\ }\textbf {\bibinfo {volume} {76}},\ \bibinfo {pages} {323}
  (\bibinfo {year} {2004})}\BibitemShut {NoStop}%
\bibitem [{\citenamefont {Han}\ \emph {et~al.}(2014)\citenamefont {Han},
  \citenamefont {Kawakami}, \citenamefont {Gmitra},\ and\ \citenamefont
  {Fabian}}]{Han2014:NN}%
  \BibitemOpen
  \bibfield  {author} {\bibinfo {author} {\bibfnamefont {W.}~\bibnamefont
  {Han}}, \bibinfo {author} {\bibfnamefont {R.~K.}\ \bibnamefont {Kawakami}},
  \bibinfo {author} {\bibfnamefont {M.}~\bibnamefont {Gmitra}}, \ and\ \bibinfo
  {author} {\bibfnamefont {J.}~\bibnamefont {Fabian}},\ }\href {\doibase
  10.1038/nnano.2014.214} {\bibfield  {journal} {\bibinfo  {journal} {Nat.
  Nano.}\ }\textbf {\bibinfo {volume} {9}},\ \bibinfo {pages} {794} (\bibinfo
  {year} {2014})}\BibitemShut {NoStop}%
\bibitem [{\citenamefont {Fabian}\ \emph {et~al.}(2007)\citenamefont {Fabian},
  \citenamefont {Matos-Abiague}, \citenamefont {Ertler}, \citenamefont
  {Stano},\ and\ \citenamefont {{\v{Z}}uti{\'{c}}}}]{Fabian2007:APS}%
  \BibitemOpen
  \bibfield  {author} {\bibinfo {author} {\bibfnamefont {J.}~\bibnamefont
  {Fabian}}, \bibinfo {author} {\bibfnamefont {A.}~\bibnamefont
  {Matos-Abiague}}, \bibinfo {author} {\bibfnamefont {C.}~\bibnamefont
  {Ertler}}, \bibinfo {author} {\bibfnamefont {P.}~\bibnamefont {Stano}}, \
  and\ \bibinfo {author} {\bibfnamefont {I.}~\bibnamefont
  {{\v{Z}}uti{\'{c}}}},\ }\href {\doibase 10.2478/v10155-010-0086-8} {\bibfield
   {journal} {\bibinfo  {journal} {Acta Phys. Slov.}\ }\textbf {\bibinfo
  {volume} {57}},\ \bibinfo {pages} {342} (\bibinfo {year} {2007})}\BibitemShut
  {NoStop}%
\bibitem [{\citenamefont {Avsar}\ \emph {et~al.}(2017)\citenamefont {Avsar},
  \citenamefont {Unuchek}, \citenamefont {Liu}, \citenamefont {Sanchez},
  \citenamefont {Watanabe}, \citenamefont {Taniguchi}, \citenamefont
  {{\"{O}}zyilmaz},\ and\ \citenamefont {Kis}}]{Avsar2017:ACS}%
  \BibitemOpen
  \bibfield  {author} {\bibinfo {author} {\bibfnamefont {A.}~\bibnamefont
  {Avsar}}, \bibinfo {author} {\bibfnamefont {D.}~\bibnamefont {Unuchek}},
  \bibinfo {author} {\bibfnamefont {J.}~\bibnamefont {Liu}}, \bibinfo {author}
  {\bibfnamefont {O.~L.}\ \bibnamefont {Sanchez}}, \bibinfo {author}
  {\bibfnamefont {K.}~\bibnamefont {Watanabe}}, \bibinfo {author}
  {\bibfnamefont {T.}~\bibnamefont {Taniguchi}}, \bibinfo {author}
  {\bibfnamefont {B.}~\bibnamefont {{\"{O}}zyilmaz}}, \ and\ \bibinfo {author}
  {\bibfnamefont {A.}~\bibnamefont {Kis}},\ }\href {\doibase
  10.1021/acsnano.7b06800} {\bibfield  {journal} {\bibinfo  {journal} {ACS
  Nano}\ }\textbf {\bibinfo {volume} {11}},\ \bibinfo {pages} {11678} (\bibinfo
  {year} {2017})}\BibitemShut {NoStop}%
\bibitem [{\citenamefont {Seyler}\ \emph {et~al.}(2018)\citenamefont {Seyler},
  \citenamefont {Zhong}, \citenamefont {Huang}, \citenamefont {Linpeng},
  \citenamefont {Wilson}, \citenamefont {Taniguchi}, \citenamefont {Watanabe},
  \citenamefont {Yao}, \citenamefont {Xiao}, \citenamefont {McGuire},
  \citenamefont {Fu},\ and\ \citenamefont {Xu}}]{Seyler2018:NL}%
  \BibitemOpen
  \bibfield  {author} {\bibinfo {author} {\bibfnamefont {K.~L.}\ \bibnamefont
  {Seyler}}, \bibinfo {author} {\bibfnamefont {D.}~\bibnamefont {Zhong}},
  \bibinfo {author} {\bibfnamefont {B.}~\bibnamefont {Huang}}, \bibinfo
  {author} {\bibfnamefont {X.}~\bibnamefont {Linpeng}}, \bibinfo {author}
  {\bibfnamefont {N.~P.}\ \bibnamefont {Wilson}}, \bibinfo {author}
  {\bibfnamefont {T.}~\bibnamefont {Taniguchi}}, \bibinfo {author}
  {\bibfnamefont {K.}~\bibnamefont {Watanabe}}, \bibinfo {author}
  {\bibfnamefont {W.}~\bibnamefont {Yao}}, \bibinfo {author} {\bibfnamefont
  {D.}~\bibnamefont {Xiao}}, \bibinfo {author} {\bibfnamefont {M.~A.}\
  \bibnamefont {McGuire}}, \bibinfo {author} {\bibfnamefont {K.-M.~C.}\
  \bibnamefont {Fu}}, \ and\ \bibinfo {author} {\bibfnamefont {X.}~\bibnamefont
  {Xu}},\ }\href {\doibase 10.1021/acs.nanolett.8b01105} {\bibfield  {journal}
  {\bibinfo  {journal} {Nano Lett.}\ }\textbf {\bibinfo {volume} {18}},\
  \bibinfo {pages} {3823} (\bibinfo {year} {2018})},\ \Eprint
  {http://arxiv.org/abs/1805.08738} {1805.08738} \BibitemShut {NoStop}%
\bibitem [{\citenamefont {Zhang}\ \emph
  {et~al.}(2009{\natexlab{a}})\citenamefont {Zhang}, \citenamefont {Liu},
  \citenamefont {Qi}, \citenamefont {Dai}, \citenamefont {Fang},\ and\
  \citenamefont {Zhang}}]{Zhang2009:NP}%
  \BibitemOpen
  \bibfield  {author} {\bibinfo {author} {\bibfnamefont {H.}~\bibnamefont
  {Zhang}}, \bibinfo {author} {\bibfnamefont {C.~X.}\ \bibnamefont {Liu}},
  \bibinfo {author} {\bibfnamefont {X.~L.}\ \bibnamefont {Qi}}, \bibinfo
  {author} {\bibfnamefont {X.}~\bibnamefont {Dai}}, \bibinfo {author}
  {\bibfnamefont {Z.}~\bibnamefont {Fang}}, \ and\ \bibinfo {author}
  {\bibfnamefont {S.~C.}\ \bibnamefont {Zhang}},\ }\href {\doibase
  10.1038/nphys1270} {\bibfield  {journal} {\bibinfo  {journal} {Nat. Phys.}\
  }\textbf {\bibinfo {volume} {5}},\ \bibinfo {pages} {438} (\bibinfo {year}
  {2009}{\natexlab{a}})}\BibitemShut {NoStop}%
\bibitem [{\citenamefont {Hsieh}\ \emph {et~al.}(2009)\citenamefont {Hsieh},
  \citenamefont {Xia}, \citenamefont {Qian}, \citenamefont {Wray},
  \citenamefont {Dil}, \citenamefont {Meier}, \citenamefont {Osterwalder},
  \citenamefont {Patthey}, \citenamefont {Checkelsky}, \citenamefont {Ong},
  \citenamefont {Fedorov}, \citenamefont {Lin}, \citenamefont {Bansil},
  \citenamefont {Grauer}, \citenamefont {Hor}, \citenamefont {Cava},\ and\
  \citenamefont {Hasan}}]{Hsieh2009:Nat}%
  \BibitemOpen
  \bibfield  {author} {\bibinfo {author} {\bibfnamefont {D.}~\bibnamefont
  {Hsieh}}, \bibinfo {author} {\bibfnamefont {Y.}~\bibnamefont {Xia}}, \bibinfo
  {author} {\bibfnamefont {D.}~\bibnamefont {Qian}}, \bibinfo {author}
  {\bibfnamefont {L.}~\bibnamefont {Wray}}, \bibinfo {author} {\bibfnamefont
  {J.~H.}\ \bibnamefont {Dil}}, \bibinfo {author} {\bibfnamefont
  {F.}~\bibnamefont {Meier}}, \bibinfo {author} {\bibfnamefont
  {J.}~\bibnamefont {Osterwalder}}, \bibinfo {author} {\bibfnamefont
  {L.}~\bibnamefont {Patthey}}, \bibinfo {author} {\bibfnamefont {J.~G.}\
  \bibnamefont {Checkelsky}}, \bibinfo {author} {\bibfnamefont {N.~P.}\
  \bibnamefont {Ong}}, \bibinfo {author} {\bibfnamefont {A.~V.}\ \bibnamefont
  {Fedorov}}, \bibinfo {author} {\bibfnamefont {H.}~\bibnamefont {Lin}},
  \bibinfo {author} {\bibfnamefont {A.}~\bibnamefont {Bansil}}, \bibinfo
  {author} {\bibfnamefont {D.}~\bibnamefont {Grauer}}, \bibinfo {author}
  {\bibfnamefont {Y.~S.}\ \bibnamefont {Hor}}, \bibinfo {author} {\bibfnamefont
  {R.~J.}\ \bibnamefont {Cava}}, \ and\ \bibinfo {author} {\bibfnamefont
  {M.~Z.}\ \bibnamefont {Hasan}},\ }\href {\doibase 10.1038/nature08234}
  {\bibfield  {journal} {\bibinfo  {journal} {Nature}\ }\textbf {\bibinfo
  {volume} {460}},\ \bibinfo {pages} {1101} (\bibinfo {year} {2009})},\ \Eprint
  {http://arxiv.org/abs/arXiv:0908.0564} {arXiv:0908.0564} \BibitemShut
  {NoStop}%
\bibitem [{\citenamefont {Zhang}\ \emph {et~al.}(2010)\citenamefont {Zhang},
  \citenamefont {He}, \citenamefont {Chang}, \citenamefont {Song},
  \citenamefont {Wang}, \citenamefont {Chen}, \citenamefont {Jia},
  \citenamefont {Fang}, \citenamefont {Dai}, \citenamefont {Shan},
  \citenamefont {Shen}, \citenamefont {Niu}, \citenamefont {Qi}, \citenamefont
  {Zhang}, \citenamefont {Ma},\ and\ \citenamefont {Xue}}]{Zhang2010:NP}%
  \BibitemOpen
  \bibfield  {author} {\bibinfo {author} {\bibfnamefont {Y.}~\bibnamefont
  {Zhang}}, \bibinfo {author} {\bibfnamefont {K.}~\bibnamefont {He}}, \bibinfo
  {author} {\bibfnamefont {C.~Z.}\ \bibnamefont {Chang}}, \bibinfo {author}
  {\bibfnamefont {C.~L.}\ \bibnamefont {Song}}, \bibinfo {author}
  {\bibfnamefont {L.~L.}\ \bibnamefont {Wang}}, \bibinfo {author}
  {\bibfnamefont {X.}~\bibnamefont {Chen}}, \bibinfo {author} {\bibfnamefont
  {J.~F.}\ \bibnamefont {Jia}}, \bibinfo {author} {\bibfnamefont
  {Z.}~\bibnamefont {Fang}}, \bibinfo {author} {\bibfnamefont {X.}~\bibnamefont
  {Dai}}, \bibinfo {author} {\bibfnamefont {W.~Y.}\ \bibnamefont {Shan}},
  \bibinfo {author} {\bibfnamefont {S.~Q.}\ \bibnamefont {Shen}}, \bibinfo
  {author} {\bibfnamefont {Q.}~\bibnamefont {Niu}}, \bibinfo {author}
  {\bibfnamefont {X.~L.}\ \bibnamefont {Qi}}, \bibinfo {author} {\bibfnamefont
  {S.~C.}\ \bibnamefont {Zhang}}, \bibinfo {author} {\bibfnamefont {X.~C.}\
  \bibnamefont {Ma}}, \ and\ \bibinfo {author} {\bibfnamefont {Q.~K.}\
  \bibnamefont {Xue}},\ }\href {\doibase 10.1038/nphys1689} {\bibfield
  {journal} {\bibinfo  {journal} {Nat. Phys.}\ }\textbf {\bibinfo {volume}
  {6}},\ \bibinfo {pages} {584} (\bibinfo {year} {2010})}\BibitemShut {NoStop}%
\bibitem [{\citenamefont {Liu}\ \emph {et~al.}(2010)\citenamefont {Liu},
  \citenamefont {Zhang}, \citenamefont {Yan}, \citenamefont {Qi}, \citenamefont
  {Frauenheim}, \citenamefont {Dai}, \citenamefont {Fang},\ and\ \citenamefont
  {Zhang}}]{Liu2010:PRB}%
  \BibitemOpen
  \bibfield  {author} {\bibinfo {author} {\bibfnamefont {C.-X.}\ \bibnamefont
  {Liu}}, \bibinfo {author} {\bibfnamefont {H.}~\bibnamefont {Zhang}}, \bibinfo
  {author} {\bibfnamefont {B.}~\bibnamefont {Yan}}, \bibinfo {author}
  {\bibfnamefont {X.-L.}\ \bibnamefont {Qi}}, \bibinfo {author} {\bibfnamefont
  {T.}~\bibnamefont {Frauenheim}}, \bibinfo {author} {\bibfnamefont
  {X.}~\bibnamefont {Dai}}, \bibinfo {author} {\bibfnamefont {Z.}~\bibnamefont
  {Fang}}, \ and\ \bibinfo {author} {\bibfnamefont {S.-C.}\ \bibnamefont
  {Zhang}},\ }\href {\doibase 10.1103/PhysRevB.81.041307} {\bibfield  {journal}
  {\bibinfo  {journal} {Phys. Rev. B}\ }\textbf {\bibinfo {volume} {81}},\
  \bibinfo {pages} {041307} (\bibinfo {year} {2010})}\BibitemShut {NoStop}%
\bibitem [{\citenamefont {Yazyev}\ \emph {et~al.}(2010)\citenamefont {Yazyev},
  \citenamefont {Moore},\ and\ \citenamefont {Louie}}]{Yazyev2010:PRL}%
  \BibitemOpen
  \bibfield  {author} {\bibinfo {author} {\bibfnamefont {O.~V.}\ \bibnamefont
  {Yazyev}}, \bibinfo {author} {\bibfnamefont {J.~E.}\ \bibnamefont {Moore}}, \
  and\ \bibinfo {author} {\bibfnamefont {S.~G.}\ \bibnamefont {Louie}},\ }\href
  {\doibase 10.1103/PhysRevLett.105.266806} {\bibfield  {journal} {\bibinfo
  {journal} {Phys. Rev. Lett.}\ }\textbf {\bibinfo {volume} {105}},\ \bibinfo
  {pages} {266806} (\bibinfo {year} {2010})}\BibitemShut {NoStop}%
\bibitem [{\citenamefont {Park}\ \emph {et~al.}(2010)\citenamefont {Park},
  \citenamefont {Heremans}, \citenamefont {Scarola},\ and\ \citenamefont
  {Minic}}]{Park2010:PRL}%
  \BibitemOpen
  \bibfield  {author} {\bibinfo {author} {\bibfnamefont {K.}~\bibnamefont
  {Park}}, \bibinfo {author} {\bibfnamefont {J.~J.}\ \bibnamefont {Heremans}},
  \bibinfo {author} {\bibfnamefont {V.~W.}\ \bibnamefont {Scarola}}, \ and\
  \bibinfo {author} {\bibfnamefont {D.}~\bibnamefont {Minic}},\ }\href
  {\doibase 10.1103/PhysRevLett.105.186801} {\bibfield  {journal} {\bibinfo
  {journal} {Phys. Rev. Lett.}\ }\textbf {\bibinfo {volume} {105}},\ \bibinfo
  {pages} {186801} (\bibinfo {year} {2010})}\BibitemShut {NoStop}%
\bibitem [{\citenamefont {Tian}\ \emph {et~al.}(2017)\citenamefont {Tian},
  \citenamefont {Yu}, \citenamefont {Shi},\ and\ \citenamefont
  {Wang}}]{Tian2017:Mat}%
  \BibitemOpen
  \bibfield  {author} {\bibinfo {author} {\bibfnamefont {W.}~\bibnamefont
  {Tian}}, \bibinfo {author} {\bibfnamefont {W.}~\bibnamefont {Yu}}, \bibinfo
  {author} {\bibfnamefont {J.}~\bibnamefont {Shi}}, \ and\ \bibinfo {author}
  {\bibfnamefont {Y.}~\bibnamefont {Wang}},\ }\href {\doibase
  10.3390/ma10070814} {\bibfield  {journal} {\bibinfo  {journal} {Materials
  (Basel).}\ }\textbf {\bibinfo {volume} {10}},\ \bibinfo {pages} {814}
  (\bibinfo {year} {2017})}\BibitemShut {NoStop}%
\bibitem [{\citenamefont {Hasan}\ and\ \citenamefont
  {Kane}(2010)}]{Hasan2010:RMP}%
  \BibitemOpen
  \bibfield  {author} {\bibinfo {author} {\bibfnamefont {M.~Z.}\ \bibnamefont
  {Hasan}}\ and\ \bibinfo {author} {\bibfnamefont {C.~L.}\ \bibnamefont
  {Kane}},\ }\href {\doibase 10.1103/RevModPhys.82.3045} {\bibfield  {journal}
  {\bibinfo  {journal} {Rev. Mod. Phys.}\ }\textbf {\bibinfo {volume} {82}},\
  \bibinfo {pages} {3045} (\bibinfo {year} {2010})}\BibitemShut {NoStop}%
\bibitem [{\citenamefont {Zhang}\ \emph
  {et~al.}(2009{\natexlab{b}})\citenamefont {Zhang}, \citenamefont {Cheng},
  \citenamefont {Chen}, \citenamefont {Jia}, \citenamefont {Ma}, \citenamefont
  {He}, \citenamefont {Wang}, \citenamefont {Zhang}, \citenamefont {Dai},
  \citenamefont {Fang}, \citenamefont {Xie},\ and\ \citenamefont
  {Xue}}]{Zhang2009:PRL}%
  \BibitemOpen
  \bibfield  {author} {\bibinfo {author} {\bibfnamefont {T.}~\bibnamefont
  {Zhang}}, \bibinfo {author} {\bibfnamefont {P.}~\bibnamefont {Cheng}},
  \bibinfo {author} {\bibfnamefont {X.}~\bibnamefont {Chen}}, \bibinfo {author}
  {\bibfnamefont {J.-F.}\ \bibnamefont {Jia}}, \bibinfo {author} {\bibfnamefont
  {X.}~\bibnamefont {Ma}}, \bibinfo {author} {\bibfnamefont {K.}~\bibnamefont
  {He}}, \bibinfo {author} {\bibfnamefont {L.}~\bibnamefont {Wang}}, \bibinfo
  {author} {\bibfnamefont {H.}~\bibnamefont {Zhang}}, \bibinfo {author}
  {\bibfnamefont {X.}~\bibnamefont {Dai}}, \bibinfo {author} {\bibfnamefont
  {Z.}~\bibnamefont {Fang}}, \bibinfo {author} {\bibfnamefont {X.}~\bibnamefont
  {Xie}}, \ and\ \bibinfo {author} {\bibfnamefont {Q.-K.}\ \bibnamefont
  {Xue}},\ }\href {\doibase 10.1103/PhysRevLett.103.266803} {\bibfield
  {journal} {\bibinfo  {journal} {Phys. Rev. Lett.}\ }\textbf {\bibinfo
  {volume} {103}},\ \bibinfo {pages} {266803} (\bibinfo {year}
  {2009}{\natexlab{b}})}\BibitemShut {NoStop}%
\bibitem [{\citenamefont {Koirala}\ \emph {et~al.}(2015)\citenamefont
  {Koirala}, \citenamefont {Brahlek}, \citenamefont {Salehi}, \citenamefont
  {Wu}, \citenamefont {Dai}, \citenamefont {Waugh}, \citenamefont {Nummy},
  \citenamefont {Han}, \citenamefont {Moon}, \citenamefont {Zhu}, \citenamefont
  {Dessau}, \citenamefont {Wu}, \citenamefont {Armitage},\ and\ \citenamefont
  {Oh}}]{Koirala2015:NL}%
  \BibitemOpen
  \bibfield  {author} {\bibinfo {author} {\bibfnamefont {N.}~\bibnamefont
  {Koirala}}, \bibinfo {author} {\bibfnamefont {M.}~\bibnamefont {Brahlek}},
  \bibinfo {author} {\bibfnamefont {M.}~\bibnamefont {Salehi}}, \bibinfo
  {author} {\bibfnamefont {L.}~\bibnamefont {Wu}}, \bibinfo {author}
  {\bibfnamefont {J.}~\bibnamefont {Dai}}, \bibinfo {author} {\bibfnamefont
  {J.}~\bibnamefont {Waugh}}, \bibinfo {author} {\bibfnamefont
  {T.}~\bibnamefont {Nummy}}, \bibinfo {author} {\bibfnamefont {M.-G.}\
  \bibnamefont {Han}}, \bibinfo {author} {\bibfnamefont {J.}~\bibnamefont
  {Moon}}, \bibinfo {author} {\bibfnamefont {Y.}~\bibnamefont {Zhu}}, \bibinfo
  {author} {\bibfnamefont {D.}~\bibnamefont {Dessau}}, \bibinfo {author}
  {\bibfnamefont {W.}~\bibnamefont {Wu}}, \bibinfo {author} {\bibfnamefont
  {N.~P.}\ \bibnamefont {Armitage}}, \ and\ \bibinfo {author} {\bibfnamefont
  {S.}~\bibnamefont {Oh}},\ }\href {\doibase 10.1021/acs.nanolett.5b03770}
  {\bibfield  {journal} {\bibinfo  {journal} {Nano Lett.}\ }\textbf {\bibinfo
  {volume} {15}},\ \bibinfo {pages} {8245} (\bibinfo {year}
  {2015})}\BibitemShut {NoStop}%
\bibitem [{\citenamefont {Song}\ \emph {et~al.}(2018)\citenamefont {Song},
  \citenamefont {Soriano}, \citenamefont {Cummings}, \citenamefont {Robles},
  \citenamefont {Ordej{\'{o}}n},\ and\ \citenamefont {Roche}}]{Song2018:NL}%
  \BibitemOpen
  \bibfield  {author} {\bibinfo {author} {\bibfnamefont {K.}~\bibnamefont
  {Song}}, \bibinfo {author} {\bibfnamefont {D.}~\bibnamefont {Soriano}},
  \bibinfo {author} {\bibfnamefont {A.~W.}\ \bibnamefont {Cummings}}, \bibinfo
  {author} {\bibfnamefont {R.}~\bibnamefont {Robles}}, \bibinfo {author}
  {\bibfnamefont {P.}~\bibnamefont {Ordej{\'{o}}n}}, \ and\ \bibinfo {author}
  {\bibfnamefont {S.}~\bibnamefont {Roche}},\ }\href {\doibase
  10.1021/acs.nanolett.7b05482} {\bibfield  {journal} {\bibinfo  {journal}
  {Nano Lett.}\ }\textbf {\bibinfo {volume} {18}},\ \bibinfo {pages} {2033}
  (\bibinfo {year} {2018})}\BibitemShut {NoStop}%
\bibitem [{\citenamefont {Khokhriakov}\ \emph {et~al.}(2018)\citenamefont
  {Khokhriakov}, \citenamefont {Cummings}, \citenamefont {Song}, \citenamefont
  {Vila}, \citenamefont {Karpiak}, \citenamefont {Dankert}, \citenamefont
  {Roche},\ and\ \citenamefont {Dash}}]{Khokhriakov2018:SA}%
  \BibitemOpen
  \bibfield  {author} {\bibinfo {author} {\bibfnamefont {D.}~\bibnamefont
  {Khokhriakov}}, \bibinfo {author} {\bibfnamefont {A.~W.}\ \bibnamefont
  {Cummings}}, \bibinfo {author} {\bibfnamefont {K.}~\bibnamefont {Song}},
  \bibinfo {author} {\bibfnamefont {M.}~\bibnamefont {Vila}}, \bibinfo {author}
  {\bibfnamefont {B.}~\bibnamefont {Karpiak}}, \bibinfo {author} {\bibfnamefont
  {A.}~\bibnamefont {Dankert}}, \bibinfo {author} {\bibfnamefont
  {S.}~\bibnamefont {Roche}}, \ and\ \bibinfo {author} {\bibfnamefont {S.~P.}\
  \bibnamefont {Dash}},\ }\href {\doibase 10.1126/sciadv.aat9349} {\bibfield
  {journal} {\bibinfo  {journal} {Sci. Adv.}\ }\textbf {\bibinfo {volume}
  {4}},\ \bibinfo {pages} {eaat9349} (\bibinfo {year} {2018})}\BibitemShut
  {NoStop}%
\bibitem [{\citenamefont {Jafarpisheh}\ \emph {et~al.}(2018)\citenamefont
  {Jafarpisheh}, \citenamefont {Cummings}, \citenamefont {Watanabe},
  \citenamefont {Taniguchi}, \citenamefont {Beschoten},\ and\ \citenamefont
  {Stampfer}}]{Jafarpisheh2018:PRB}%
  \BibitemOpen
  \bibfield  {author} {\bibinfo {author} {\bibfnamefont {S.}~\bibnamefont
  {Jafarpisheh}}, \bibinfo {author} {\bibfnamefont {A.~W.}\ \bibnamefont
  {Cummings}}, \bibinfo {author} {\bibfnamefont {K.}~\bibnamefont {Watanabe}},
  \bibinfo {author} {\bibfnamefont {T.}~\bibnamefont {Taniguchi}}, \bibinfo
  {author} {\bibfnamefont {B.}~\bibnamefont {Beschoten}}, \ and\ \bibinfo
  {author} {\bibfnamefont {C.}~\bibnamefont {Stampfer}},\ }\href {\doibase
  10.1103/PhysRevB.98.241402} {\bibfield  {journal} {\bibinfo  {journal} {Phys.
  Rev. B}\ }\textbf {\bibinfo {volume} {98}},\ \bibinfo {pages} {241402(R)}
  (\bibinfo {year} {2018})}\BibitemShut {NoStop}%
\bibitem [{\citenamefont {Zhang}\ \emph {et~al.}(2014)\citenamefont {Zhang},
  \citenamefont {Triola},\ and\ \citenamefont {Rossi}}]{Zhang2014:PRL}%
  \BibitemOpen
  \bibfield  {author} {\bibinfo {author} {\bibfnamefont {J.}~\bibnamefont
  {Zhang}}, \bibinfo {author} {\bibfnamefont {C.}~\bibnamefont {Triola}}, \
  and\ \bibinfo {author} {\bibfnamefont {E.}~\bibnamefont {Rossi}},\ }\href
  {\doibase 10.1103/PhysRevLett.112.096802} {\bibfield  {journal} {\bibinfo
  {journal} {Phys. Rev. Lett.}\ }\textbf {\bibinfo {volume} {112}},\ \bibinfo
  {pages} {096802} (\bibinfo {year} {2014})}\BibitemShut {NoStop}%
\bibitem [{\citenamefont {Hesjedal}\ and\ \citenamefont
  {Chen}(2017)}]{Hesjedal2016:NM}%
  \BibitemOpen
  \bibfield  {author} {\bibinfo {author} {\bibfnamefont {T.}~\bibnamefont
  {Hesjedal}}\ and\ \bibinfo {author} {\bibfnamefont {Y.}~\bibnamefont
  {Chen}},\ }\href {\doibase 10.1038/nmat4835} {\bibfield  {journal} {\bibinfo
  {journal} {Nat. Mater.}\ }\textbf {\bibinfo {volume} {16}},\ \bibinfo {pages}
  {3} (\bibinfo {year} {2017})}\BibitemShut {NoStop}%
\bibitem [{\citenamefont {Zhao}\ \emph {et~al.}(2017)\citenamefont {Zhao},
  \citenamefont {Xu}, \citenamefont {Pan}, \citenamefont {Zhou}, \citenamefont
  {Zhou},\ and\ \citenamefont {Chai}}]{Zhao2017:AFM}%
  \BibitemOpen
  \bibfield  {author} {\bibinfo {author} {\bibfnamefont {Y.}~\bibnamefont
  {Zhao}}, \bibinfo {author} {\bibfnamefont {K.}~\bibnamefont {Xu}}, \bibinfo
  {author} {\bibfnamefont {F.}~\bibnamefont {Pan}}, \bibinfo {author}
  {\bibfnamefont {C.}~\bibnamefont {Zhou}}, \bibinfo {author} {\bibfnamefont
  {F.}~\bibnamefont {Zhou}}, \ and\ \bibinfo {author} {\bibfnamefont
  {Y.}~\bibnamefont {Chai}},\ }\href {\doibase 10.1002/adfm.201603484}
  {\bibfield  {journal} {\bibinfo  {journal} {Adv. Funct. Mater.}\ }\textbf
  {\bibinfo {volume} {27}},\ \bibinfo {pages} {1603484} (\bibinfo {year}
  {2017})}\BibitemShut {NoStop}%
\bibitem [{\citenamefont {Banszerus}\ \emph {et~al.}(2015)\citenamefont
  {Banszerus}, \citenamefont {Schmitz}, \citenamefont {Engels}, \citenamefont
  {Dauber}, \citenamefont {Oellers}, \citenamefont {Haupt}, \citenamefont
  {Watanabe}, \citenamefont {Taniguchi}, \citenamefont {Beschoten},\ and\
  \citenamefont {Stampfer}}]{Banszerus2015:SA}%
  \BibitemOpen
  \bibfield  {author} {\bibinfo {author} {\bibfnamefont {L.}~\bibnamefont
  {Banszerus}}, \bibinfo {author} {\bibfnamefont {M.}~\bibnamefont {Schmitz}},
  \bibinfo {author} {\bibfnamefont {S.}~\bibnamefont {Engels}}, \bibinfo
  {author} {\bibfnamefont {J.}~\bibnamefont {Dauber}}, \bibinfo {author}
  {\bibfnamefont {M.}~\bibnamefont {Oellers}}, \bibinfo {author} {\bibfnamefont
  {F.}~\bibnamefont {Haupt}}, \bibinfo {author} {\bibfnamefont
  {K.}~\bibnamefont {Watanabe}}, \bibinfo {author} {\bibfnamefont
  {T.}~\bibnamefont {Taniguchi}}, \bibinfo {author} {\bibfnamefont
  {B.}~\bibnamefont {Beschoten}}, \ and\ \bibinfo {author} {\bibfnamefont
  {C.}~\bibnamefont {Stampfer}},\ }\href {\doibase 10.1126/sciadv.1500222}
  {\bibfield  {journal} {\bibinfo  {journal} {Sci. Adv.}\ }\textbf {\bibinfo
  {volume} {1}},\ \bibinfo {pages} {1} (\bibinfo {year} {2015})}\BibitemShut
  {NoStop}%
\bibitem [{\citenamefont {Gmitra}\ \emph {et~al.}(2009)\citenamefont {Gmitra},
  \citenamefont {Konschuh}, \citenamefont {Ertler}, \citenamefont
  {Ambrosch-Draxl},\ and\ \citenamefont {Fabian}}]{Gmitra2009:PRB}%
  \BibitemOpen
  \bibfield  {author} {\bibinfo {author} {\bibfnamefont {M.}~\bibnamefont
  {Gmitra}}, \bibinfo {author} {\bibfnamefont {S.}~\bibnamefont {Konschuh}},
  \bibinfo {author} {\bibfnamefont {C.}~\bibnamefont {Ertler}}, \bibinfo
  {author} {\bibfnamefont {C.}~\bibnamefont {Ambrosch-Draxl}}, \ and\ \bibinfo
  {author} {\bibfnamefont {J.}~\bibnamefont {Fabian}},\ }\href {\doibase
  10.1103/PhysRevB.80.235431} {\bibfield  {journal} {\bibinfo  {journal} {Phys.
  Rev. B}\ }\textbf {\bibinfo {volume} {80}},\ \bibinfo {pages} {235431}
  (\bibinfo {year} {2009})}\BibitemShut {NoStop}%
\bibitem [{\citenamefont {Gmitra}\ and\ \citenamefont
  {Fabian}(2015)}]{Gmitra2015:PRB}%
  \BibitemOpen
  \bibfield  {author} {\bibinfo {author} {\bibfnamefont {M.}~\bibnamefont
  {Gmitra}}\ and\ \bibinfo {author} {\bibfnamefont {J.}~\bibnamefont
  {Fabian}},\ }\href {\doibase 10.1103/PhysRevB.92.155403} {\bibfield
  {journal} {\bibinfo  {journal} {Phys. Rev. B}\ }\textbf {\bibinfo {volume}
  {92}},\ \bibinfo {pages} {155403} (\bibinfo {year} {2015})}\BibitemShut
  {NoStop}%
\bibitem [{\citenamefont {Gmitra}\ \emph {et~al.}(2016)\citenamefont {Gmitra},
  \citenamefont {Kochan}, \citenamefont {H\"ogl},\ and\ \citenamefont
  {Fabian}}]{Gmitra2016:PRB}%
  \BibitemOpen
  \bibfield  {author} {\bibinfo {author} {\bibfnamefont {M.}~\bibnamefont
  {Gmitra}}, \bibinfo {author} {\bibfnamefont {D.}~\bibnamefont {Kochan}},
  \bibinfo {author} {\bibfnamefont {P.}~\bibnamefont {H\"ogl}}, \ and\ \bibinfo
  {author} {\bibfnamefont {J.}~\bibnamefont {Fabian}},\ }\href {\doibase
  10.1103/PhysRevB.93.155104} {\bibfield  {journal} {\bibinfo  {journal} {Phys.
  Rev. B}\ }\textbf {\bibinfo {volume} {93}},\ \bibinfo {pages} {155104}
  (\bibinfo {year} {2016})}\BibitemShut {NoStop}%
\bibitem [{\citenamefont {Zollner}\ \emph {et~al.}(2016)\citenamefont
  {Zollner}, \citenamefont {Gmitra}, \citenamefont {Frank},\ and\ \citenamefont
  {Fabian}}]{Zollner2016:PRB}%
  \BibitemOpen
  \bibfield  {author} {\bibinfo {author} {\bibfnamefont {K.}~\bibnamefont
  {Zollner}}, \bibinfo {author} {\bibfnamefont {M.}~\bibnamefont {Gmitra}},
  \bibinfo {author} {\bibfnamefont {T.}~\bibnamefont {Frank}}, \ and\ \bibinfo
  {author} {\bibfnamefont {J.}~\bibnamefont {Fabian}},\ }\href {\doibase
  10.1103/PhysRevB.94.155441} {\bibfield  {journal} {\bibinfo  {journal} {Phys.
  Rev. B}\ }\textbf {\bibinfo {volume} {94}},\ \bibinfo {pages} {155441}
  (\bibinfo {year} {2016})}\BibitemShut {NoStop}%
\bibitem [{\citenamefont {Cummings}\ \emph {et~al.}(2017)\citenamefont
  {Cummings}, \citenamefont {Garcia}, \citenamefont {Fabian},\ and\
  \citenamefont {Roche}}]{Cummings2017:PRL}%
  \BibitemOpen
  \bibfield  {author} {\bibinfo {author} {\bibfnamefont {A.~W.}\ \bibnamefont
  {Cummings}}, \bibinfo {author} {\bibfnamefont {J.~H.}\ \bibnamefont
  {Garcia}}, \bibinfo {author} {\bibfnamefont {J.}~\bibnamefont {Fabian}}, \
  and\ \bibinfo {author} {\bibfnamefont {S.}~\bibnamefont {Roche}},\ }\href
  {\doibase 10.1103/PhysRevLett.119.206601} {\bibfield  {journal} {\bibinfo
  {journal} {Phys. Rev. Lett.}\ }\textbf {\bibinfo {volume} {119}},\ \bibinfo
  {pages} {206601} (\bibinfo {year} {2017})}\BibitemShut {NoStop}%
\bibitem [{\citenamefont {Zollner}\ \emph {et~al.}(2018)\citenamefont
  {Zollner}, \citenamefont {Gmitra},\ and\ \citenamefont
  {Fabian}}]{Zollner2017:NJP}%
  \BibitemOpen
  \bibfield  {author} {\bibinfo {author} {\bibfnamefont {K.}~\bibnamefont
  {Zollner}}, \bibinfo {author} {\bibfnamefont {M.}~\bibnamefont {Gmitra}}, \
  and\ \bibinfo {author} {\bibfnamefont {J.}~\bibnamefont {Fabian}},\ }\href
  {\doibase 10.1088/1367-2630/aace51} {\bibfield  {journal} {\bibinfo
  {journal} {New J. Phys.}\ }\textbf {\bibinfo {volume} {20}},\ \bibinfo
  {pages} {073007} (\bibinfo {year} {2018})},\ \Eprint
  {http://arxiv.org/abs/1710.08117} {1710.08117} \BibitemShut {NoStop}%
\bibitem [{\citenamefont {Gmitra}\ and\ \citenamefont
  {Fabian}(2017)}]{Gmitra2017:PRL}%
  \BibitemOpen
  \bibfield  {author} {\bibinfo {author} {\bibfnamefont {M.}~\bibnamefont
  {Gmitra}}\ and\ \bibinfo {author} {\bibfnamefont {J.}~\bibnamefont
  {Fabian}},\ }\href {\doibase 10.1103/PhysRevLett.119.146401} {\bibfield
  {journal} {\bibinfo  {journal} {Phys. Rev. Lett.}\ }\textbf {\bibinfo
  {volume} {119}},\ \bibinfo {pages} {146401} (\bibinfo {year}
  {2017})}\BibitemShut {NoStop}%
\bibitem [{\citenamefont {Khoo}\ \emph {et~al.}(2017)\citenamefont {Khoo},
  \citenamefont {Morpurgo},\ and\ \citenamefont {Levitov}}]{Khoo2017:NL}%
  \BibitemOpen
  \bibfield  {author} {\bibinfo {author} {\bibfnamefont {J.~Y.}\ \bibnamefont
  {Khoo}}, \bibinfo {author} {\bibfnamefont {A.~F.}\ \bibnamefont {Morpurgo}},
  \ and\ \bibinfo {author} {\bibfnamefont {L.}~\bibnamefont {Levitov}},\ }\href
  {\doibase 10.1021/acs.nanolett.7b03604} {\bibfield  {journal} {\bibinfo
  {journal} {Nano Lett.}\ }\textbf {\bibinfo {volume} {17}},\ \bibinfo {pages}
  {7003} (\bibinfo {year} {2017})}\BibitemShut {NoStop}%
\bibitem [{\citenamefont {Zalic}\ \emph {et~al.}(2017)\citenamefont {Zalic},
  \citenamefont {Dvir},\ and\ \citenamefont {Steinberg}}]{Zalic2017:PRB}%
  \BibitemOpen
  \bibfield  {author} {\bibinfo {author} {\bibfnamefont {A.}~\bibnamefont
  {Zalic}}, \bibinfo {author} {\bibfnamefont {T.}~\bibnamefont {Dvir}}, \ and\
  \bibinfo {author} {\bibfnamefont {H.}~\bibnamefont {Steinberg}},\ }\href
  {\doibase 10.1103/PhysRevB.96.075104} {\bibfield  {journal} {\bibinfo
  {journal} {Phys. Rev. B}\ }\textbf {\bibinfo {volume} {96}},\ \bibinfo
  {pages} {075104} (\bibinfo {year} {2017})}\BibitemShut {NoStop}%
\bibitem [{\citenamefont {Song}\ \emph {et~al.}(2010)\citenamefont {Song},
  \citenamefont {Wang}, \citenamefont {Jiang}, \citenamefont {Zhang},
  \citenamefont {Chang}, \citenamefont {Wang}, \citenamefont {He},
  \citenamefont {Chen}, \citenamefont {Jia}, \citenamefont {Wang},
  \citenamefont {Fang}, \citenamefont {Dai}, \citenamefont {Xie}, \citenamefont
  {Qi}, \citenamefont {Zhang}, \citenamefont {Xue},\ and\ \citenamefont
  {Ma}}]{Song2010:APL}%
  \BibitemOpen
  \bibfield  {author} {\bibinfo {author} {\bibfnamefont {C.-L.}\ \bibnamefont
  {Song}}, \bibinfo {author} {\bibfnamefont {Y.-L.}\ \bibnamefont {Wang}},
  \bibinfo {author} {\bibfnamefont {Y.-P.}\ \bibnamefont {Jiang}}, \bibinfo
  {author} {\bibfnamefont {Y.}~\bibnamefont {Zhang}}, \bibinfo {author}
  {\bibfnamefont {C.-Z.}\ \bibnamefont {Chang}}, \bibinfo {author}
  {\bibfnamefont {L.}~\bibnamefont {Wang}}, \bibinfo {author} {\bibfnamefont
  {K.}~\bibnamefont {He}}, \bibinfo {author} {\bibfnamefont {X.}~\bibnamefont
  {Chen}}, \bibinfo {author} {\bibfnamefont {J.-F.}\ \bibnamefont {Jia}},
  \bibinfo {author} {\bibfnamefont {Y.}~\bibnamefont {Wang}}, \bibinfo {author}
  {\bibfnamefont {Z.}~\bibnamefont {Fang}}, \bibinfo {author} {\bibfnamefont
  {X.}~\bibnamefont {Dai}}, \bibinfo {author} {\bibfnamefont {X.-C.}\
  \bibnamefont {Xie}}, \bibinfo {author} {\bibfnamefont {X.-L.}\ \bibnamefont
  {Qi}}, \bibinfo {author} {\bibfnamefont {S.-C.}\ \bibnamefont {Zhang}},
  \bibinfo {author} {\bibfnamefont {Q.-K.}\ \bibnamefont {Xue}}, \ and\
  \bibinfo {author} {\bibfnamefont {X.}~\bibnamefont {Ma}},\ }\href {\doibase
  10.1063/1.3494595} {\bibfield  {journal} {\bibinfo  {journal} {Applied
  Physics Letters}\ }\textbf {\bibinfo {volume} {97}},\ \bibinfo {pages}
  {143118} (\bibinfo {year} {2010})}\BibitemShut {NoStop}%
\bibitem [{\citenamefont {Dang}\ \emph {et~al.}(2010)\citenamefont {Dang},
  \citenamefont {Peng}, \citenamefont {Li}, \citenamefont {Wang},\ and\
  \citenamefont {Liu}}]{Dang2010:NL}%
  \BibitemOpen
  \bibfield  {author} {\bibinfo {author} {\bibfnamefont {W.}~\bibnamefont
  {Dang}}, \bibinfo {author} {\bibfnamefont {H.}~\bibnamefont {Peng}}, \bibinfo
  {author} {\bibfnamefont {H.}~\bibnamefont {Li}}, \bibinfo {author}
  {\bibfnamefont {P.}~\bibnamefont {Wang}}, \ and\ \bibinfo {author}
  {\bibfnamefont {Z.}~\bibnamefont {Liu}},\ }\href {\doibase 10.1021/nl100938e}
  {\bibfield  {journal} {\bibinfo  {journal} {Nano Lett.}\ }\textbf {\bibinfo
  {volume} {10}},\ \bibinfo {pages} {2870} (\bibinfo {year}
  {2010})}\BibitemShut {NoStop}%
\bibitem [{\citenamefont {Steinberg}\ \emph {et~al.}(2015)\citenamefont
  {Steinberg}, \citenamefont {Orona}, \citenamefont {Fatemi}, \citenamefont
  {Sanchez-Yamagishi}, \citenamefont {Watanabe}, \citenamefont {Taniguchi},\
  and\ \citenamefont {Jarillo-Herrero}}]{Steinberg2015:PRB}%
  \BibitemOpen
  \bibfield  {author} {\bibinfo {author} {\bibfnamefont {H.}~\bibnamefont
  {Steinberg}}, \bibinfo {author} {\bibfnamefont {L.~A.}\ \bibnamefont
  {Orona}}, \bibinfo {author} {\bibfnamefont {V.}~\bibnamefont {Fatemi}},
  \bibinfo {author} {\bibfnamefont {J.~D.}\ \bibnamefont {Sanchez-Yamagishi}},
  \bibinfo {author} {\bibfnamefont {K.}~\bibnamefont {Watanabe}}, \bibinfo
  {author} {\bibfnamefont {T.}~\bibnamefont {Taniguchi}}, \ and\ \bibinfo
  {author} {\bibfnamefont {P.}~\bibnamefont {Jarillo-Herrero}},\ }\href
  {\doibase 10.1103/PhysRevB.92.241409} {\bibfield  {journal} {\bibinfo
  {journal} {Phys. Rev. B}\ }\textbf {\bibinfo {volume} {92}},\ \bibinfo
  {pages} {241409} (\bibinfo {year} {2015})}\BibitemShut {NoStop}%
\bibitem [{\citenamefont {Bahn}\ and\ \citenamefont {Jacobsen}(2002)}]{ASE}%
  \BibitemOpen
  \bibfield  {author} {\bibinfo {author} {\bibfnamefont {S.~R.}\ \bibnamefont
  {Bahn}}\ and\ \bibinfo {author} {\bibfnamefont {K.~W.}\ \bibnamefont
  {Jacobsen}},\ }\href {\doibase 10.1109/5992.998641} {\bibfield  {journal}
  {\bibinfo  {journal} {Comput. Sci. Eng.}\ }\textbf {\bibinfo {volume} {4}},\
  \bibinfo {pages} {56} (\bibinfo {year} {2002})}\BibitemShut {NoStop}%
\bibitem [{\citenamefont {Nakajima}(1963)}]{Nakajima1963:JPCS}%
  \BibitemOpen
  \bibfield  {author} {\bibinfo {author} {\bibfnamefont {S.}~\bibnamefont
  {Nakajima}},\ }\href {\doibase https://doi.org/10.1016/0022-3697(63)90207-5}
  {\bibfield  {journal} {\bibinfo  {journal} {Journal of Physics and Chemistry
  of Solids}\ }\textbf {\bibinfo {volume} {24}},\ \bibinfo {pages} {479 }
  (\bibinfo {year} {1963})}\BibitemShut {NoStop}%
\bibitem [{\citenamefont {Baskin}\ and\ \citenamefont
  {Meyer}(1955)}]{Baskin1955:PR}%
  \BibitemOpen
  \bibfield  {author} {\bibinfo {author} {\bibfnamefont {Y.}~\bibnamefont
  {Baskin}}\ and\ \bibinfo {author} {\bibfnamefont {L.}~\bibnamefont {Meyer}},\
  }\href {\doibase 10.1103/PhysRev.100.544} {\bibfield  {journal} {\bibinfo
  {journal} {Phys. Rev.}\ }\textbf {\bibinfo {volume} {100}},\ \bibinfo {pages}
  {544} (\bibinfo {year} {1955})}\BibitemShut {NoStop}%
\bibitem [{\citenamefont {Momma}\ and\ \citenamefont {Izumi}(2011)}]{VESTA}%
  \BibitemOpen
  \bibfield  {author} {\bibinfo {author} {\bibfnamefont {K.}~\bibnamefont
  {Momma}}\ and\ \bibinfo {author} {\bibfnamefont {F.}~\bibnamefont {Izumi}},\
  }\href {\doibase 10.1107/S0021889811038970} {\bibfield  {journal} {\bibinfo
  {journal} {Journal of Applied Crystallography}\ }\textbf {\bibinfo {volume}
  {44}},\ \bibinfo {pages} {1272} (\bibinfo {year} {2011})}\BibitemShut
  {NoStop}%
\bibitem [{\citenamefont {Hohenberg}\ and\ \citenamefont
  {Kohn}(1964)}]{Hohenberg1964:PRB}%
  \BibitemOpen
  \bibfield  {author} {\bibinfo {author} {\bibfnamefont {P.}~\bibnamefont
  {Hohenberg}}\ and\ \bibinfo {author} {\bibfnamefont {W.}~\bibnamefont
  {Kohn}},\ }\href {\doibase 10.1103/PhysRev.136.B864} {\bibfield  {journal}
  {\bibinfo  {journal} {Phys. Rev.}\ }\textbf {\bibinfo {volume} {136}},\
  \bibinfo {pages} {B864} (\bibinfo {year} {1964})}\BibitemShut {NoStop}%
\bibitem [{\citenamefont {Giannozzi}\ and\ \citenamefont
  {et~al.}(2009)}]{Giannozzi2009:JPCM}%
  \BibitemOpen
  \bibfield  {author} {\bibinfo {author} {\bibfnamefont {P.}~\bibnamefont
  {Giannozzi}}\ and\ \bibinfo {author} {\bibnamefont {et~al.}},\ }\href@noop {}
  {\bibfield  {journal} {\bibinfo  {journal} {J. Phys.: Condens. Mat.}\
  }\textbf {\bibinfo {volume} {21}},\ \bibinfo {pages} {395502} (\bibinfo
  {year} {2009})}\BibitemShut {NoStop}%
\bibitem [{\citenamefont {Kresse}\ and\ \citenamefont
  {Joubert}(1999)}]{Kresse1999:PRB}%
  \BibitemOpen
  \bibfield  {author} {\bibinfo {author} {\bibfnamefont {G.}~\bibnamefont
  {Kresse}}\ and\ \bibinfo {author} {\bibfnamefont {D.}~\bibnamefont
  {Joubert}},\ }\href {\doibase 10.1103/PhysRevB.59.1758} {\bibfield  {journal}
  {\bibinfo  {journal} {Phys. Rev. B}\ }\textbf {\bibinfo {volume} {59}},\
  \bibinfo {pages} {1758} (\bibinfo {year} {1999})}\BibitemShut {NoStop}%
\bibitem [{\citenamefont {Perdew}\ \emph {et~al.}(1996)\citenamefont {Perdew},
  \citenamefont {Burke},\ and\ \citenamefont {Ernzerhof}}]{Perdew1996:PRL}%
  \BibitemOpen
  \bibfield  {author} {\bibinfo {author} {\bibfnamefont {J.~P.}\ \bibnamefont
  {Perdew}}, \bibinfo {author} {\bibfnamefont {K.}~\bibnamefont {Burke}}, \
  and\ \bibinfo {author} {\bibfnamefont {M.}~\bibnamefont {Ernzerhof}},\ }\href
  {\doibase 10.1103/PhysRevLett.77.3865} {\bibfield  {journal} {\bibinfo
  {journal} {Phys. Rev. Lett.}\ }\textbf {\bibinfo {volume} {77}},\ \bibinfo
  {pages} {3865} (\bibinfo {year} {1996})}\BibitemShut {NoStop}%
\bibitem [{\citenamefont {Grimme}(2006)}]{Grimme2006:JCC}%
  \BibitemOpen
  \bibfield  {author} {\bibinfo {author} {\bibfnamefont {S.}~\bibnamefont
  {Grimme}},\ }\href {\doibase 10.1002/jcc.20495} {\bibfield  {journal}
  {\bibinfo  {journal} {J. Comput. Chem.}\ }\textbf {\bibinfo {volume} {27}},\
  \bibinfo {pages} {1787} (\bibinfo {year} {2006})}\BibitemShut {NoStop}%
\bibitem [{\citenamefont {Barone}\ \emph {et~al.}(2009)\citenamefont {Barone},
  \citenamefont {Casarin}, \citenamefont {Forrer}, \citenamefont {Pavone},
  \citenamefont {Sambi},\ and\ \citenamefont {Vittadini}}]{Barone2009:JCC}%
  \BibitemOpen
  \bibfield  {author} {\bibinfo {author} {\bibfnamefont {V.}~\bibnamefont
  {Barone}}, \bibinfo {author} {\bibfnamefont {M.}~\bibnamefont {Casarin}},
  \bibinfo {author} {\bibfnamefont {D.}~\bibnamefont {Forrer}}, \bibinfo
  {author} {\bibfnamefont {M.}~\bibnamefont {Pavone}}, \bibinfo {author}
  {\bibfnamefont {M.}~\bibnamefont {Sambi}}, \ and\ \bibinfo {author}
  {\bibfnamefont {A.}~\bibnamefont {Vittadini}},\ }\href {\doibase
  10.1002/jcc.21112} {\bibfield  {journal} {\bibinfo  {journal} {J. Comput.
  Chem.}\ }\textbf {\bibinfo {volume} {30}},\ \bibinfo {pages} {934} (\bibinfo
  {year} {2009})}\BibitemShut {NoStop}%
\bibitem [{\citenamefont {Bengtsson}(1999)}]{Bengtsson1999:PRB}%
  \BibitemOpen
  \bibfield  {author} {\bibinfo {author} {\bibfnamefont {L.}~\bibnamefont
  {Bengtsson}},\ }\href {\doibase 10.1103/PhysRevB.59.12301} {\bibfield
  {journal} {\bibinfo  {journal} {Phys. Rev. B}\ }\textbf {\bibinfo {volume}
  {59}},\ \bibinfo {pages} {12301} (\bibinfo {year} {1999})}\BibitemShut
  {NoStop}%
\bibitem [{\citenamefont {Aguilera}\ \emph {et~al.}(2013)\citenamefont
  {Aguilera}, \citenamefont {Friedrich}, \citenamefont {Bihlmayer},\ and\
  \citenamefont {Bl\"ugel}}]{Aguilera2013:PRB}%
  \BibitemOpen
  \bibfield  {author} {\bibinfo {author} {\bibfnamefont {I.}~\bibnamefont
  {Aguilera}}, \bibinfo {author} {\bibfnamefont {C.}~\bibnamefont {Friedrich}},
  \bibinfo {author} {\bibfnamefont {G.}~\bibnamefont {Bihlmayer}}, \ and\
  \bibinfo {author} {\bibfnamefont {S.}~\bibnamefont {Bl\"ugel}},\ }\href
  {\doibase 10.1103/PhysRevB.88.045206} {\bibfield  {journal} {\bibinfo
  {journal} {Phys. Rev. B}\ }\textbf {\bibinfo {volume} {88}},\ \bibinfo
  {pages} {045206} (\bibinfo {year} {2013})}\BibitemShut {NoStop}%
\bibitem [{\citenamefont {Nechaev}\ \emph {et~al.}(2013)\citenamefont
  {Nechaev}, \citenamefont {Hatch}, \citenamefont {Bianchi}, \citenamefont
  {Guan}, \citenamefont {Friedrich}, \citenamefont {Aguilera}, \citenamefont
  {Mi}, \citenamefont {Iversen}, \citenamefont {Bl\"ugel}, \citenamefont
  {Hofmann},\ and\ \citenamefont {Chulkov}}]{Nechaev2013:PRB}%
  \BibitemOpen
  \bibfield  {author} {\bibinfo {author} {\bibfnamefont {I.~A.}\ \bibnamefont
  {Nechaev}}, \bibinfo {author} {\bibfnamefont {R.~C.}\ \bibnamefont {Hatch}},
  \bibinfo {author} {\bibfnamefont {M.}~\bibnamefont {Bianchi}}, \bibinfo
  {author} {\bibfnamefont {D.}~\bibnamefont {Guan}}, \bibinfo {author}
  {\bibfnamefont {C.}~\bibnamefont {Friedrich}}, \bibinfo {author}
  {\bibfnamefont {I.}~\bibnamefont {Aguilera}}, \bibinfo {author}
  {\bibfnamefont {J.~L.}\ \bibnamefont {Mi}}, \bibinfo {author} {\bibfnamefont
  {B.~B.}\ \bibnamefont {Iversen}}, \bibinfo {author} {\bibfnamefont
  {S.}~\bibnamefont {Bl\"ugel}}, \bibinfo {author} {\bibfnamefont
  {P.}~\bibnamefont {Hofmann}}, \ and\ \bibinfo {author} {\bibfnamefont
  {E.~V.}\ \bibnamefont {Chulkov}},\ }\href {\doibase
  10.1103/PhysRevB.87.121111} {\bibfield  {journal} {\bibinfo  {journal} {Phys.
  Rev. B}\ }\textbf {\bibinfo {volume} {87}},\ \bibinfo {pages} {121111}
  (\bibinfo {year} {2013})}\BibitemShut {NoStop}%
\bibitem [{\citenamefont {Yazyev}\ \emph {et~al.}(2012)\citenamefont {Yazyev},
  \citenamefont {Kioupakis}, \citenamefont {Moore},\ and\ \citenamefont
  {Louie}}]{Yazyev2012:PRB}%
  \BibitemOpen
  \bibfield  {author} {\bibinfo {author} {\bibfnamefont {O.~V.}\ \bibnamefont
  {Yazyev}}, \bibinfo {author} {\bibfnamefont {E.}~\bibnamefont {Kioupakis}},
  \bibinfo {author} {\bibfnamefont {J.~E.}\ \bibnamefont {Moore}}, \ and\
  \bibinfo {author} {\bibfnamefont {S.~G.}\ \bibnamefont {Louie}},\ }\href
  {\doibase 10.1103/PhysRevB.85.161101} {\bibfield  {journal} {\bibinfo
  {journal} {Phys. Rev. B}\ }\textbf {\bibinfo {volume} {85}},\ \bibinfo
  {pages} {161101} (\bibinfo {year} {2012})}\BibitemShut {NoStop}%
\bibitem [{\citenamefont {Kochan}\ \emph {et~al.}(2017)\citenamefont {Kochan},
  \citenamefont {Irmer},\ and\ \citenamefont {Fabian}}]{Kochan2017:PRB}%
  \BibitemOpen
  \bibfield  {author} {\bibinfo {author} {\bibfnamefont {D.}~\bibnamefont
  {Kochan}}, \bibinfo {author} {\bibfnamefont {S.}~\bibnamefont {Irmer}}, \
  and\ \bibinfo {author} {\bibfnamefont {J.}~\bibnamefont {Fabian}},\ }\href
  {\doibase 10.1103/PhysRevB.95.165415} {\bibfield  {journal} {\bibinfo
  {journal} {Phys. Rev. B}\ }\textbf {\bibinfo {volume} {95}},\ \bibinfo
  {pages} {165415} (\bibinfo {year} {2017})}\BibitemShut {NoStop}%
\bibitem [{\citenamefont {Zollner}\ \emph {et~al.}(2019)\citenamefont
  {Zollner}, \citenamefont {Gmitra},\ and\ \citenamefont
  {Fabian}}]{Zollner2019:PRB}%
  \BibitemOpen
  \bibfield  {author} {\bibinfo {author} {\bibfnamefont {K.}~\bibnamefont
  {Zollner}}, \bibinfo {author} {\bibfnamefont {M.}~\bibnamefont {Gmitra}}, \
  and\ \bibinfo {author} {\bibfnamefont {J.}~\bibnamefont {Fabian}},\ }\href
  {\doibase 10.1103/PhysRevB.99.125151} {\bibfield  {journal} {\bibinfo
  {journal} {Phys. Rev. B}\ }\textbf {\bibinfo {volume} {99}},\ \bibinfo
  {pages} {125151} (\bibinfo {year} {2019})}\BibitemShut {NoStop}%
\bibitem [{\citenamefont {Di~Sante}\ \emph {et~al.}(2019)\citenamefont
  {Di~Sante}, \citenamefont {Eck}, \citenamefont {Bauernfeind}, \citenamefont
  {Will}, \citenamefont {Thomale}, \citenamefont {Sch\"afer}, \citenamefont
  {Claessen},\ and\ \citenamefont {Sangiovanni}}]{Sante2019:PRB}%
  \BibitemOpen
  \bibfield  {author} {\bibinfo {author} {\bibfnamefont {D.}~\bibnamefont
  {Di~Sante}}, \bibinfo {author} {\bibfnamefont {P.}~\bibnamefont {Eck}},
  \bibinfo {author} {\bibfnamefont {M.}~\bibnamefont {Bauernfeind}}, \bibinfo
  {author} {\bibfnamefont {M.}~\bibnamefont {Will}}, \bibinfo {author}
  {\bibfnamefont {R.}~\bibnamefont {Thomale}}, \bibinfo {author} {\bibfnamefont
  {J.}~\bibnamefont {Sch\"afer}}, \bibinfo {author} {\bibfnamefont
  {R.}~\bibnamefont {Claessen}}, \ and\ \bibinfo {author} {\bibfnamefont
  {G.}~\bibnamefont {Sangiovanni}},\ }\href {\doibase
  10.1103/PhysRevB.99.035145} {\bibfield  {journal} {\bibinfo  {journal} {Phys.
  Rev. B}\ }\textbf {\bibinfo {volume} {99}},\ \bibinfo {pages} {035145}
  (\bibinfo {year} {2019})}\BibitemShut {NoStop}%
\bibitem [{\citenamefont {Kane}\ and\ \citenamefont
  {Mele}(2005)}]{Kane2005:PRL}%
  \BibitemOpen
  \bibfield  {author} {\bibinfo {author} {\bibfnamefont {C.~L.}\ \bibnamefont
  {Kane}}\ and\ \bibinfo {author} {\bibfnamefont {E.~J.}\ \bibnamefont
  {Mele}},\ }\href {\doibase 10.1103/PhysRevLett.95.146802} {\bibfield
  {journal} {\bibinfo  {journal} {Phys. Rev. Lett.}\ }\textbf {\bibinfo
  {volume} {95}},\ \bibinfo {pages} {146802} (\bibinfo {year}
  {2005})}\BibitemShut {NoStop}%
\bibitem [{\citenamefont {Konschuh}\ \emph {et~al.}(2012)\citenamefont
  {Konschuh}, \citenamefont {Gmitra}, \citenamefont {Kochan},\ and\
  \citenamefont {Fabian}}]{Konschuh2012:PRB}%
  \BibitemOpen
  \bibfield  {author} {\bibinfo {author} {\bibfnamefont {S.}~\bibnamefont
  {Konschuh}}, \bibinfo {author} {\bibfnamefont {M.}~\bibnamefont {Gmitra}},
  \bibinfo {author} {\bibfnamefont {D.}~\bibnamefont {Kochan}}, \ and\ \bibinfo
  {author} {\bibfnamefont {J.}~\bibnamefont {Fabian}},\ }\href {\doibase
  10.1103/PhysRevB.85.115423} {\bibfield  {journal} {\bibinfo  {journal} {Phys.
  Rev. B}\ }\textbf {\bibinfo {volume} {85}},\ \bibinfo {pages} {115423}
  (\bibinfo {year} {2012})}\BibitemShut {NoStop}%
\bibitem [{\citenamefont {Jin}\ and\ \citenamefont {Jhi}(2013)}]{Jin2013:PRB}%
  \BibitemOpen
  \bibfield  {author} {\bibinfo {author} {\bibfnamefont {K.-H.}\ \bibnamefont
  {Jin}}\ and\ \bibinfo {author} {\bibfnamefont {S.-H.}\ \bibnamefont {Jhi}},\
  }\href {\doibase 10.1103/PhysRevB.87.075442} {\bibfield  {journal} {\bibinfo
  {journal} {Phys. Rev. B}\ }\textbf {\bibinfo {volume} {87}},\ \bibinfo
  {pages} {075442} (\bibinfo {year} {2013})}\BibitemShut {NoStop}%
\bibitem [{\citenamefont {Frank}\ \emph {et~al.}(2018)\citenamefont {Frank},
  \citenamefont {H\"ogl}, \citenamefont {Gmitra}, \citenamefont {Kochan},\ and\
  \citenamefont {Fabian}}]{Frank2018:PRL}%
  \BibitemOpen
  \bibfield  {author} {\bibinfo {author} {\bibfnamefont {T.}~\bibnamefont
  {Frank}}, \bibinfo {author} {\bibfnamefont {P.}~\bibnamefont {H\"ogl}},
  \bibinfo {author} {\bibfnamefont {M.}~\bibnamefont {Gmitra}}, \bibinfo
  {author} {\bibfnamefont {D.}~\bibnamefont {Kochan}}, \ and\ \bibinfo {author}
  {\bibfnamefont {J.}~\bibnamefont {Fabian}},\ }\href {\doibase
  10.1103/PhysRevLett.120.156402} {\bibfield  {journal} {\bibinfo  {journal}
  {Phys. Rev. Lett.}\ }\textbf {\bibinfo {volume} {120}},\ \bibinfo {pages}
  {156402} (\bibinfo {year} {2018})}\BibitemShut {NoStop}%
\bibitem [{\citenamefont {Alsharari}\ \emph {et~al.}(2018)\citenamefont
  {Alsharari}, \citenamefont {Asmar},\ and\ \citenamefont
  {Ulloa}}]{Alsharari2018:PRB}%
  \BibitemOpen
  \bibfield  {author} {\bibinfo {author} {\bibfnamefont {A.~M.}\ \bibnamefont
  {Alsharari}}, \bibinfo {author} {\bibfnamefont {M.~M.}\ \bibnamefont
  {Asmar}}, \ and\ \bibinfo {author} {\bibfnamefont {S.~E.}\ \bibnamefont
  {Ulloa}},\ }\href {\doibase 10.1103/PhysRevB.97.241104} {\bibfield  {journal}
  {\bibinfo  {journal} {Phys. Rev. B}\ }\textbf {\bibinfo {volume} {97}},\
  \bibinfo {pages} {241104(R)} (\bibinfo {year} {2018})}\BibitemShut {NoStop}%
\bibitem [{\citenamefont {Frank}\ \emph {et~al.}(2016)\citenamefont {Frank},
  \citenamefont {Gmitra},\ and\ \citenamefont {Fabian}}]{Frank2016:PRB}%
  \BibitemOpen
  \bibfield  {author} {\bibinfo {author} {\bibfnamefont {T.}~\bibnamefont
  {Frank}}, \bibinfo {author} {\bibfnamefont {M.}~\bibnamefont {Gmitra}}, \
  and\ \bibinfo {author} {\bibfnamefont {J.}~\bibnamefont {Fabian}},\ }\href
  {\doibase 10.1103/PhysRevB.93.155142} {\bibfield  {journal} {\bibinfo
  {journal} {Phys. Rev. B}\ }\textbf {\bibinfo {volume} {93}},\ \bibinfo
  {pages} {155142} (\bibinfo {year} {2016})}\BibitemShut {NoStop}%
\bibitem [{\citenamefont {Zhang}\ \emph {et~al.}(2018)\citenamefont {Zhang},
  \citenamefont {Zhao}, \citenamefont {Zhou}, \citenamefont {Xue},
  \citenamefont {Ma},\ and\ \citenamefont {Yang}}]{Zhang2018:PRB}%
  \BibitemOpen
  \bibfield  {author} {\bibinfo {author} {\bibfnamefont {J.}~\bibnamefont
  {Zhang}}, \bibinfo {author} {\bibfnamefont {B.}~\bibnamefont {Zhao}},
  \bibinfo {author} {\bibfnamefont {T.}~\bibnamefont {Zhou}}, \bibinfo {author}
  {\bibfnamefont {Y.}~\bibnamefont {Xue}}, \bibinfo {author} {\bibfnamefont
  {C.}~\bibnamefont {Ma}}, \ and\ \bibinfo {author} {\bibfnamefont
  {Z.}~\bibnamefont {Yang}},\ }\href {\doibase 10.1103/PhysRevB.97.085401}
  {\bibfield  {journal} {\bibinfo  {journal} {Phys. Rev. B}\ }\textbf {\bibinfo
  {volume} {97}},\ \bibinfo {pages} {085401} (\bibinfo {year}
  {2018})}\BibitemShut {NoStop}%
\bibitem [{\citenamefont {Zhang}\ \emph {et~al.}(2015)\citenamefont {Zhang},
  \citenamefont {Zhao}, \citenamefont {Yao},\ and\ \citenamefont
  {Yang}}]{Zhang2015:PRB}%
  \BibitemOpen
  \bibfield  {author} {\bibinfo {author} {\bibfnamefont {J.}~\bibnamefont
  {Zhang}}, \bibinfo {author} {\bibfnamefont {B.}~\bibnamefont {Zhao}},
  \bibinfo {author} {\bibfnamefont {Y.}~\bibnamefont {Yao}}, \ and\ \bibinfo
  {author} {\bibfnamefont {Z.}~\bibnamefont {Yang}},\ }\href {\doibase
  10.1103/PhysRevB.92.165418} {\bibfield  {journal} {\bibinfo  {journal} {Phys.
  Rev. B}\ }\textbf {\bibinfo {volume} {92}},\ \bibinfo {pages} {165418}
  (\bibinfo {year} {2015})}\BibitemShut {NoStop}%
\bibitem [{\citenamefont {Ren}\ \emph {et~al.}(2011)\citenamefont {Ren},
  \citenamefont {Taskin}, \citenamefont {Sasaki}, \citenamefont {Segawa},\ and\
  \citenamefont {Ando}}]{Ren2011:PRB}%
  \BibitemOpen
  \bibfield  {author} {\bibinfo {author} {\bibfnamefont {Z.}~\bibnamefont
  {Ren}}, \bibinfo {author} {\bibfnamefont {A.~A.}\ \bibnamefont {Taskin}},
  \bibinfo {author} {\bibfnamefont {S.}~\bibnamefont {Sasaki}}, \bibinfo
  {author} {\bibfnamefont {K.}~\bibnamefont {Segawa}}, \ and\ \bibinfo {author}
  {\bibfnamefont {Y.}~\bibnamefont {Ando}},\ }\href {\doibase
  10.1103/PhysRevB.84.165311} {\bibfield  {journal} {\bibinfo  {journal} {Phys.
  Rev. B}\ }\textbf {\bibinfo {volume} {84}},\ \bibinfo {pages} {165311}
  (\bibinfo {year} {2011})}\BibitemShut {NoStop}%
\bibitem [{\citenamefont {Arakane}\ \emph {et~al.}(2012)\citenamefont
  {Arakane}, \citenamefont {Sato}, \citenamefont {Souma}, \citenamefont
  {Kosaka}, \citenamefont {Nakayama}, \citenamefont {Komatsu}, \citenamefont
  {Takahashi}, \citenamefont {Ren}, \citenamefont {Segawa},\ and\ \citenamefont
  {Ando}}]{Arakane2012:NC}%
  \BibitemOpen
  \bibfield  {author} {\bibinfo {author} {\bibfnamefont {T.}~\bibnamefont
  {Arakane}}, \bibinfo {author} {\bibfnamefont {T.}~\bibnamefont {Sato}},
  \bibinfo {author} {\bibfnamefont {S.}~\bibnamefont {Souma}}, \bibinfo
  {author} {\bibfnamefont {K.}~\bibnamefont {Kosaka}}, \bibinfo {author}
  {\bibfnamefont {K.}~\bibnamefont {Nakayama}}, \bibinfo {author}
  {\bibfnamefont {M.}~\bibnamefont {Komatsu}}, \bibinfo {author} {\bibfnamefont
  {T.}~\bibnamefont {Takahashi}}, \bibinfo {author} {\bibfnamefont
  {Z.}~\bibnamefont {Ren}}, \bibinfo {author} {\bibfnamefont {K.}~\bibnamefont
  {Segawa}}, \ and\ \bibinfo {author} {\bibfnamefont {Y.}~\bibnamefont
  {Ando}},\ }\href {\doibase 10.1038/ncomms1639} {\bibfield  {journal}
  {\bibinfo  {journal} {Nat. Commun.}\ }\textbf {\bibinfo {volume} {3}},\
  \bibinfo {pages} {636} (\bibinfo {year} {2012})}\BibitemShut {NoStop}%
\end{thebibliography}%

\end{document}